\documentclass[a4paper,11pt]{article}
\pdfoutput=1
\usepackage{jheppub}

\usepackage{amsmath}
\usepackage{mathtools}
\usepackage{multirow}
\usepackage{array}
\usepackage{bm}

\usepackage{graphicx}
\usepackage{lscape}
\usepackage{subfig}

\usepackage{hyperref}
\usepackage{booktabs}
\usepackage{float}

\usepackage{xpunctuate}
\newcommand{\be}{\begin{equation}}
\newcommand{\ee}{\end{equation}}
\newcommand{\bea}{\begin{eqnarray}}
\newcommand{\eea}{\end{eqnarray}}
\newcommand{\bei}{\begin{itemize}}
\newcommand{\eei}{\end{itemize}}
\newcommand{\bean}{\begin{eqnarray*}}
\newcommand{\eean}{\\end{eqnarray*}}
\newcommand{\nn}{\nonumber \\}

\def\eps{\epsilon}

\def\top #1{\mathcal{T}_{#1}}

\allowdisplaybreaks[1]

\newcommand{\gammaAB}[2]{\ensuremath\gamma}

\newcommand{\dA}{\ensuremath d\AA}

\newcommand{\dAk}[1]{\ensuremath \underbrace{\dA \ldots \dA}_{\text{#1 times}}}
\newcommand{\Den}{\ensuremath D}
\newcommand{\dd}{\ensuremath \mathrm{d}}
\DeclareMathOperator{\dlog}{\mathit{d}log}

\newcommand{\FF}{\ensuremath \text{F}}

\newcommand{\GG}{\ensuremath \text{I}}
\newcommand{\GGvec}{\ensuremath \mathbf{I}}

\newcommand{\MM}{\ensuremath \mathbb{M}}
\renewcommand{\AA}{\ensuremath \mathbb{A}}

\newcommand{\unipd}{Dipartimento di Fisica ed Astronomia, Universit\`a di Padova, Via Marzolo 8, 35131 Padova, Italy}
\newcommand{\uzh}{Department of Physics, University of Z{\"u}rich, CH-8057 Z{\"u}rich, Switzerland}
\newcommand{\pdinfn}{INFN, Sezione di Padova, Via Marzolo 8, 35131 Padova, Italy}
\newcommand{\miinfn}{INFN, Sezione di Milano, Via Celoria 16, 20133 Milano, Italy}
\newcommand{\ub}{Department of Physics, University at Buffalo, The State University of New York, Buffalo 14260, USA}

\newcommand{\dG}{dG}

\newcommand{\eg}{e.g\xperiod}
\newcommand{\ie}{i.e\xperiod}
\newcommand{\feyneps}{\varepsilon}
\newcommand{\ieps}{i\feyneps}

\renewcommand{\Im}{\operatorname{Im}}
\newcommand{\heaviside}{\theta}

\allowdisplaybreaks

\author[a]{Stefano Di Vita,}
\author[b]{Thomas Gehrmann,}
\author[c,d]{Stefano Laporta,}
\author[c,d]{Pierpaolo Mastrolia,}
\author[b]{Amedeo Primo,}
\author[e]{Ulrich Schubert}

\affiliation[a]{\miinfn}
\affiliation[b]{\uzh}
\affiliation[c]{\unipd}
\affiliation[d]{\pdinfn}
\affiliation[e]{\ub}

\emailAdd{stefano.divita@mi.infn.it}
\emailAdd{thomas.gehrmann@uzh.ch}
\emailAdd{stefano.laporta@pd.infn.it}
\emailAdd{pierpaolo.mastrolia@pd.infn.it}
\emailAdd{aprimo@physik.uzh.ch}
\emailAdd{ulrichsc@buffalo.edu}

\preprint{\texttt{ZU-TH 20/19}}

\title{Master integrals for the NNLO virtual corrections to $q \bar{q} \rightarrow t \bar{t}$ scattering in QCD: the non-planar graphs}


\abstract{
We complete the analytic evaluation of the master integrals for the two-loop
non-planar box diagrams contributing to the top-pair production in
the quark-initiated channel, at next-to-next-to-leading order in
QCD. The integrals are determined from their 
differential equations, which are cast into 
a canonical form using the Magnus exponential. The
analytic expressions of the 
Laurent series coefficients of the integrals
 are expressed  as combinations of generalized
polylogarithms, which we validate with several numerical checks. 
We discuss the analytic continuation of the planar and the non-planar master
integrals, which contribute to $q {\bar q} \to t {\bar t}$ in QCD, as well as to
the companion QED scattering processes $ e e \to \mu \mu$ and  $e \mu \to e \mu$. 
}

\begin{document}

\maketitle
\flushbottom
\section{Introduction}

The top quark is the heaviest elementary matter particle. Its large production rate at the CERN LHC enables precision studies, thereby 
testing the Standard Model of particle physics at an unprecedented level, and potentially uncovering indirect 
evidence for new physics effects. Precision measurements of top quark observables~\cite{Aad:2015mbv,Aaboud:2017fha,Khachatryan:2015oqa,Sirunyan:2018wem} must be confronted with 
equally precise theory predictions, thereby requiring the perturbation theory descriptions of these 
observables to be extended to high orders. 

The available second-order (next-to-next-to-leading order, NNLO) QCD corrections to the top quark pair production process 
(initially for the total cross section~\cite{Czakon:2013goa} and subsequently for differential distributions~\cite{Czakon:2014xsa,Czakon:2015owf,Czakon:2016dgf,Catani:2019iny}) 
deliver a competitive level of theory predictions, and are enabling a multitude of precision studies with 
top quark pairs. A key ingredient to these calculations are the two-loop QCD corrections to the matrix elements 
for top pair production in quark-antiquark annihilation and gluon fusion. While numerical 
representations of these two-loop matrix elements were derived already some time ago~\cite{Baernreuther:2013caa,Chen:2017jvi}, only partial 
analytical results are available for them up to now~\cite{Bonciani:2008az,Bonciani:2009nb,Bonciani:2010mn,Bonciani:2013ywa} (and used 
in partial validations~\cite{Abelof:2014fza,Abelof:2015lna} of the total NNLO 
cross section calculation). Such analytical results allow in-depth 
investigations into the structure of the matrix elements, enabling to investigate  limiting behaviors and 
analyticity structures, as well as providing an independent approach to their numerical evaluation. 
Up to now, full results for the two-loop top quark production matrix elements 
could not be derived due to incomplete knowledge on the 
relevant two-loop Feynman integrals. Important steps towards the evaluation of the missing two-loop functions have been recently undertaken in~\cite{vonManteuffel:2017hms,Adams:2018bsn,Adams:2018kez,Chen:2019zoy}.

In this work, we complete the task of determining all the non-planar two-loop functions that are
needed for the evaluation of the scattering amplitudes for the quark initiated channel 
$q {\bar q} \to t {\bar t}$ at NNLO in QCD. 
Owing to the large value of the top mass compared
to the mass of the two incoming quarks, which we suppose to have different flavor, we treat the latter as massless. 

The results of this paper represent an additional milestone of the research
program dedicated to the two-loop QCD and QED corrections to the interaction of two fermionic currents, that was initiated with the 
study of the muon-electron scattering, in the context of the {\sc MUonE}
experiment~\cite{Calame:2015fva,Abbiendi:2016xup}, which is currently under evaluation at CERN. 

By following the same the path of the calculation of two-loop integrals for $\mu e \to \mu e$
and the crossing-related processes considered in~\cite{Mastrolia:2017pfy,DiVita:2018nnh}, we adopt a consolidated strategy 
\cite{Henn:2013pwa,Argeri:2014qva}, which was
proven to be particularly effective in the context of multi-loop integrals
that involve multiple kinematic scales 
\cite{Argeri:2014qva,DiVita:2014pza,Bonciani:2016ypc,DiVita:2017xlr,Mastrolia:2017pfy,DiVita:2018nnh,Primo:2018zby}.
By means of integration-by-parts identities (IBPs)
\cite{Tkachov:1981wb,Chetyrkin:1981qh,Laporta:2001dd}, 
we identify a set
of 52 master integrals (MIs) that we evaluate analytically, through the
differential equations method 
\cite{Barucchi:1973zm,Kotikov:1990kg,Remiddi:1997ny,Gehrmann:1999as}. 
In particular, we consider an
initial set of MIs that obey a system of first-order
differential equations (DEQs) in two independent kinematic variables
that is {\it linear} in the space-time dimensions $d$. The system is
subsequently cast in {\it canonical form} \cite{Henn:2013pwa} by means of the Magnus
exponential matrix \cite{Argeri:2014qva}. The matrix associated to the canonical
system, where the dependence on $(d-4)$ is factorized from the
kinematics, is a logarithmic differential form, which -- once parametrized in terms of suitable
variables -- has a polynomial alphabet, constituted of eleven letters.
Therefore, the canonical MIs are found to admit a Taylor series representation
around $d = 4$, 
whose coefficients are combinations of generalized polylogarithms
(GPLs) \cite{Goncharov:polylog,Remiddi:1999ew,Gehrmann:2001pz,Vollinga:2004sn}. 
The otherwise unknown integration constants are determined either from
the knowledge of the analytic expression of the MIs in special
kinematic configurations or by imposing their regularity at
the pseudo-thresholds of the DEQs. 
Finally, we show how the MIs, that we initially compute in an unphysical region, can be analytically continued to the
top-pair production region.  Due to the non-trivial structure of their branch-cuts, the analytic continuation of the two-loop functions
considered in this paper represents a paradigmatic case, that can be useful for the study of other planar and non-planar diagrams that involve massive
particles.
As a byproduct of the current analysis, we obtain the analytic continuation to the physical region of the functions required for the
$\mu e \to \mu e$ and $e e \to \mu \mu$ scattering in QED
\cite{Mastrolia:2017pfy,DiVita:2018nnh}\footnote{The evaluation of the
master integrals for the di-muon production in lepton-pair scattering,
within the physical region, has been recently considered in \cite{Lee:2019lno}.}.

In the completion of this calculation, we
benefited from publicly available software dedicated to multi-loop calculus.
The IBPs decomposition and the generation of the dimension-shifting identities and DEQs obeyed by the MIs
have been performed with the packages {\tt Kira} \cite{Maierhoefer2018a}, {\tt LiteRed} \cite{Lee:2012cn,Lee:2013mka} and {\tt Reduze} 
\cite{vonManteuffel:2012np,vonManteuffel:2014qoa}.
The analytic expressions of the MIs, which we evaluate numerically
with the help of {\tt GiNaC} \cite{Bauer:2000cp}, were successfully tested against the
numerical values provided either by {\tt SecDec}
\cite{Borowka:2015mxa} or, for the most complicated 7-denominator topologies, by an in-house algorithm. 

Beside these important validations, our results have been successfully compared
against the computation of the master integrals relevant to the same integral topologies expressed in a 
different basis set, 
independently obtained by Becchetti, Bonciani, Casconi, Ferroglia, Lavacca and
von Manteuffel~\cite{Becchetti:2019tjy}, published in tandem to the current manuscript.

The remainder of the paper is organized as follows: in section~\ref{sec:Notation} we set our notation and conventions for the non-planar two loop functions.
In section~\ref{sec:DEQ}, we present the system of DEQs obeyed by the MIs and their solutions in the unphysical region. In section~\ref{sec:Analytic}, we study the analytic continuation of the MIs to the physical region. Finally, in section~\ref{sec:Numerics}, we discuss the numerical evaluation of the 7-denominator integrals. Appendix~\ref{sec:gMIcoeff} contains the coefficients of the linear combinations of MIs that satisfy a canonical system of DEQs.

The analytic expressions of the considered MIs are given in the ancillary files accompanying the \texttt{arXiv} version of this publication.

%%%%%%%%%%%%%%%%%%
%%% Local Variables:
%%% TeX-master: "../main"
%%% End:

\section{The non-planar four-point topology}
\label{sec:Notation}

\begin{figure}[t]
  \centering
  \includegraphics{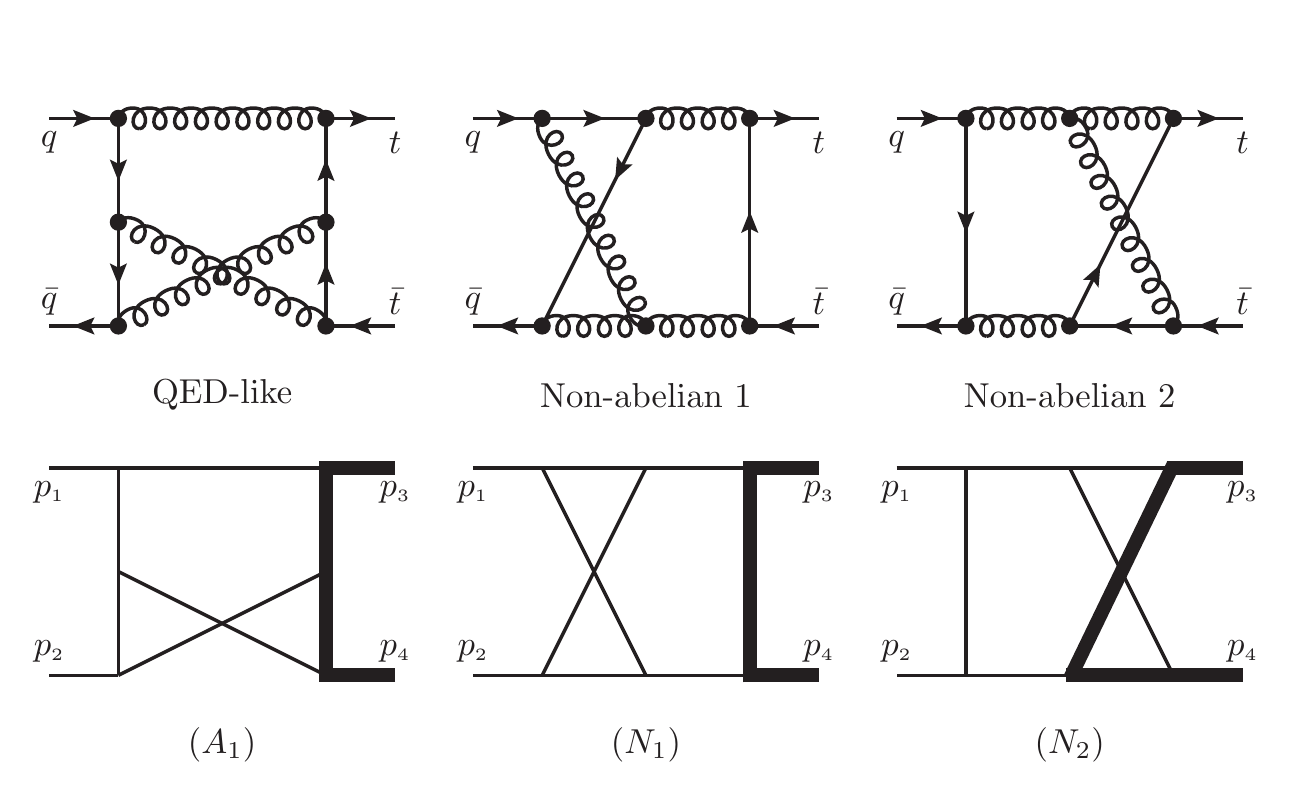}
  \caption{Representative non-planar diagrams contributing at
    two-loops to $q(p_1) + \bar{q}(p_2) \to t(p_3) + \bar{t}(p_4)$
    (top row), and the associated integral families (bottom
    row). Massive propagators and external legs are depicted using
    thick lines. The diagrams have been drawn with
    \texttt{axodraw2}~\cite{Collins2016}.}
  \label{fig:npgraphs}
\end{figure}

In this paper, we complete the determination of the Feynman integrals for the $q \bar{q} \rightarrow t \bar{t}$ scattering process
\begin{align}
  q(p_1) + \bar{q}(p_2)  \to    t(p_3) +  \bar{t}(p_4)\,,
\end{align}
with kinematics specified by
\begin{align}
p_1^2&=p_2^2=0\,,\quad p_3^2=p_4^2=m^2\,,\nn
  s = (p_1+p_2)^2\,,  \quad t&= (p_2-p_3)^2\,, \quad u=(p_1-p_3)^2=2m^2 -t-s  \, ,
\end{align}
where $m$ is the top quark mass. 
The full set of two-loop three-point integrals that are involved in the process has been known for some time~\cite{Gehrmann:2001ck,Bonciani:2003te,Bonciani:2003hc,Bonciani:2008wf,Asatrian:2008uk,Beneke:2008ei,Bell:2008ws,Huber:2009se}, as well all the relevant planar four-point functions~\cite{Bonciani:2008az,Bonciani:2009nb,Mastrolia:2017pfy}.
On top of these contributions,  the evaluation of the full double-virtual matrix element requires the computation of two-loop non-planar Feynman diagrams.
Representative non-planar diagrams are shown in the top row of figure~\ref{fig:npgraphs}. In the
bottom row, we also show the integral families onto which we map those
diagrams. Massive propagators and external legs are depicted using
thick lines.
The MIs for the QED-like family $A_1$ are already available in the
literature, as they have been studied in the context of the NNLO QED corrections to
$ee\mu\mu$ processes (with suitable redefinitions of the momenta and
of the Mandelstam invariants). In particular, they have been first
computed in~\cite{DiVita:2018nnh} (in an unphysical region, to be
analytically continued), and later in~\cite{Lee:2019lno} (directly in
the heavy-fermion-production kinematic region). As for the genuine QCD
contributions, the MIs for family $N_1$ have been computed
in~\cite{Manteuffel2013}, while the MIs for family $N_2$ are the subject of the
present publication.

In arbitrary $d$ dimensions, we parametrize the integrals of family $N_2$ as
\begin{gather}
I^{[d]}(n_1,\ldots,n_9)
\equiv
  \int \widetilde{\dd^d k_1}\widetilde{\dd^d k_2}\,
  \frac{1}{\Den_{1}^{n_1} \ldots \Den_{9}^{n_9}}\,,
\label{eq:def:ourintegrals}
\end{gather}
where $D_{i}$ are the inverse scalar propagators
\begin{gather}
\Den_1 = k_1^2,\quad
\Den_2 = k_2^2-m^2,\quad
\Den_3 = (k_2-p_3)^2,\quad
\Den_4 = (k_1-p_2)^2, \nn
\Den_5 = (k_1-p_3-p_4)^2,\quad
\Den_6 = (k_1-k_2)^2-m^2,\quad
\Den_7 = (k_1-k_2-p_4)^2, \nn
\Den_8 = (k_1-p_3)^2,\quad
\Den_9 = (k_2-p_2)^2\, ,
\label{eq:2Lfamily}
\end{gather}
with $k_1$ and $k_2$ denoting the loop momenta. 

The analytic calculation described in section~\ref{sec:DEQ} is performed by expanding the MIs around $d=4$, while the numerical evaluation presented in section~\ref{sec:Numerics} as a check of our results is carried over around $d=6$. Therefore, we set $\eps\equiv(d_*-d)/2$, where $d_*=4$ and $d_*=6$ according to the case. In addition, we define our integration measure as
\begin{align}
  \widetilde{\dd^{d}k} = {} & \frac{\dd^{d}k}{i \pi^{d/2}\,\Gamma\left(1+\epsilon\right) } \left(\frac{m^2}{\mu^2}\right)^\epsilon\,,
                                      \label{eq:intmeasure1}
\end{align}
where $\mu$ is the 't Hooft scale of dimensional regularization.
In this convention, the two-loop tadpole integral $\eps^2 I^{[4-2\eps]}(0,2,0,0,0,2,0,0,0)$ is normalized to $1$.
%%% Local Variables:
%%% TeX-master: "../main"
%%% End:

\section{Solution of the system of differential equations}
\label{sec:DEQ}
From the IBP reduction of the two-loop integrals belonging to the family defined in eqs.~\eqref{eq:def:ourintegrals}-\eqref{eq:2Lfamily} we identify a basis of 52 MIs. We determine the analytic expressions of the MIs by solving their DEQs in the
Mandelstam invariants $s$, $t$ and the top quark mass $m^2$. IBPs and DEQs have been derived by requiring the external legs to be \emph{on-shell}. 
The solution of the DEQs in terms of known functions is facilitated by the following reparametrization of the kinematics:
\begin{equation}
\frac{u-m^2}{t-m^2}=-\frac{  z^2}{w}, \qquad  \frac{s}{m^2} = -\frac{(1-w)^2}{w}\, ,
  \label{eq:varswz}
\end{equation}
where the constraint on the Mandelstam invariants $s+t+u=2m^2$ is understood. The dimensionless variables  ${w}$ and ${z}$ appearing in eq.~\eqref{eq:varswz} were already introduced in~\cite{DiVita:2018nnh} for the computation of the MIs of the non-planar family $A_1$ (modulo the relabeling $s\leftrightarrow t$). Also in this case, the above change of variables rationalizes the coefficients of the canonical DEQs, hence allowing to express the MIs in terms of GPLs.

 \begin{figure}[t]
  \centering
  \captionsetup[subfigure]{labelformat=empty}
  \subfloat[$\mathcal{T}_1$]{%
    \includegraphics[width=0.15\textwidth]{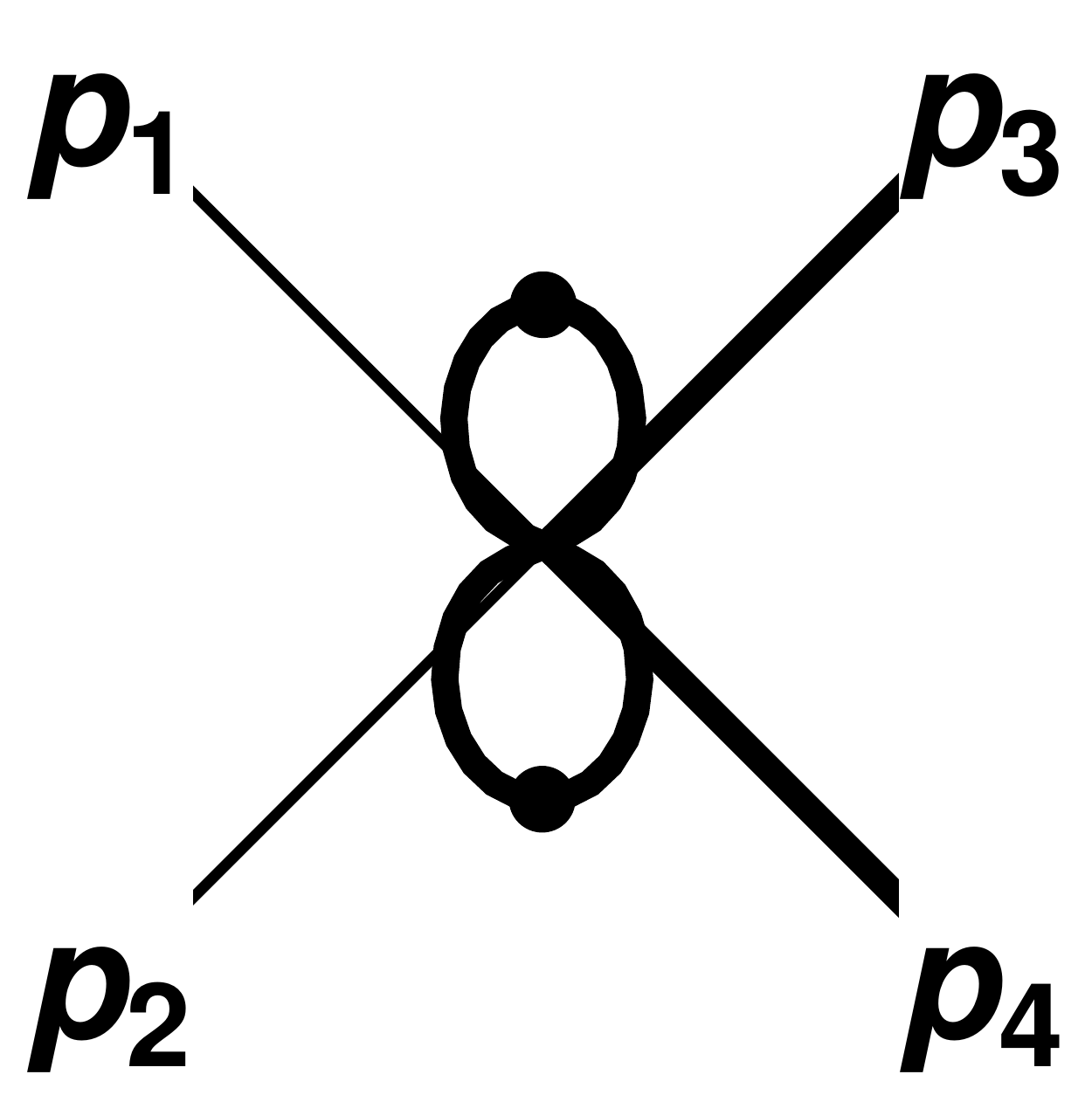}
  }
  \subfloat[$\mathcal{T}_2$]{%
    \includegraphics[width=0.15\textwidth]{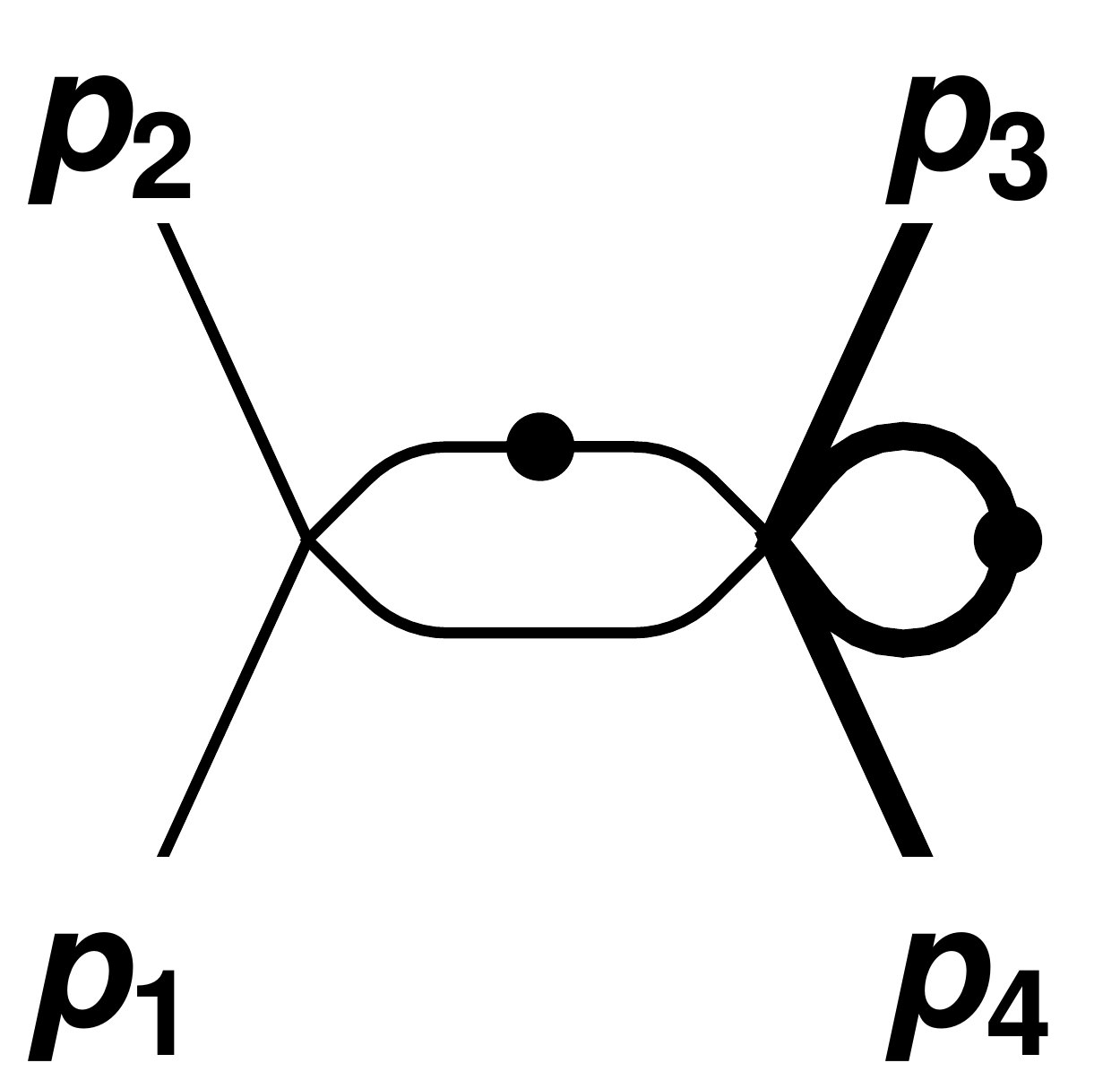}
  }
  \subfloat[$\mathcal{T}_3$]{%
    \includegraphics[width=0.15\textwidth]{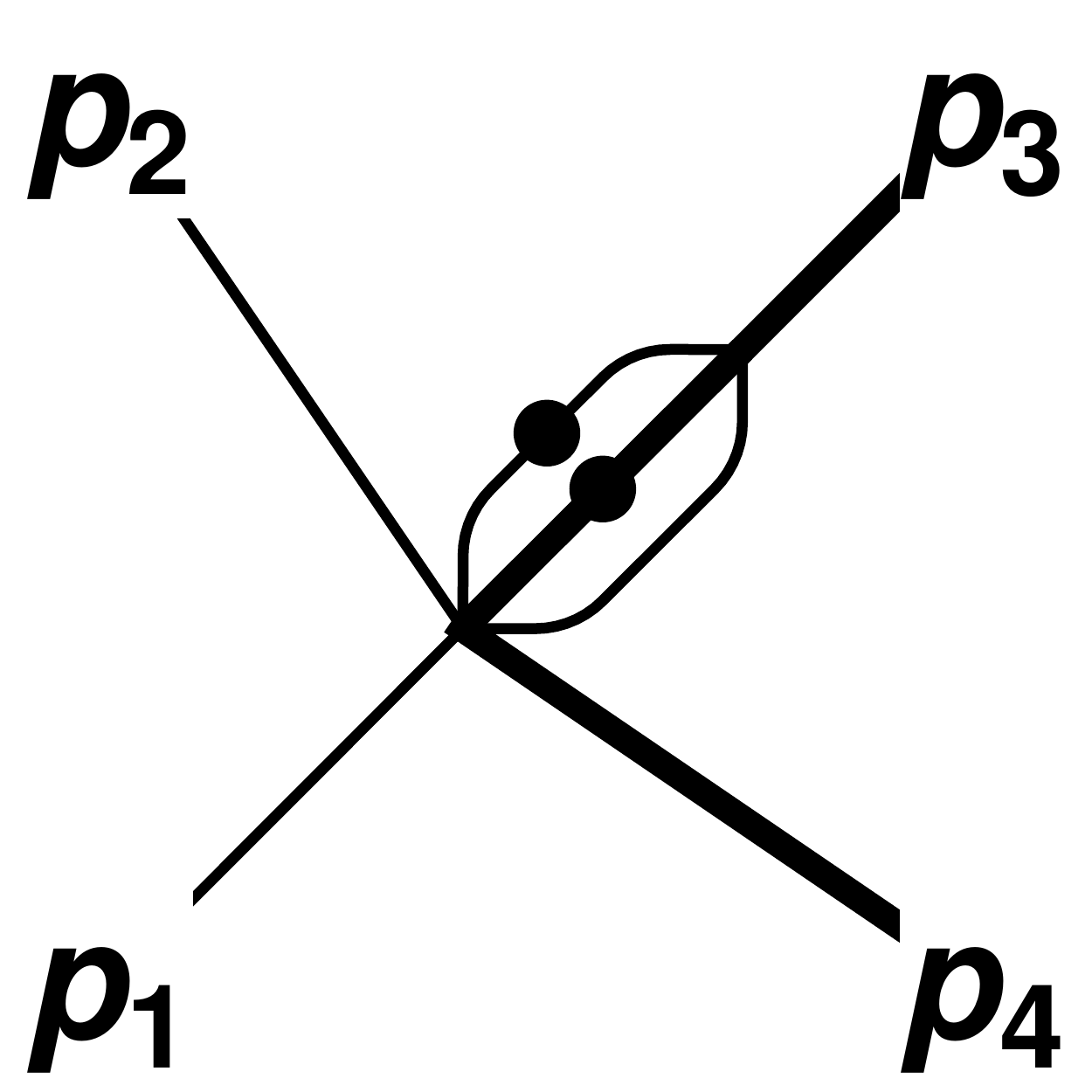}
  }
  \subfloat[$\mathcal{T}_4$]{%
    \includegraphics[width=0.15\textwidth]{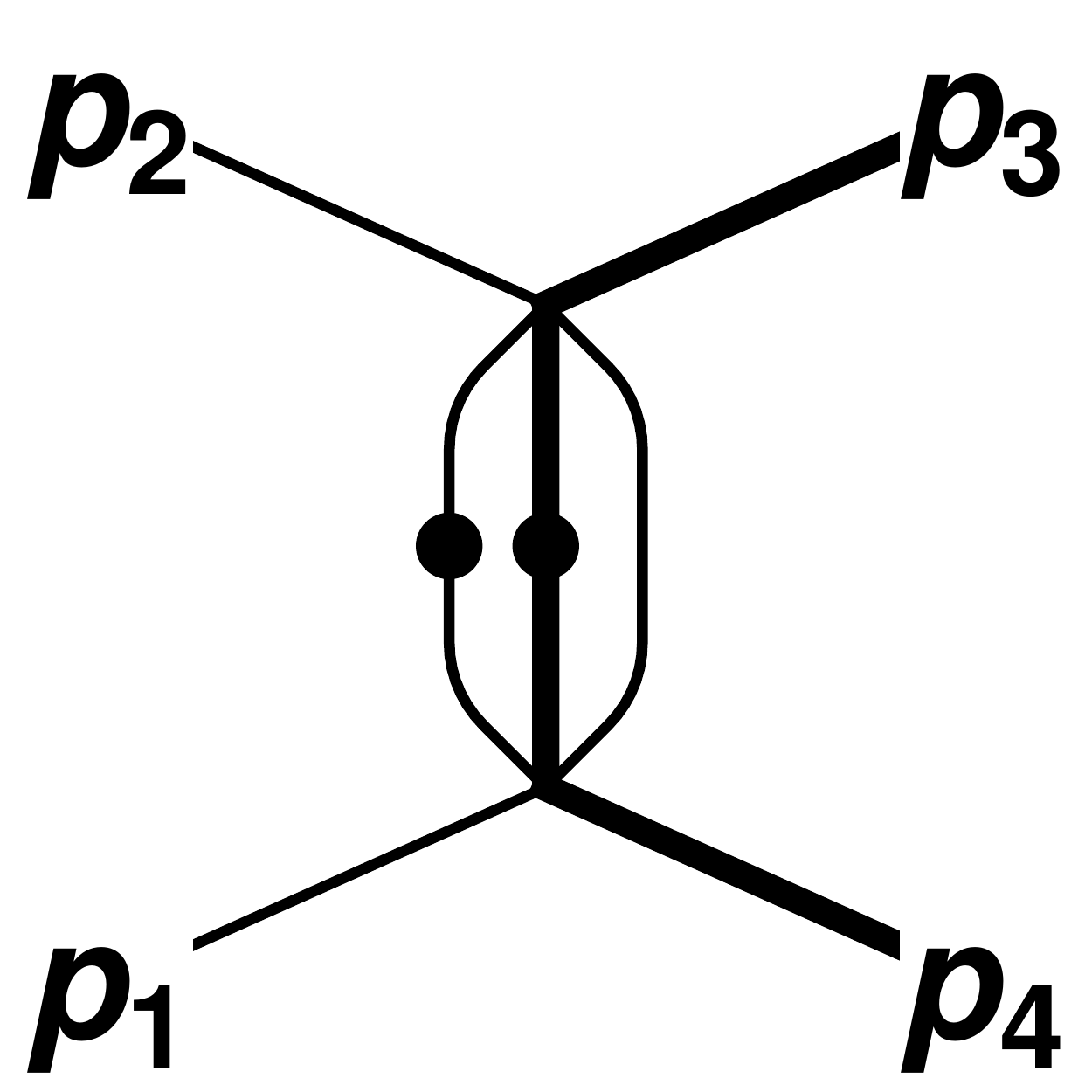}
  }
  \subfloat[$\mathcal{T}_5$]{%
    \includegraphics[width=0.15\textwidth]{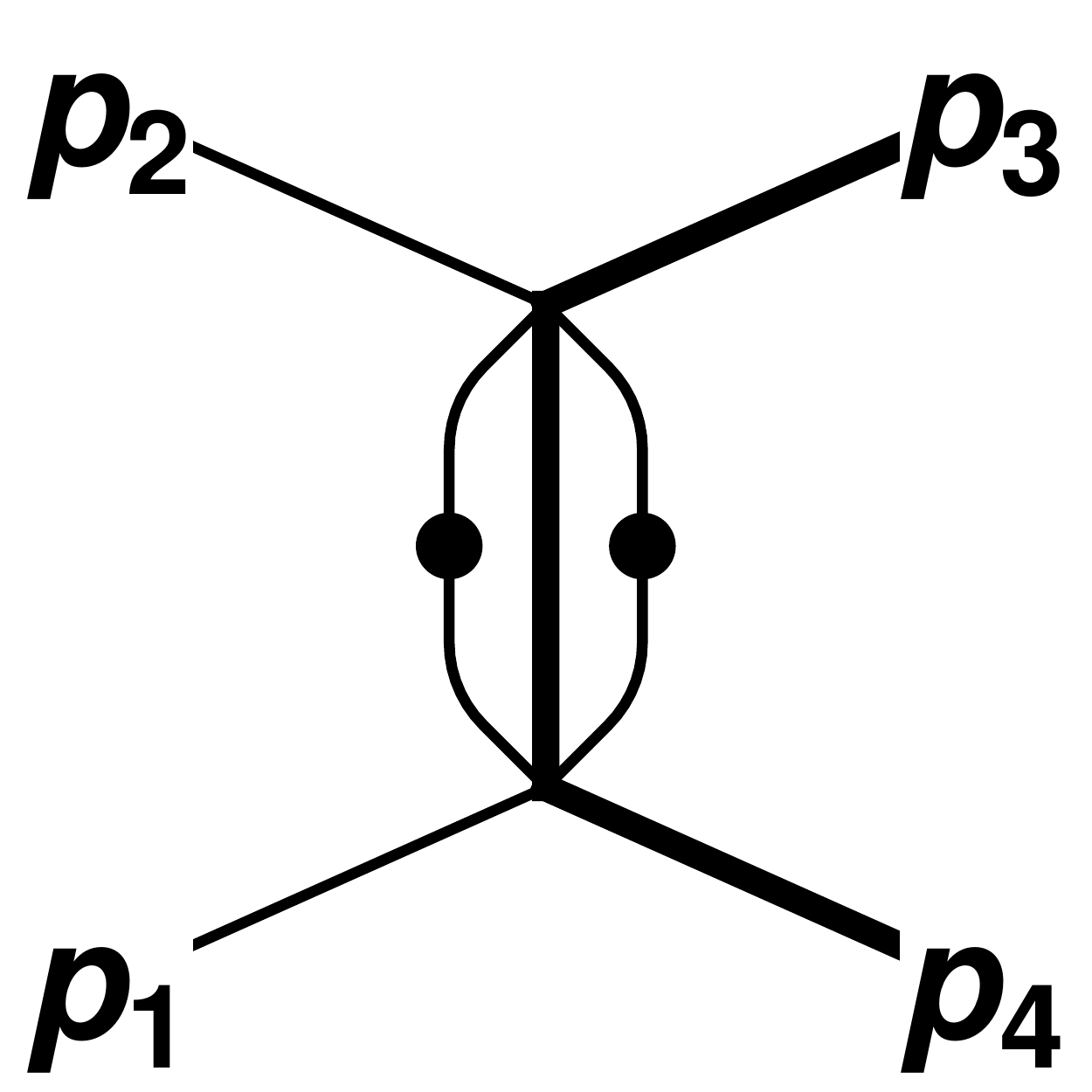}
  }\\
  \subfloat[$\mathcal{T}_6$]{%
    \includegraphics[width=0.15\textwidth]{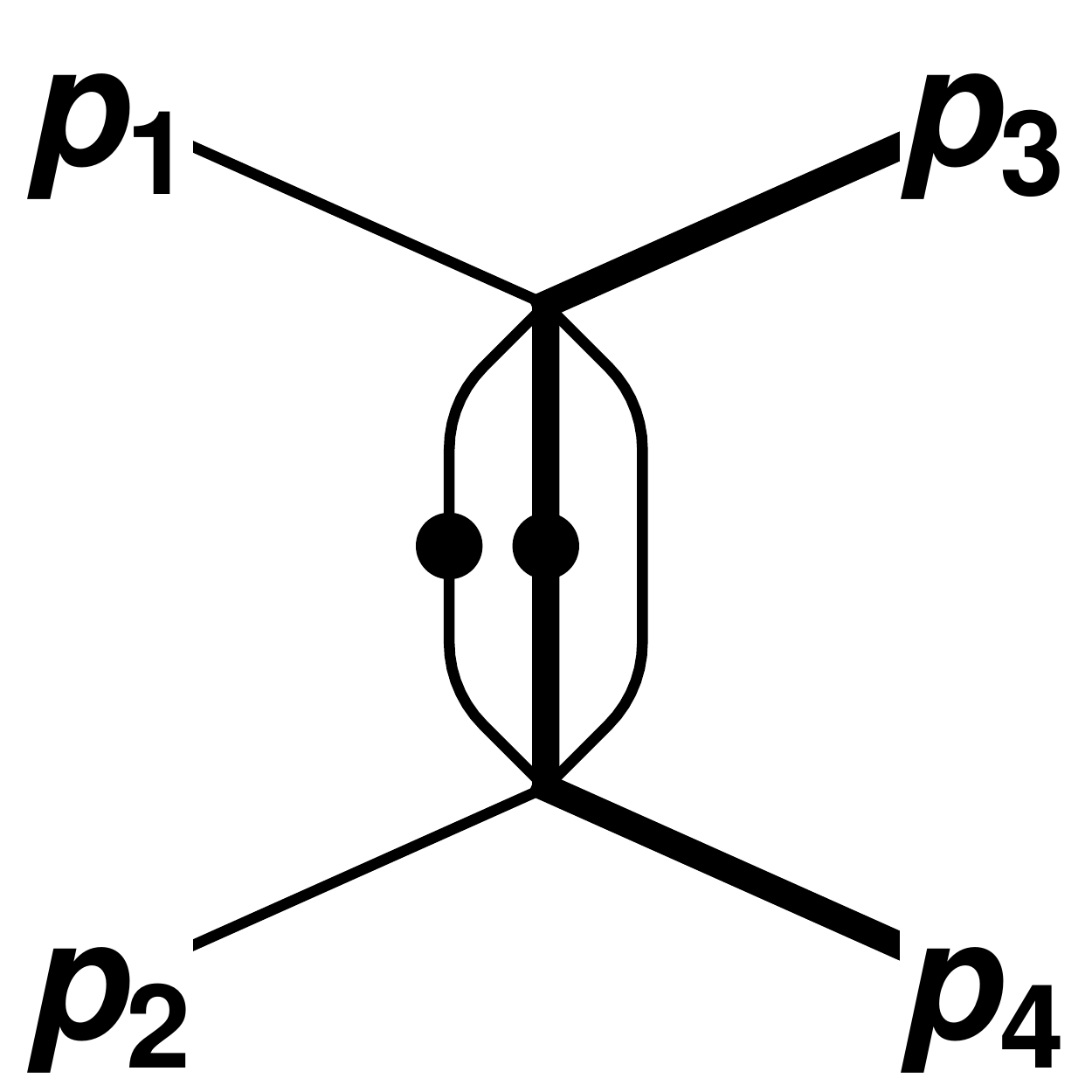}
  }
  \subfloat[$\mathcal{T}_7$]{%
    \includegraphics[width=0.15\textwidth]{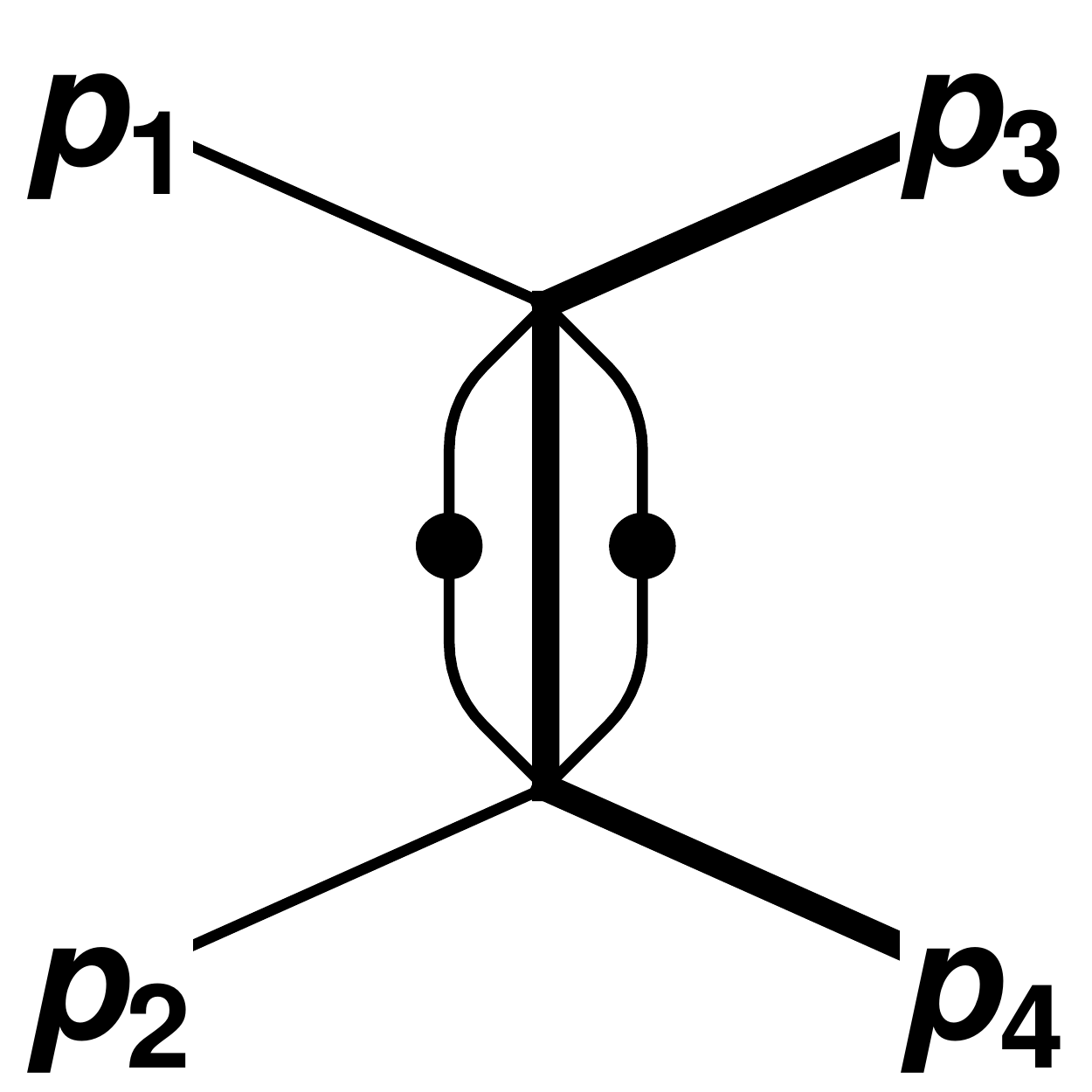}
  }
  \subfloat[$\mathcal{T}_8$]{%
    \includegraphics[width=0.15\textwidth]{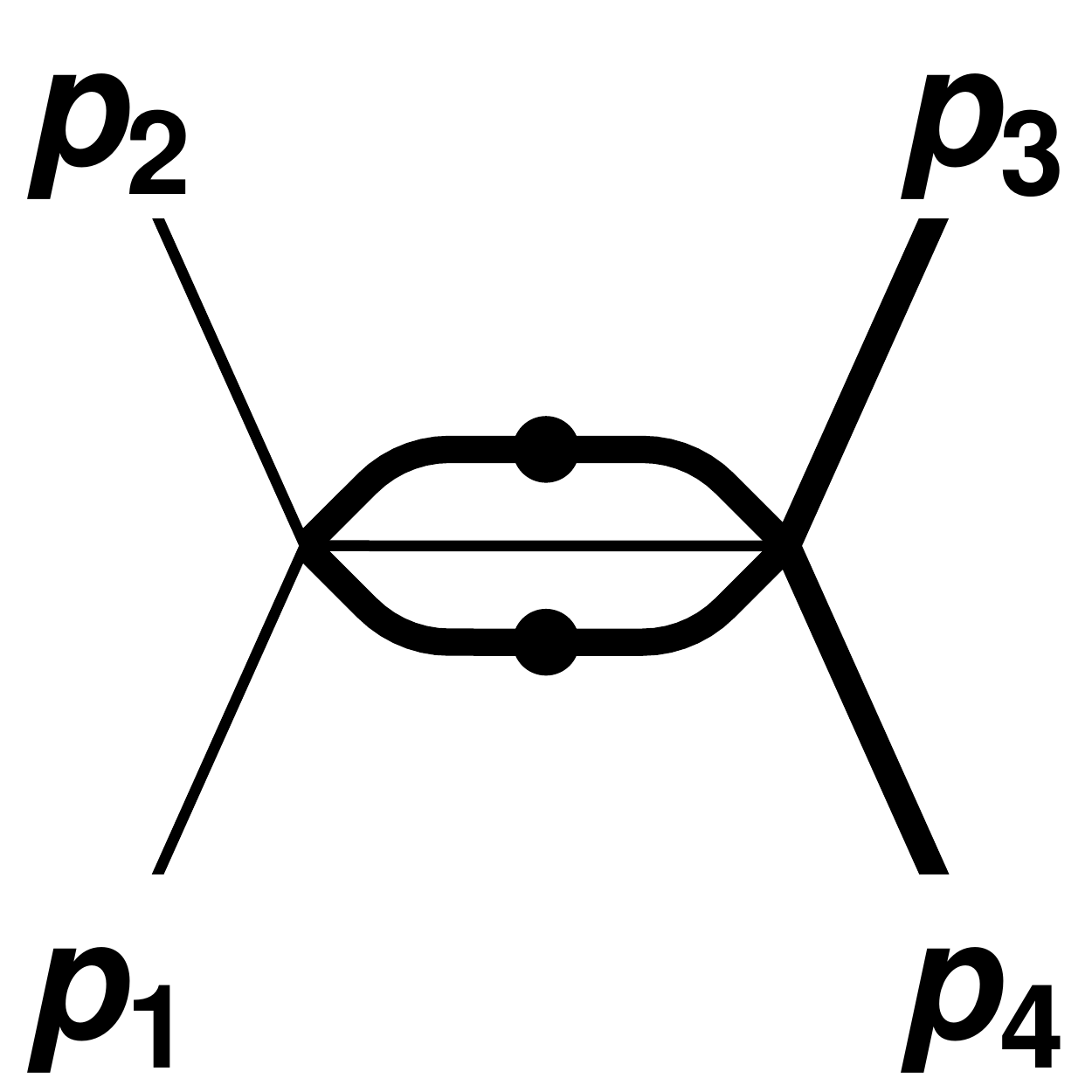}
  }
  \subfloat[$\mathcal{T}_9$]{%
    \includegraphics[width=0.15\textwidth]{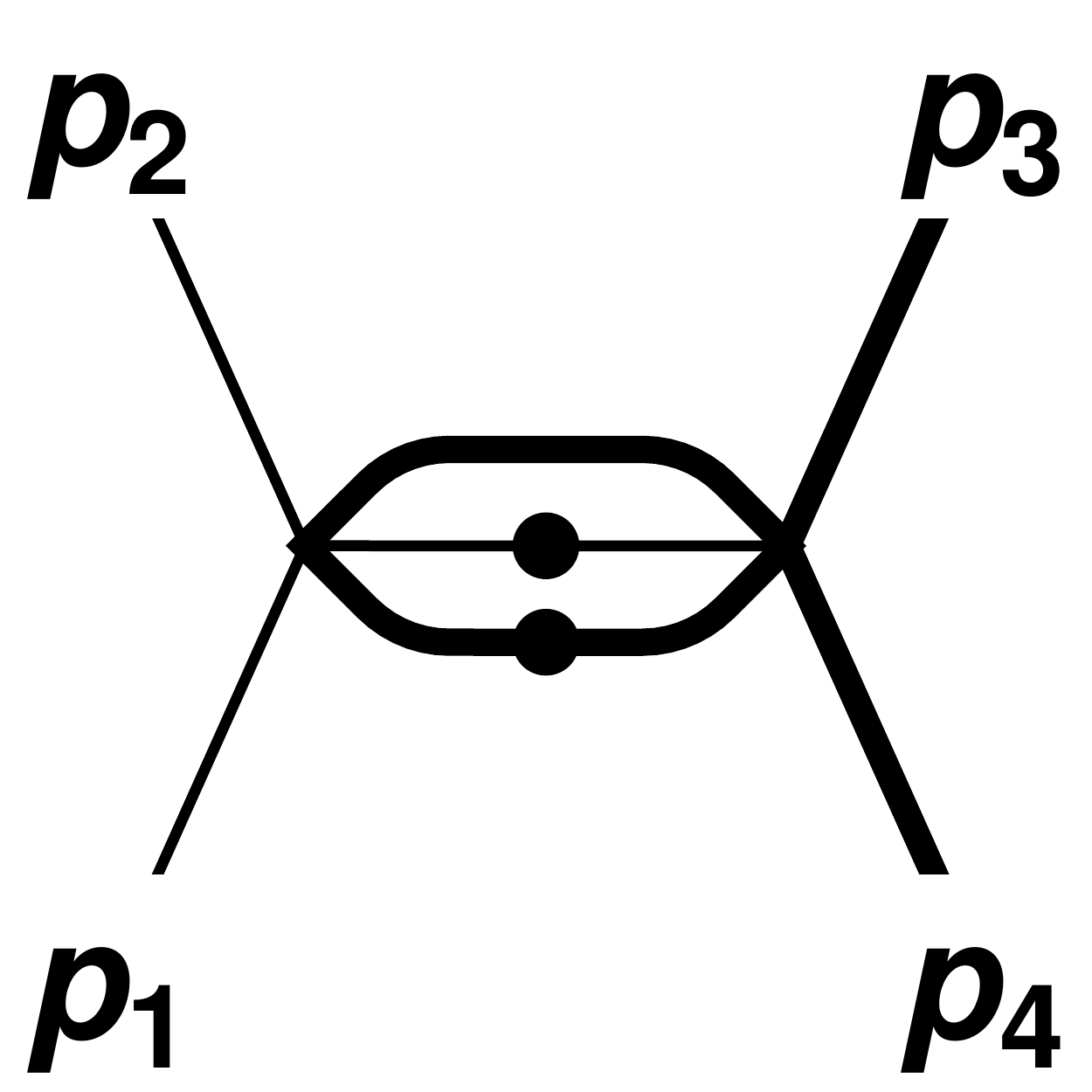}
  }
  \subfloat[$\mathcal{T}_{10}$]{%
    \includegraphics[width=0.15\textwidth]{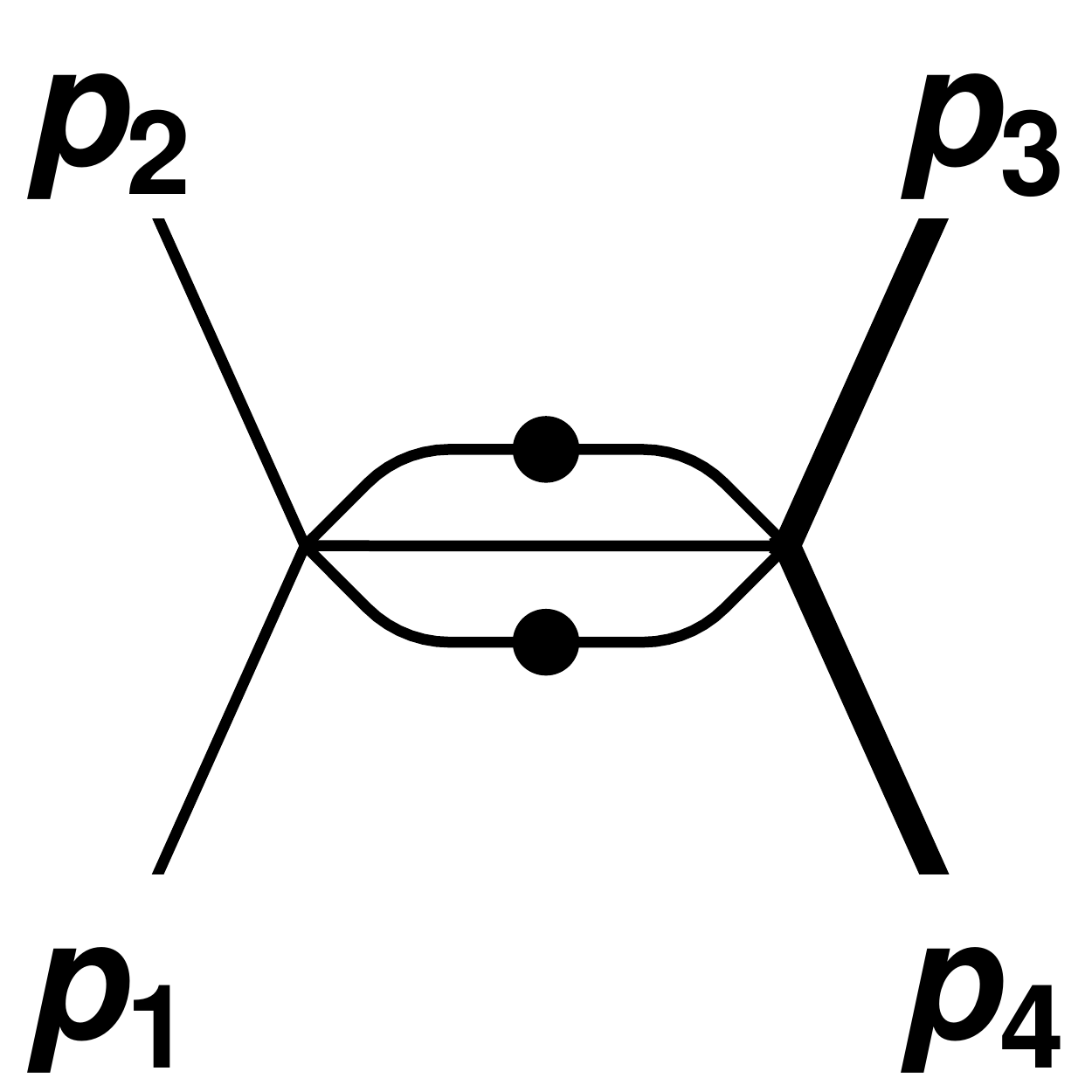}
  }\\
  \subfloat[$\mathcal{T}_{11}$]{%
    \includegraphics[width=0.15\textwidth]{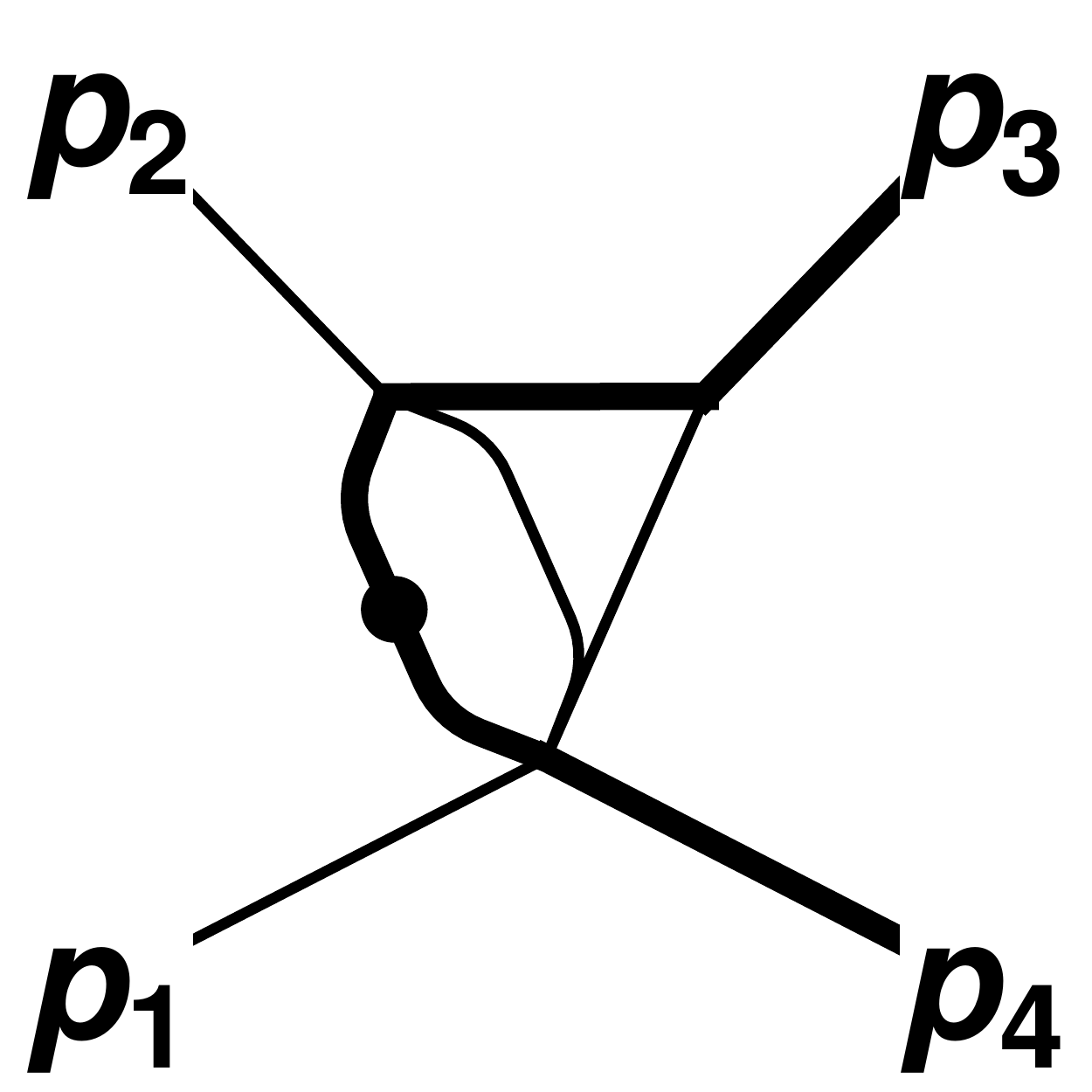}
  }
  \subfloat[$\mathcal{T}_{12}$]{%
    \includegraphics[width=0.15\textwidth]{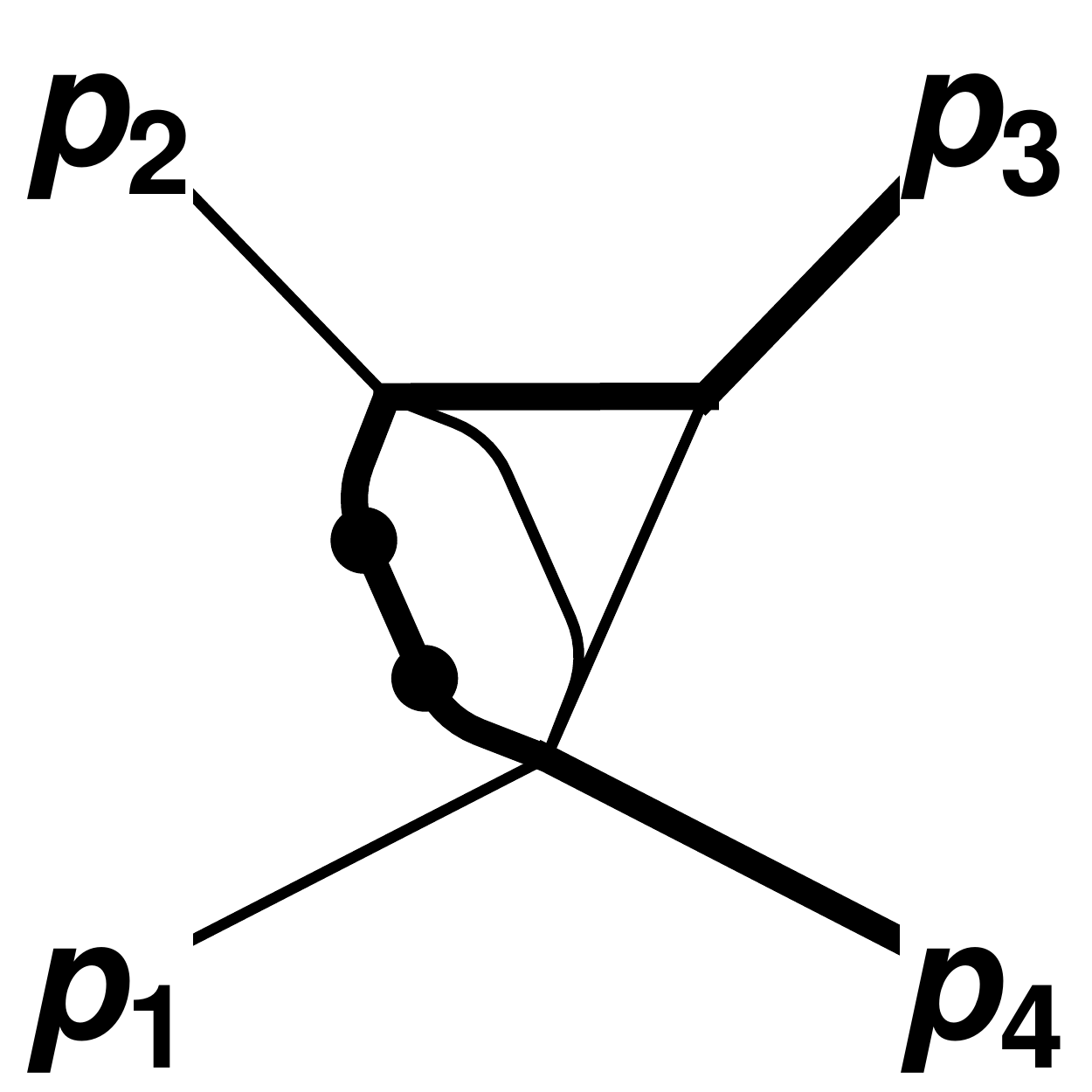}
  }
  \subfloat[$\mathcal{T}_{13}$]{%
    \includegraphics[width=0.15\textwidth]{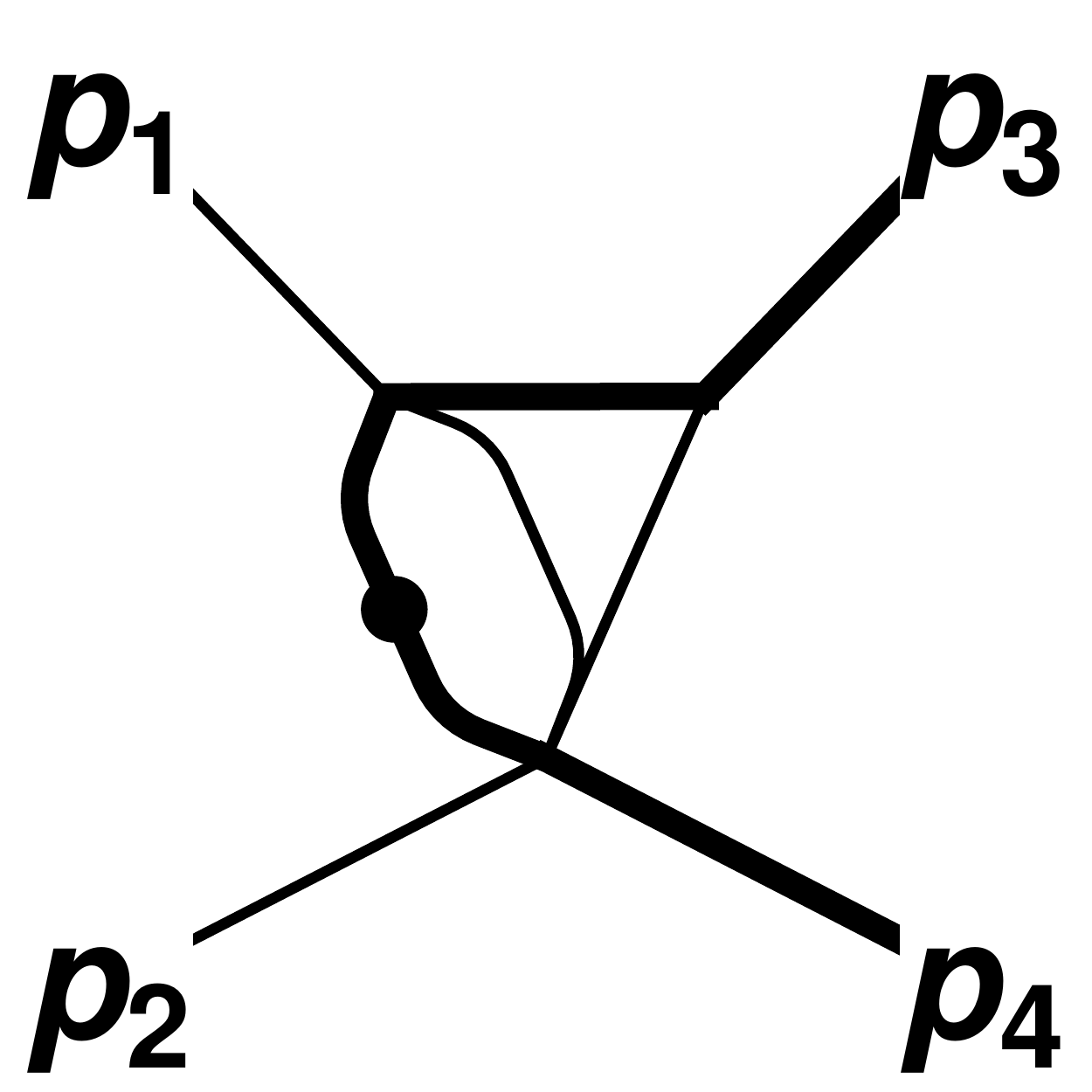}
  }
  \subfloat[$\mathcal{T}_{14}$]{%
    \includegraphics[width=0.15\textwidth]{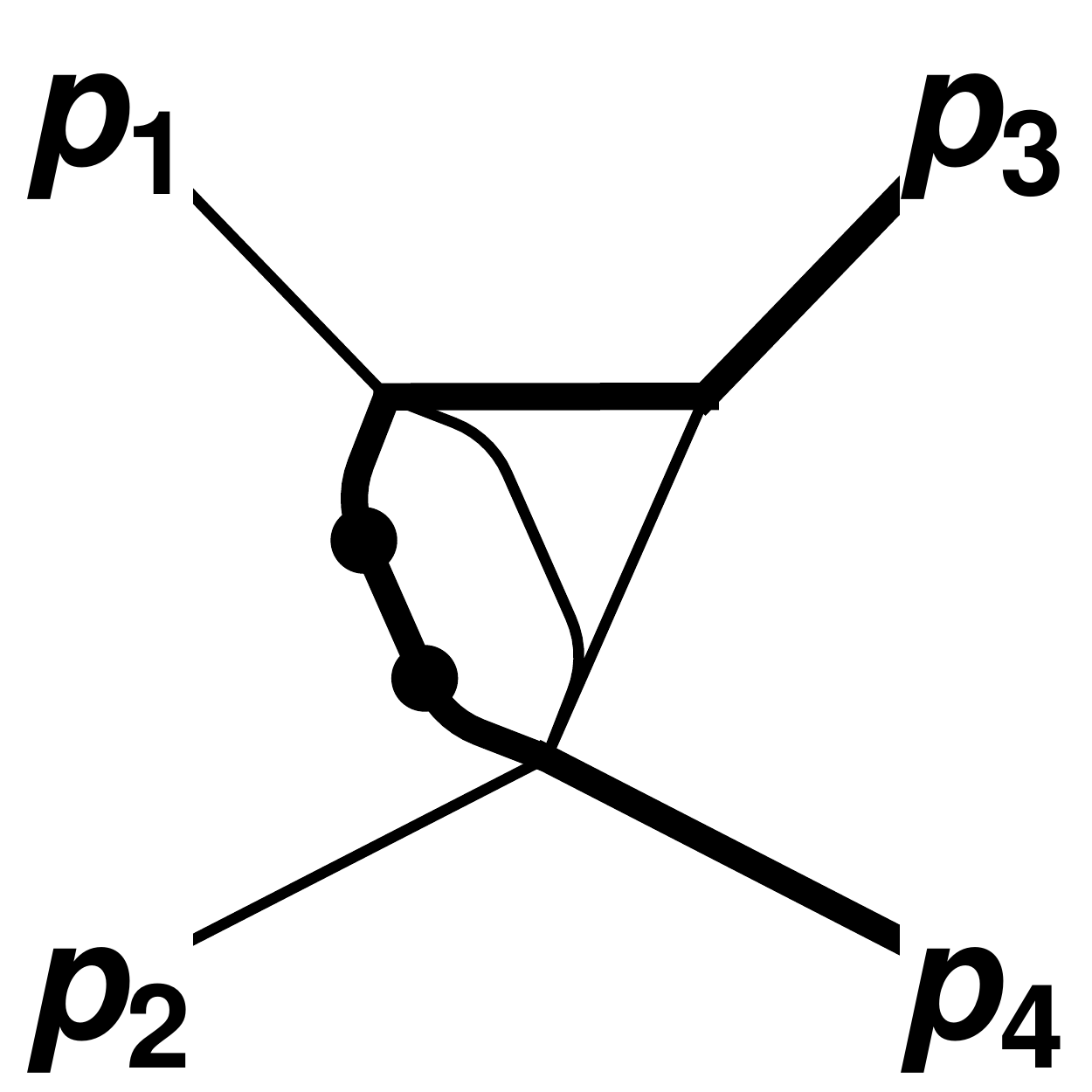}
  }
  \subfloat[$\mathcal{T}_{15}$]{%
    \includegraphics[width=0.15\textwidth]{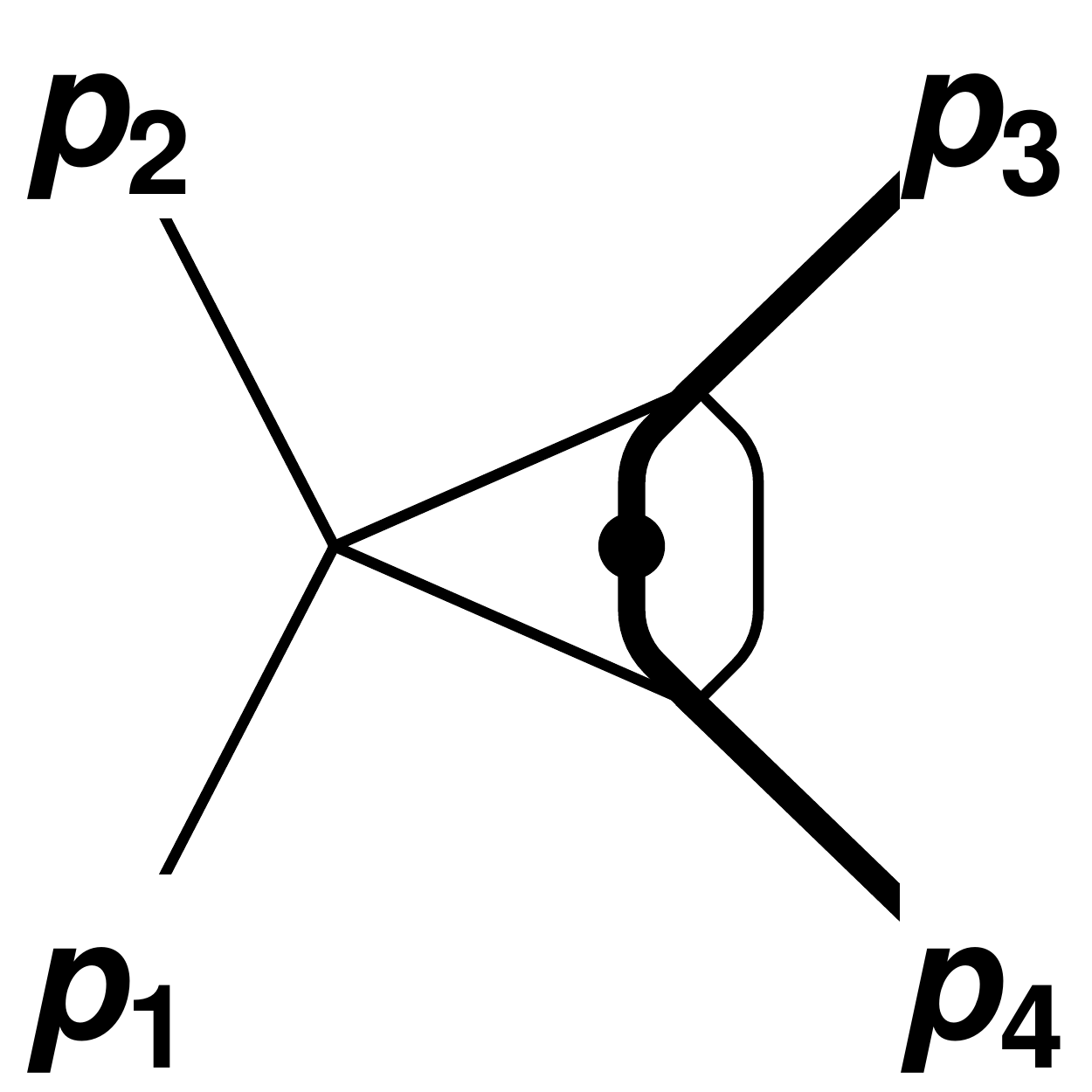}
  }\\
  \subfloat[$\mathcal{T}_{16}$]{%
    \includegraphics[width=0.15\textwidth]{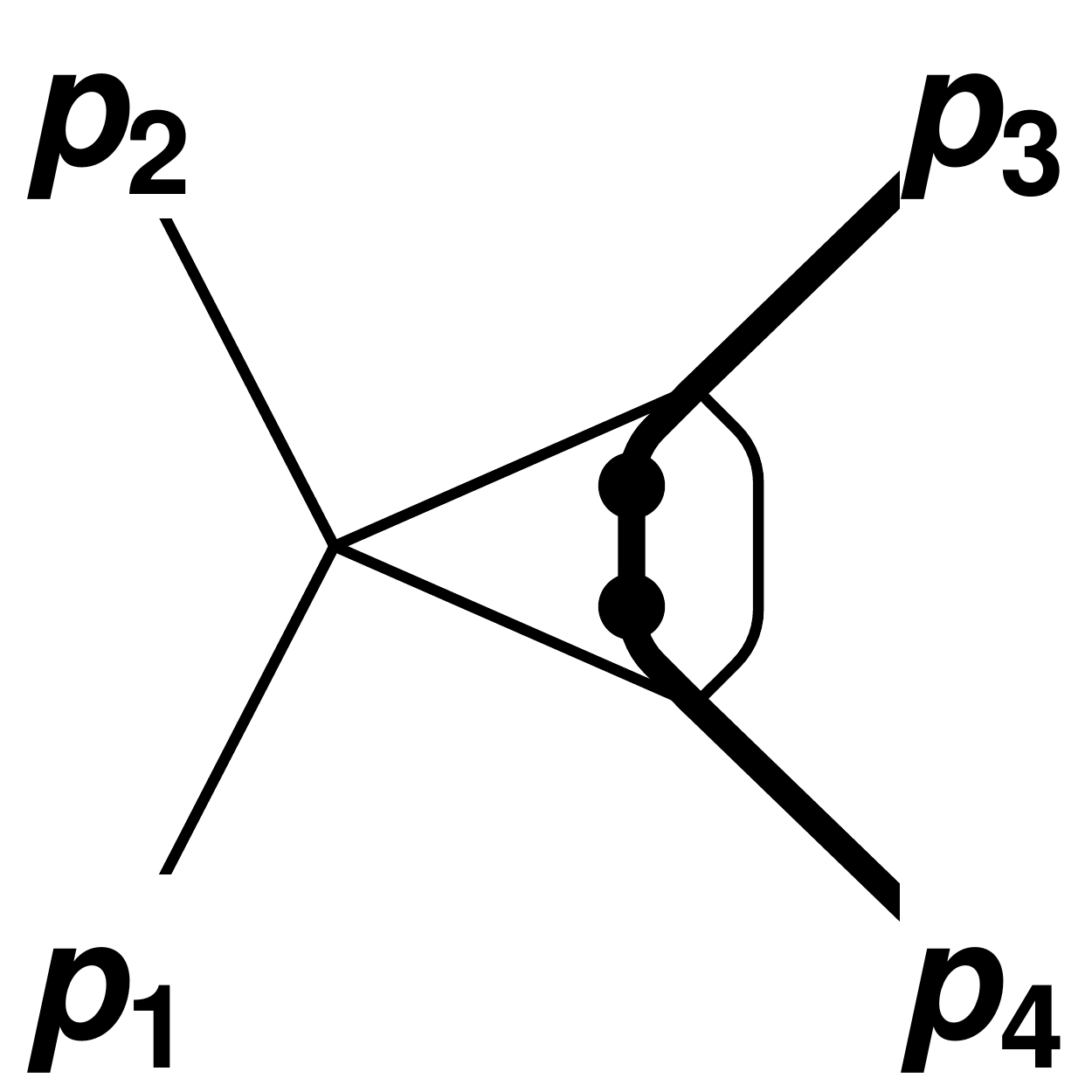}
  }
  \subfloat[$\mathcal{T}_{17}$]{%
    \includegraphics[width=0.15\textwidth]{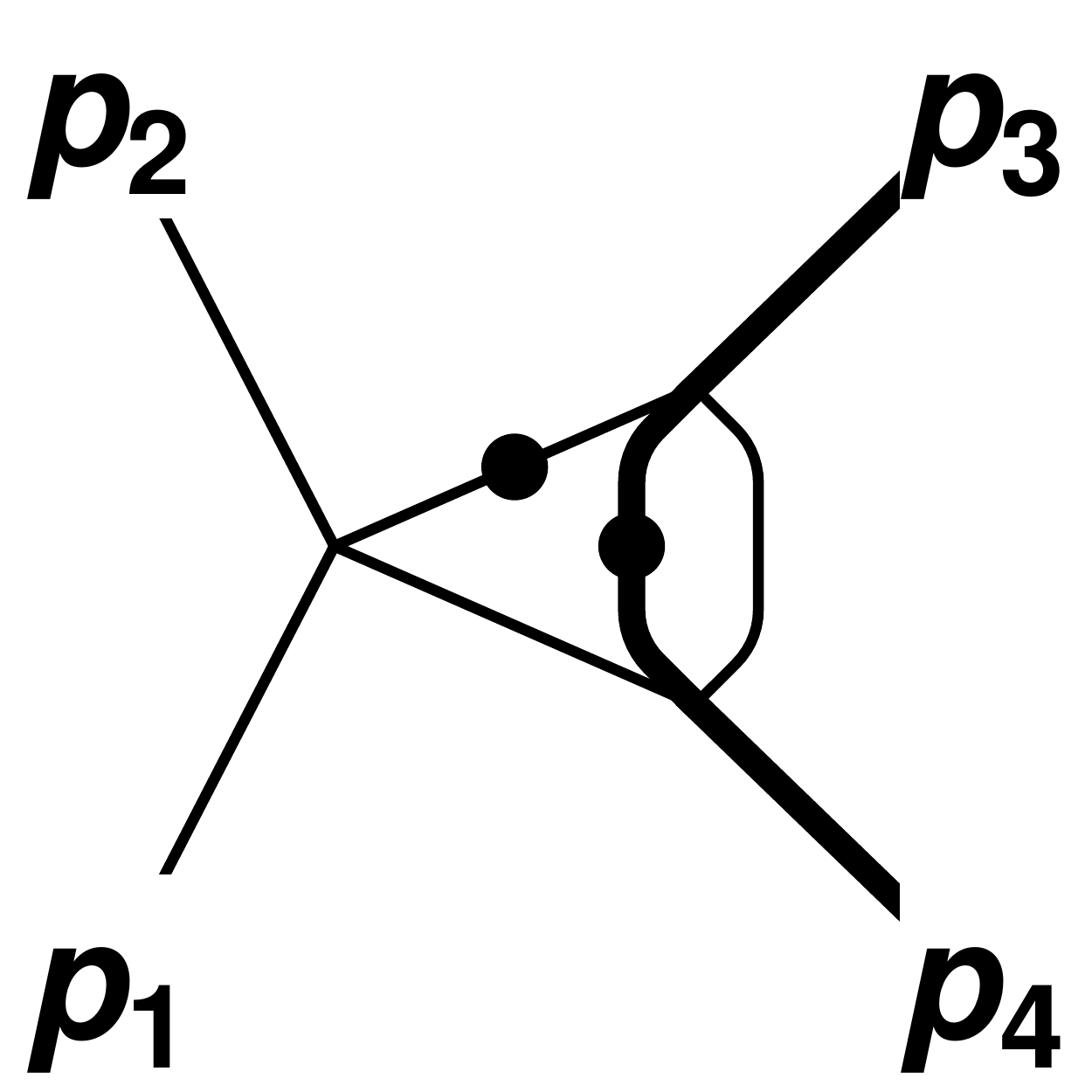}
  }
  \subfloat[$\mathcal{T}_{18}$]{%
    \includegraphics[width=0.15\textwidth]{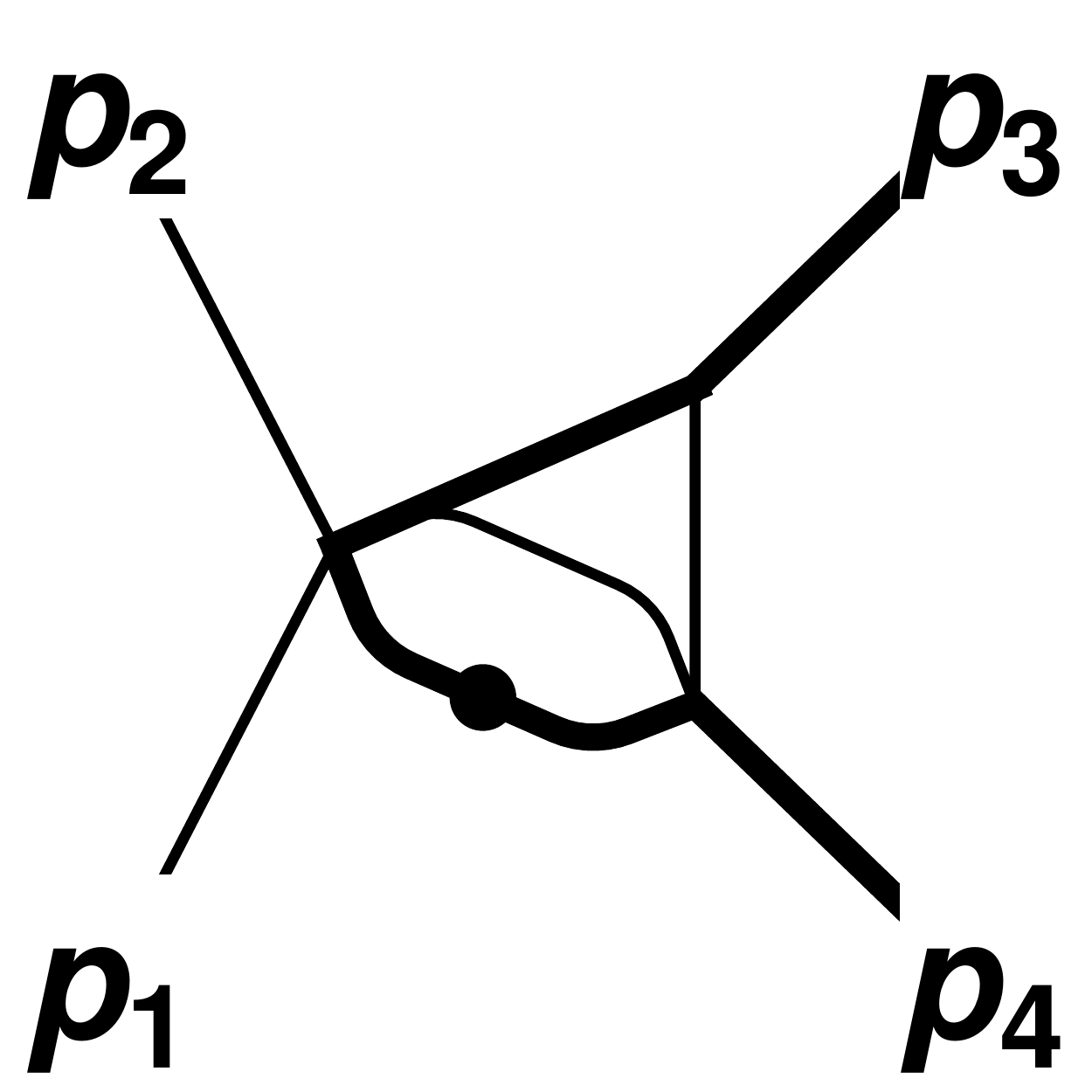}
  }
    \subfloat[$\mathcal{T}_{19}$]{%
    \includegraphics[width=0.15\textwidth]{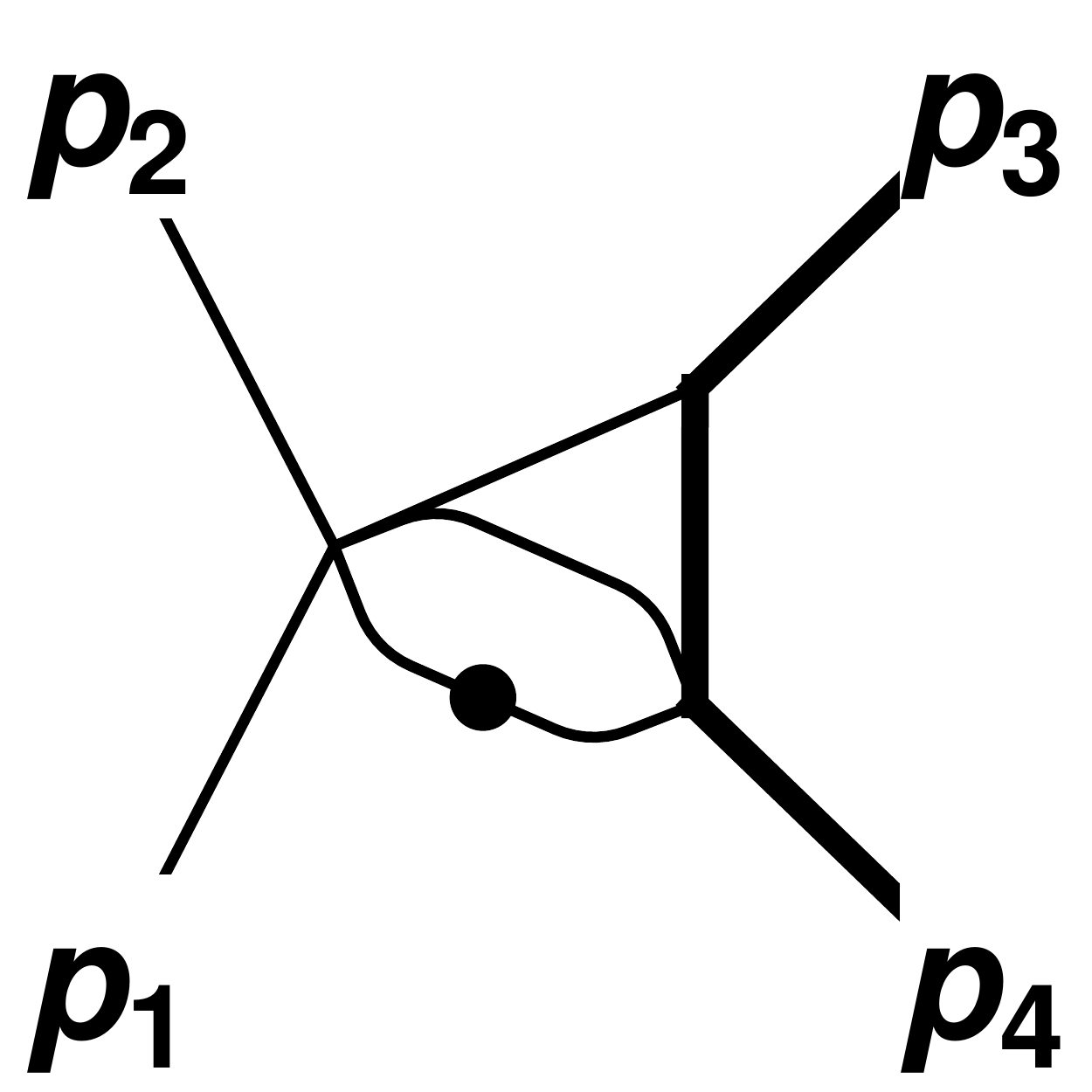}
  }
  \subfloat[$\mathcal{T}_{20}$]{%
    \includegraphics[width=0.15\textwidth]{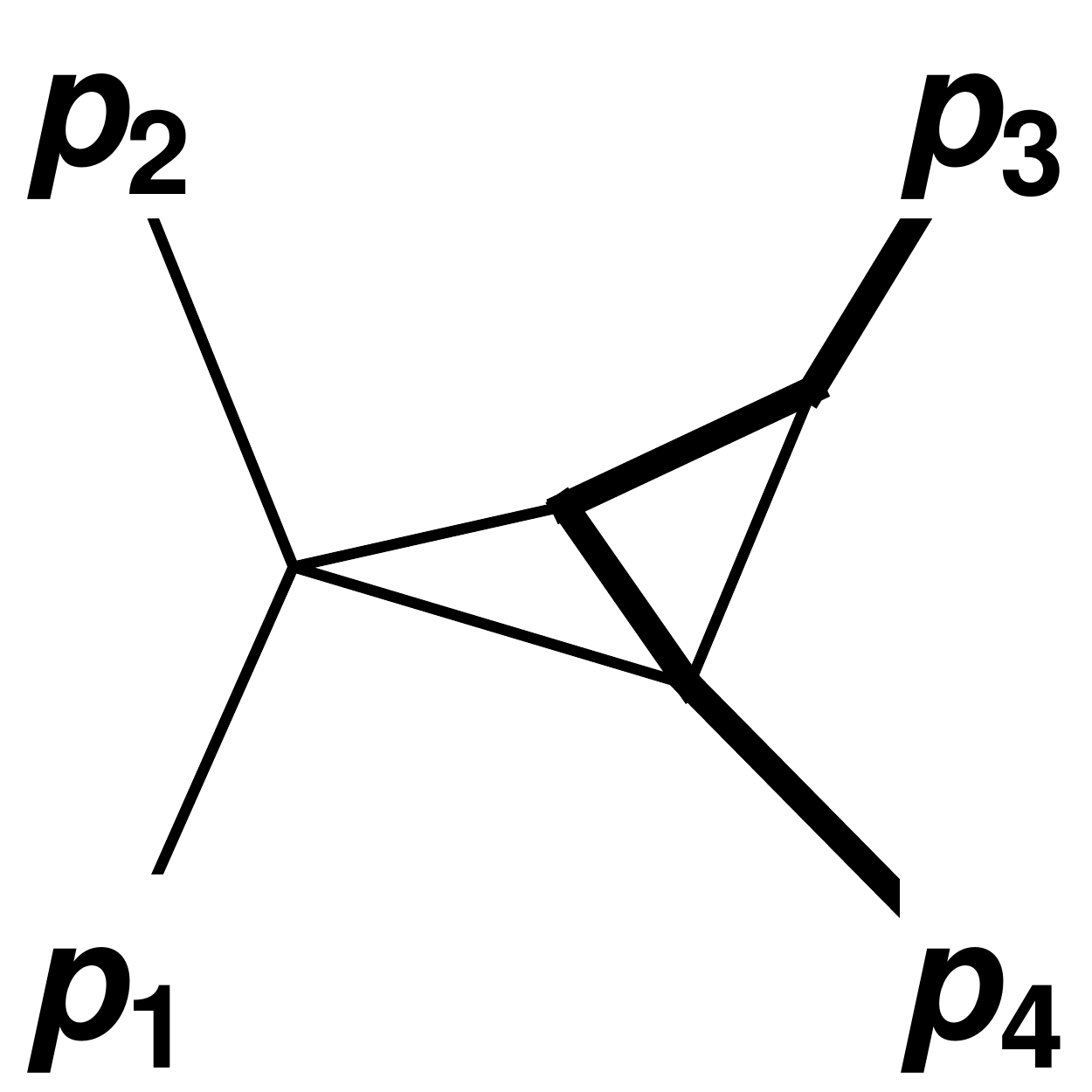}
  }\\
  \subfloat[$\mathcal{T}_{21}$]{%
    \includegraphics[width=0.15\textwidth]{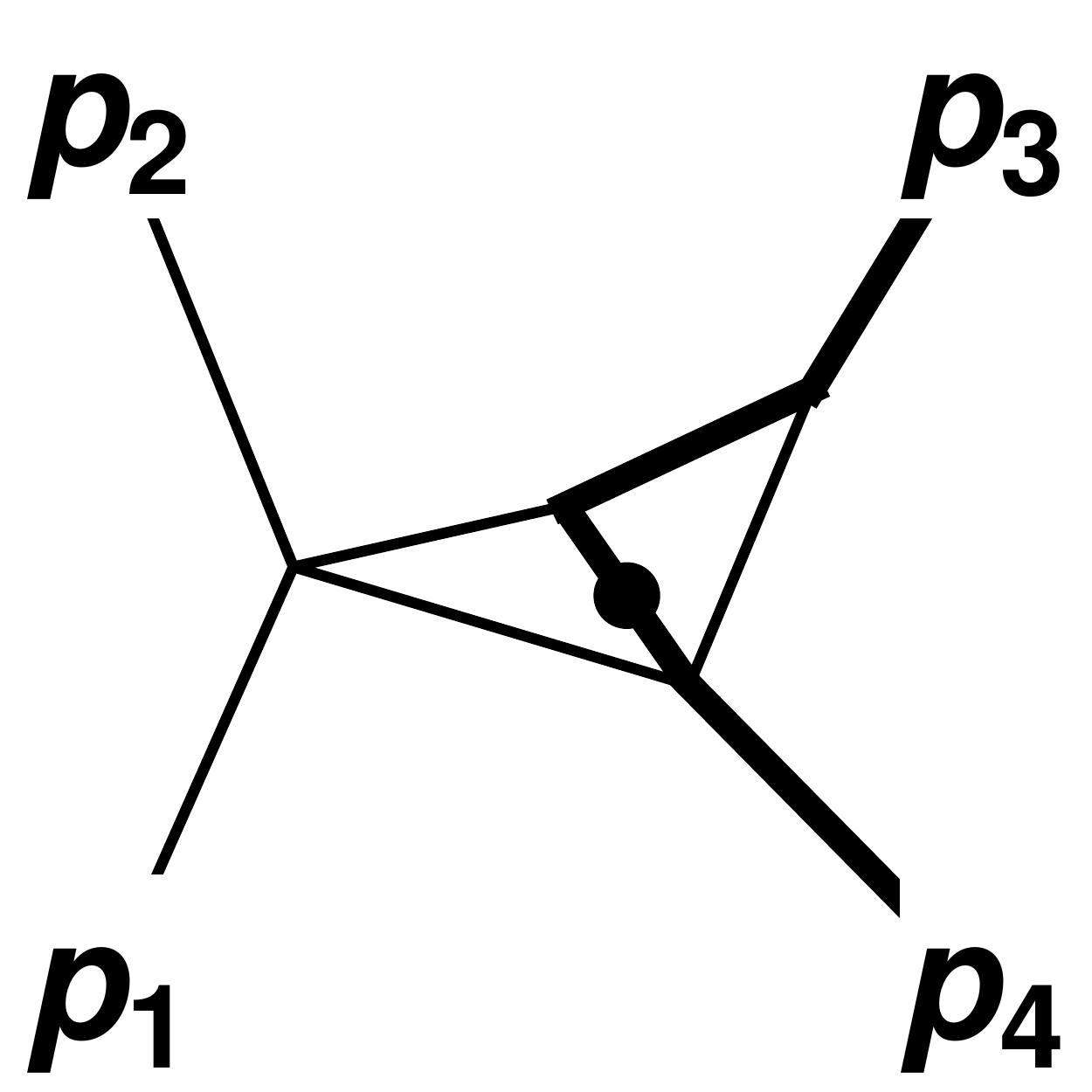}
  }
  \subfloat[$\mathcal{T}_{22}$]{%
    \includegraphics[width=0.15\textwidth]{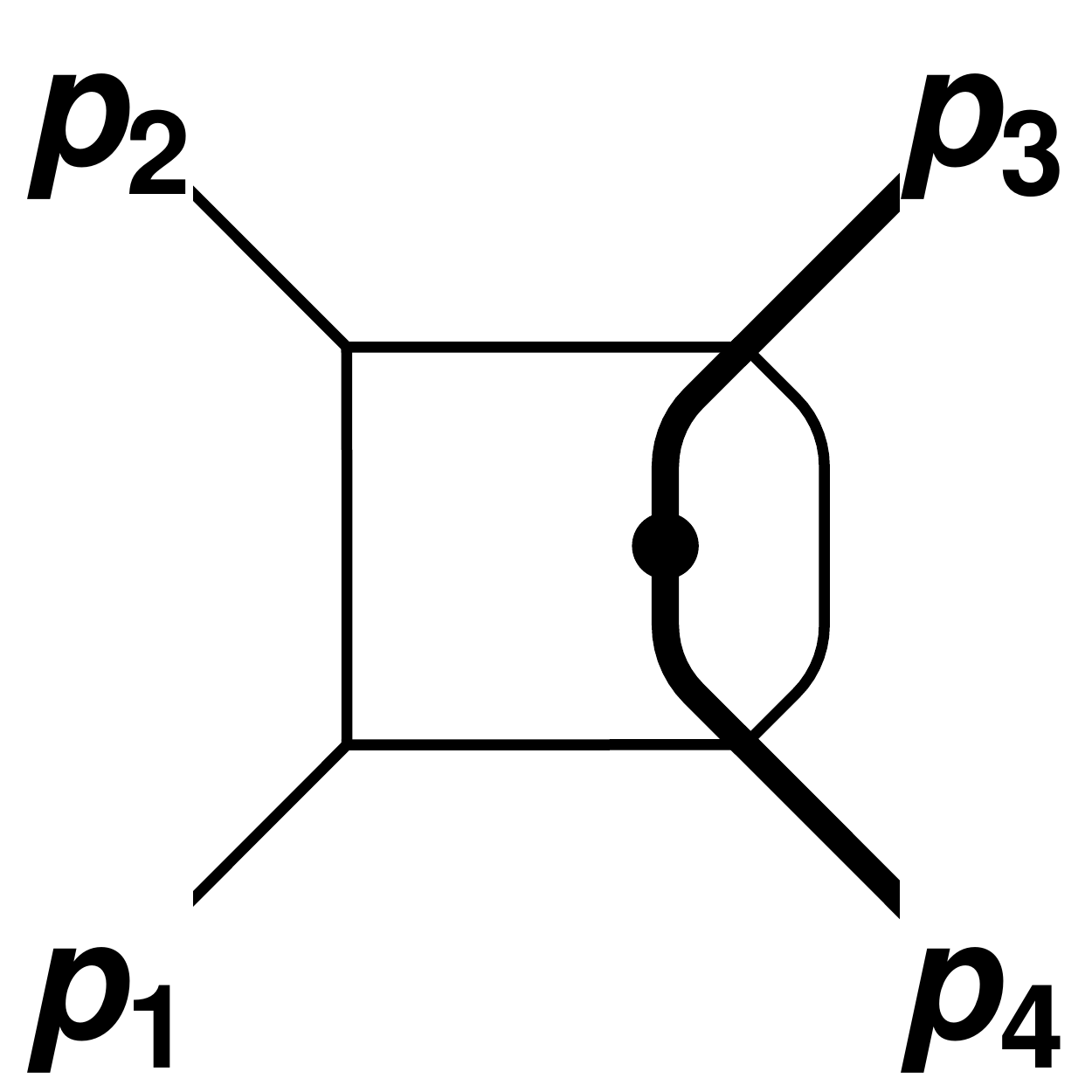}
  }
  \subfloat[$\mathcal{T}_{23}$]{%
    \includegraphics[width=0.15\textwidth]{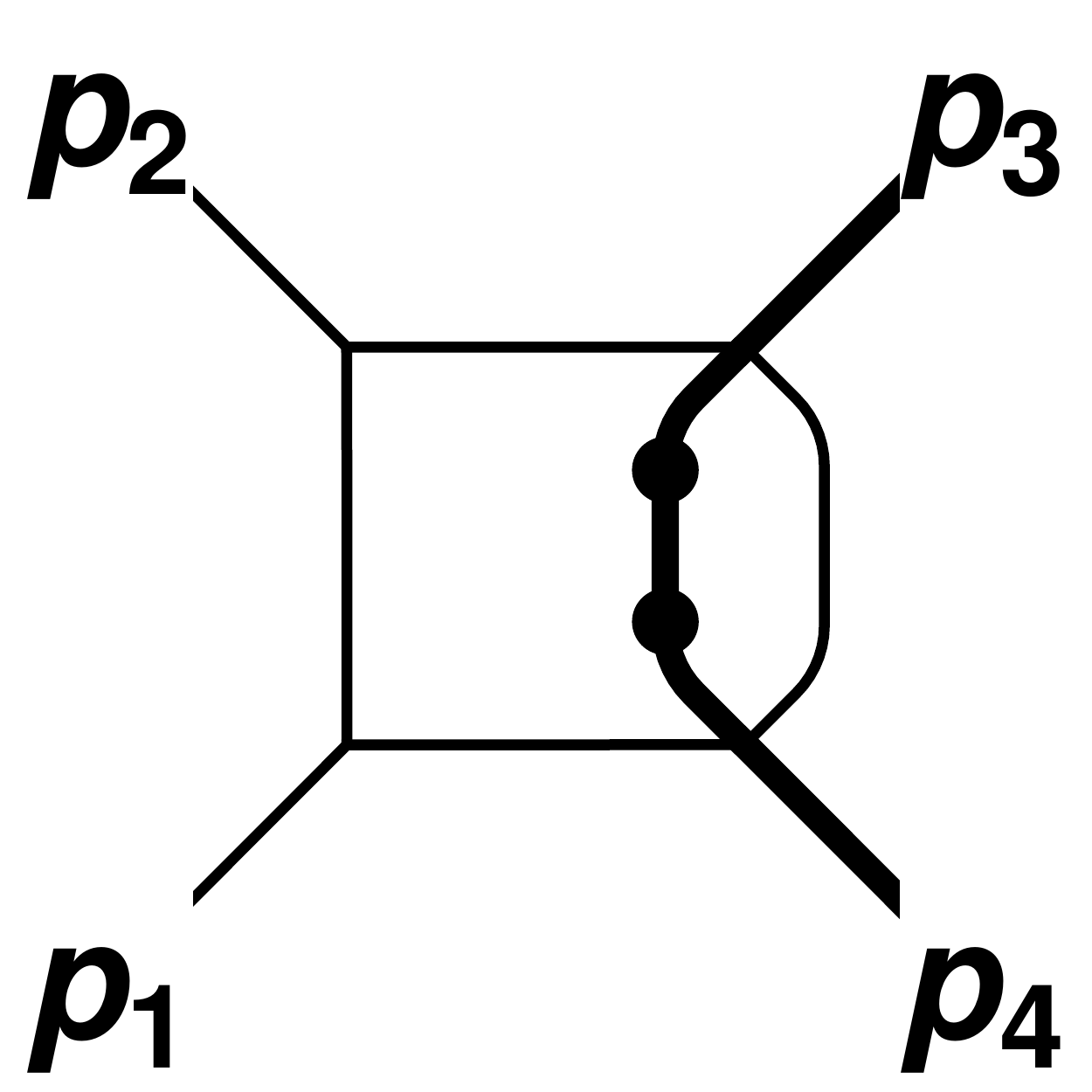}
  }
  \subfloat[$\mathcal{T}_{24}$]{%
    \includegraphics[width=0.15\textwidth]{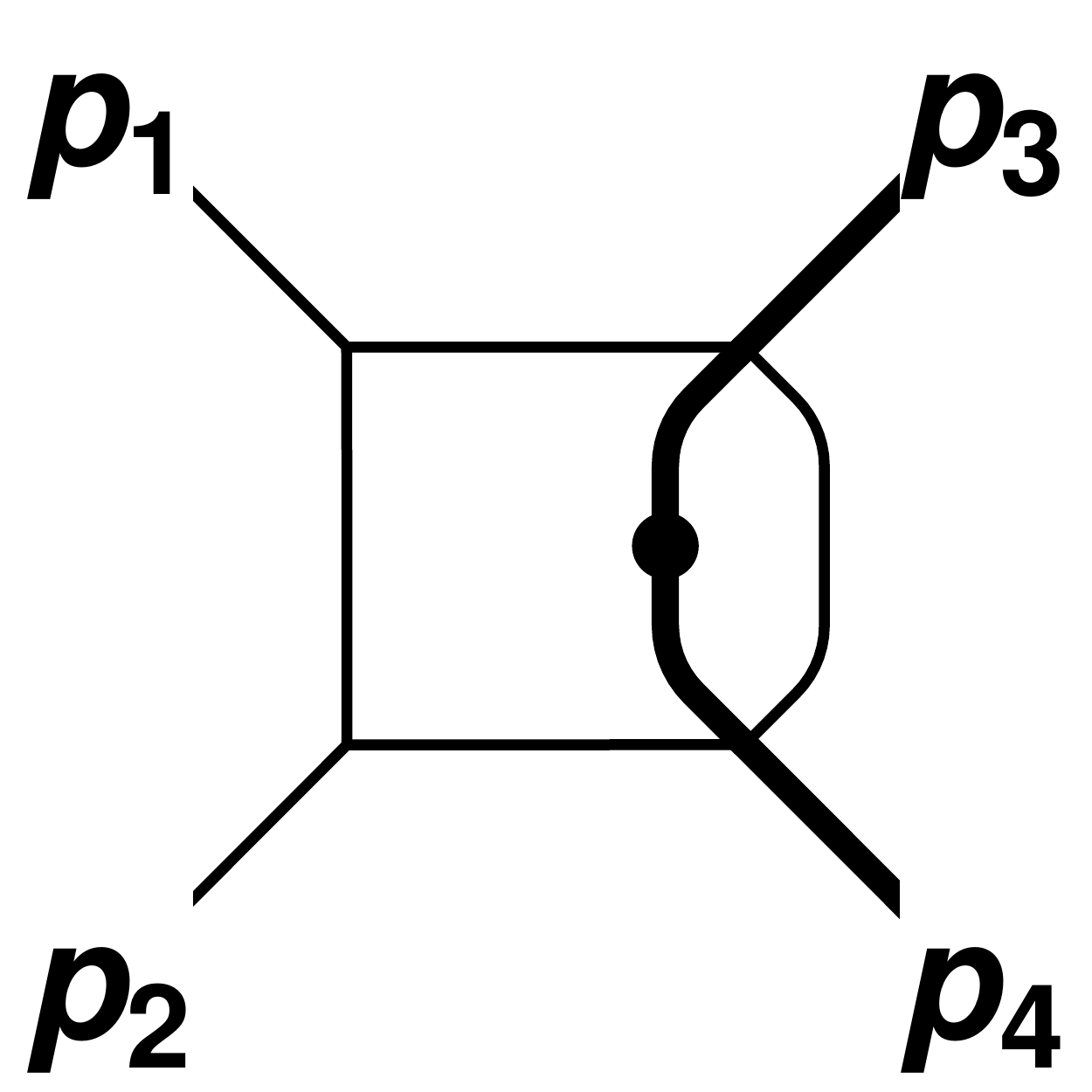}
  }
    \subfloat[$\mathcal{T}_{25}$]{%
      \includegraphics[width=0.15\textwidth]{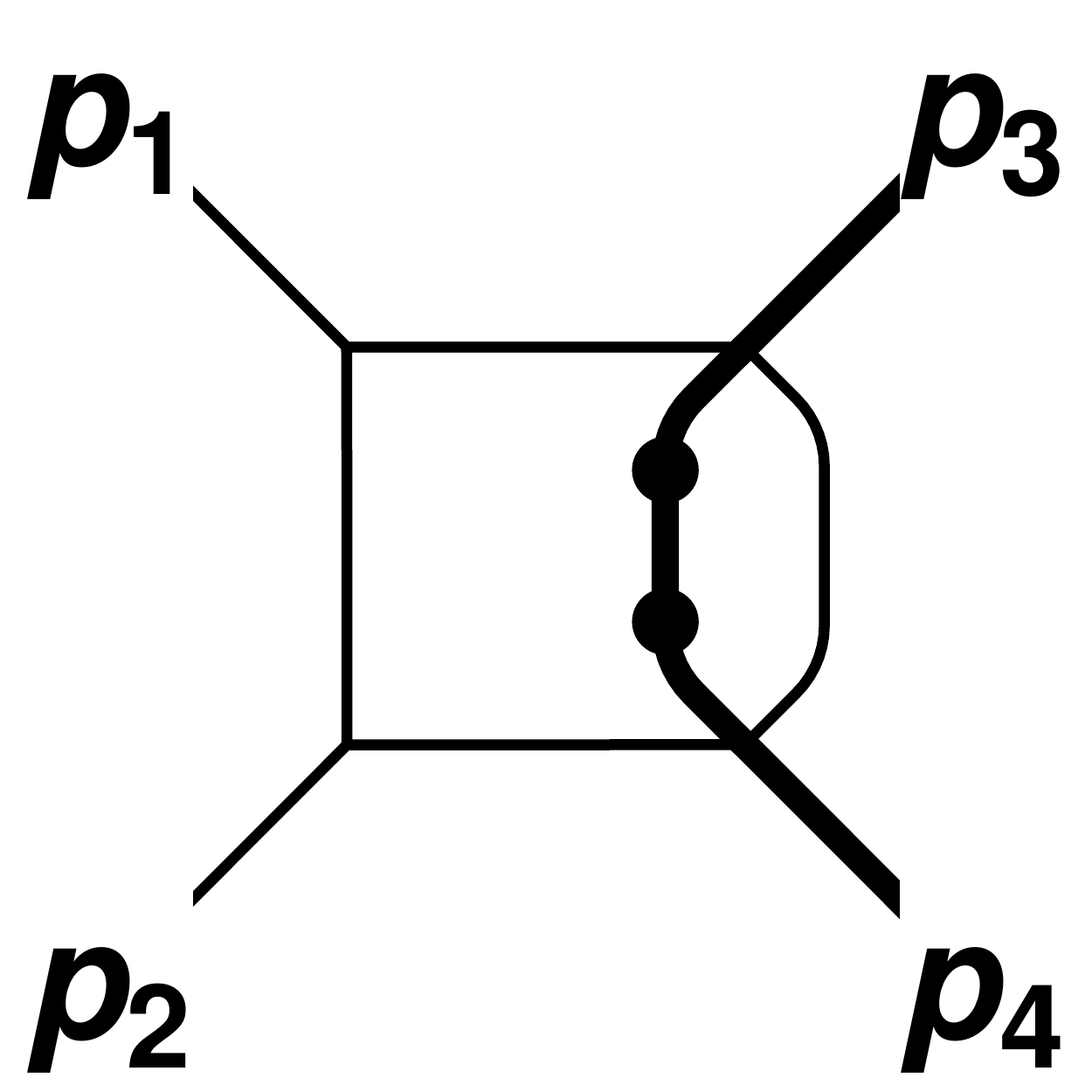}
      }
 \caption{
The first 25 MIs $\mathcal{T}_{1,\ldots,25}$ for the two-loop non-planar topology $N_2$ of figure~\ref{fig:npgraphs}.
Massless propagators  are represented by think lines and massive propagators by thick ones. Each dot indicates an additional power of the corresponding propagator. Numerator insertions are indicated explicitly on top of each diagram.}
 \label{fig:MIsT6a}
\end{figure}

%%%%%%%%%%%%%%%%%%
%%% Local Variables:
%%% TeX-master: "../main"
%%% End:
 \begin{figure}[t]
  \centering
  \captionsetup[subfigure]{labelformat=empty}
  \subfloat[$\mathcal{T}_{26}$]{%
    \includegraphics[width=0.15\textwidth]{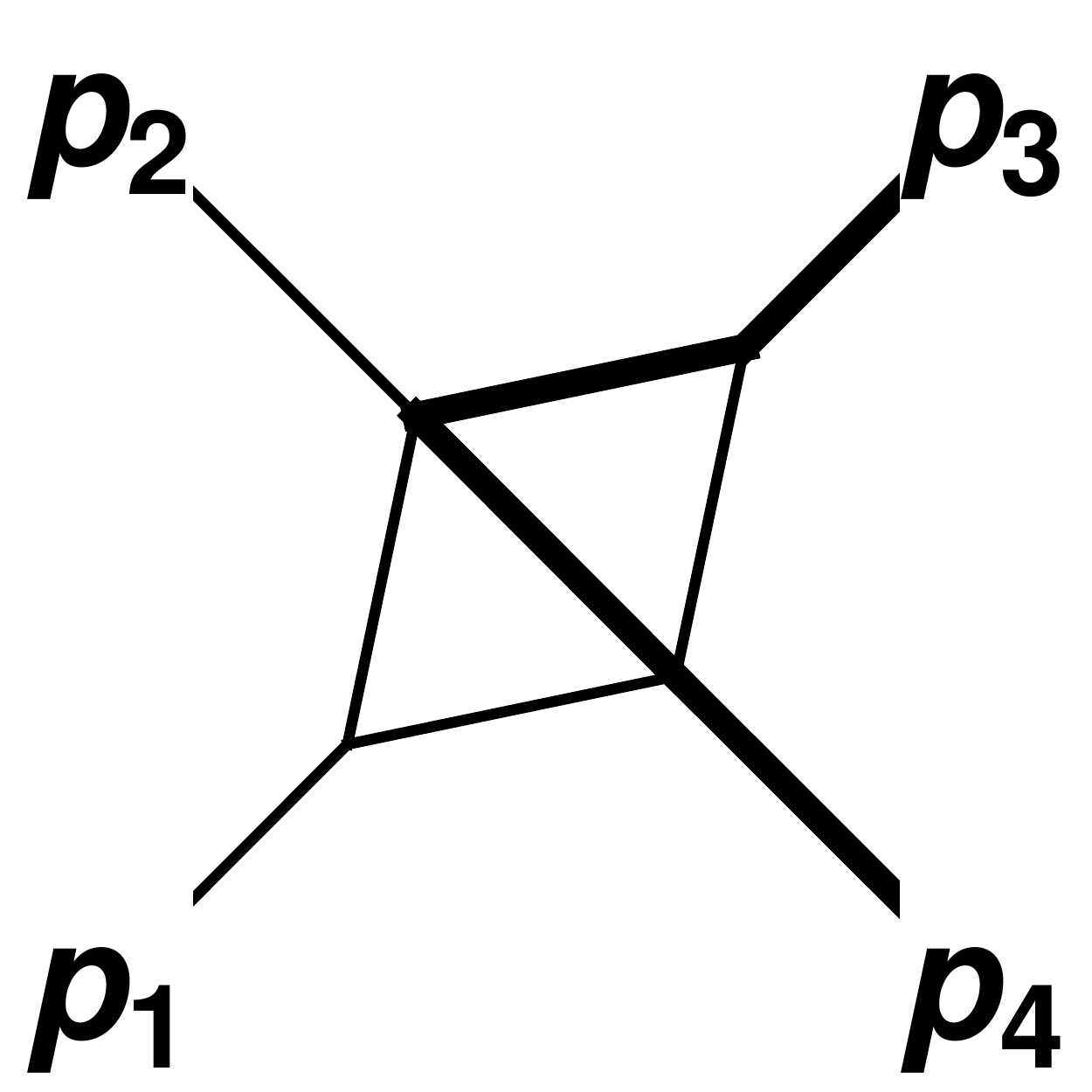}
  }
  \subfloat[$\mathcal{T}_{27}$]{%
    \includegraphics[width=0.15\textwidth]{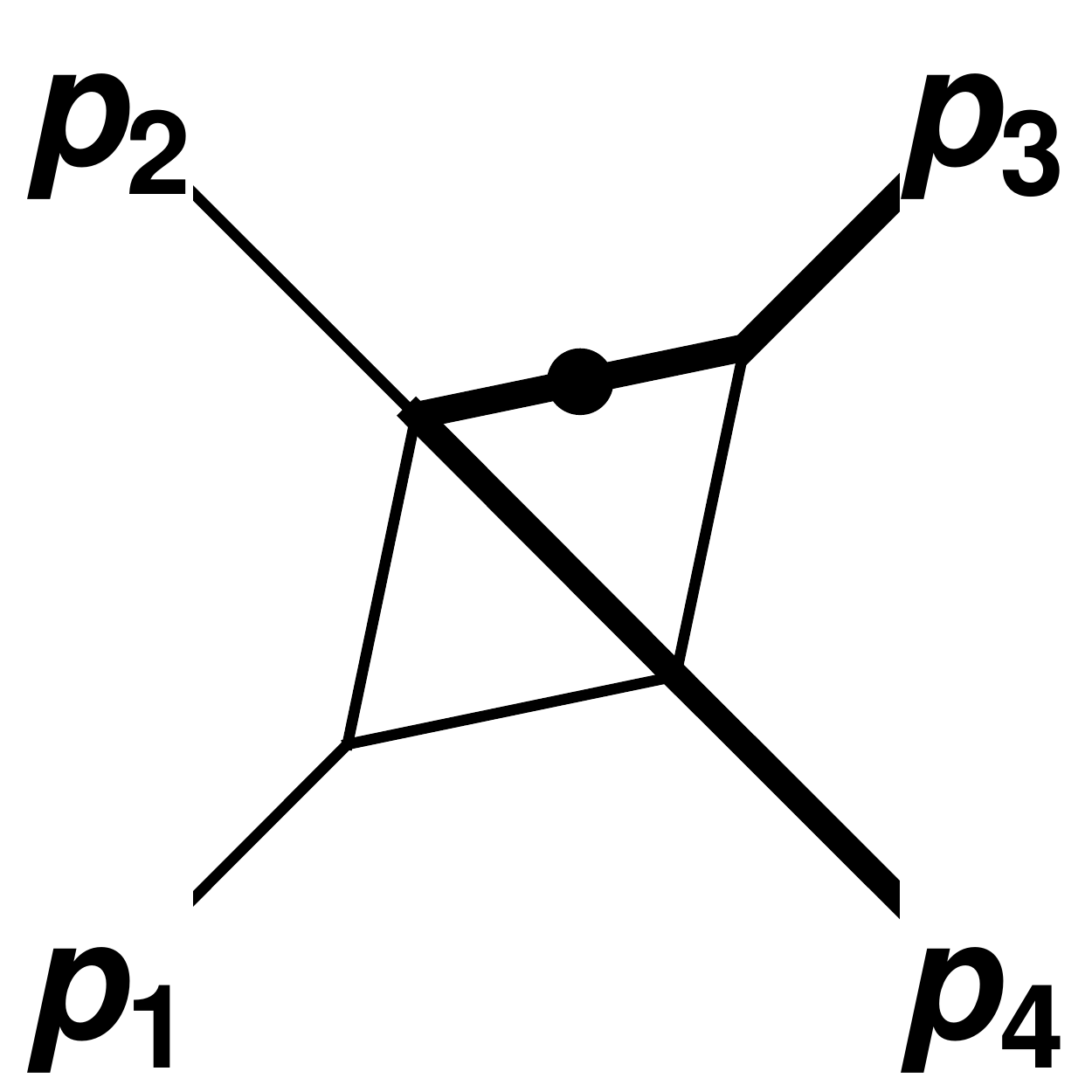}
  }
  \subfloat[$\mathcal{T}_{28}$]{%
    \includegraphics[width=0.15\textwidth]{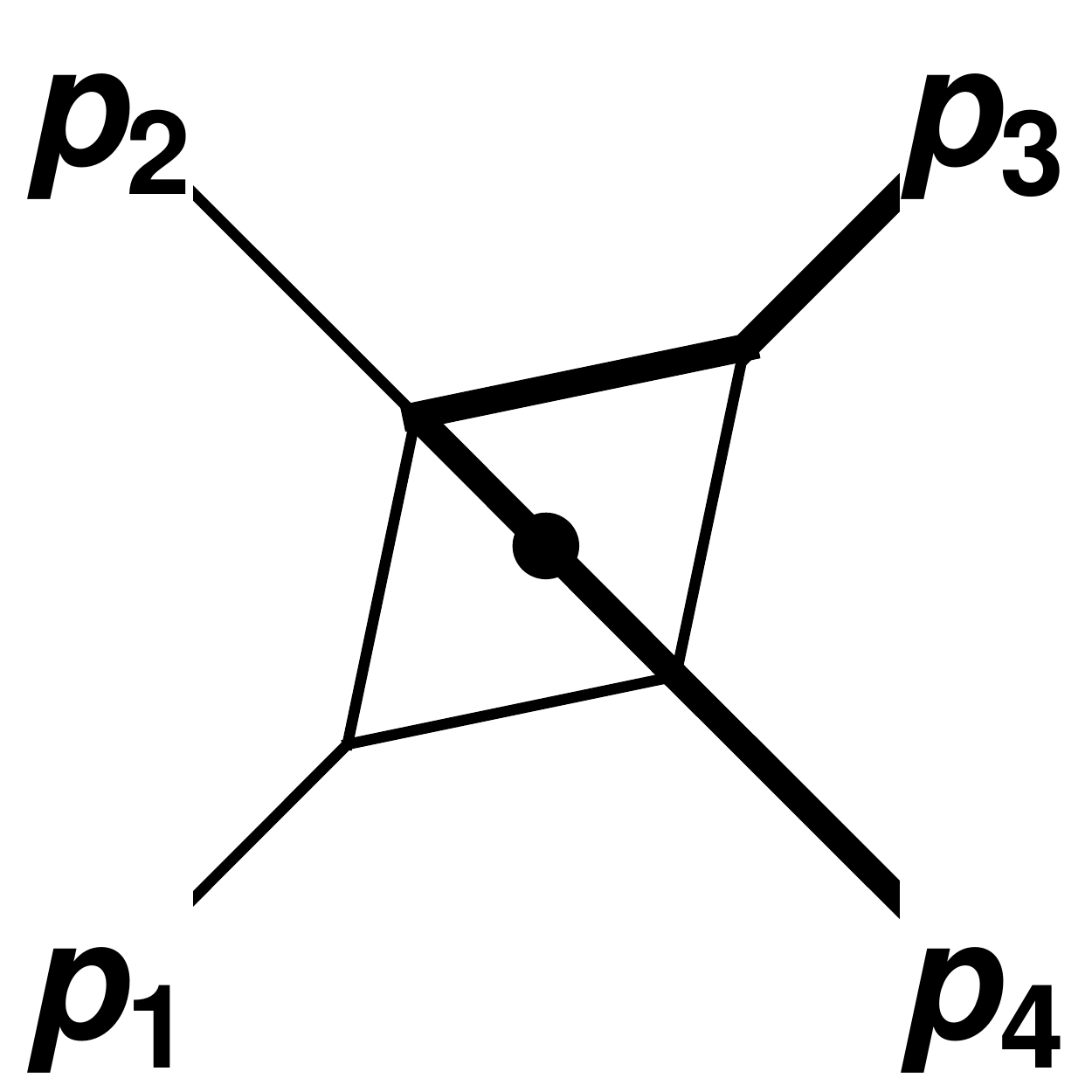}
  }
  \subfloat[$\mathcal{T}_{29}$]{%
    \includegraphics[width=0.15\textwidth]{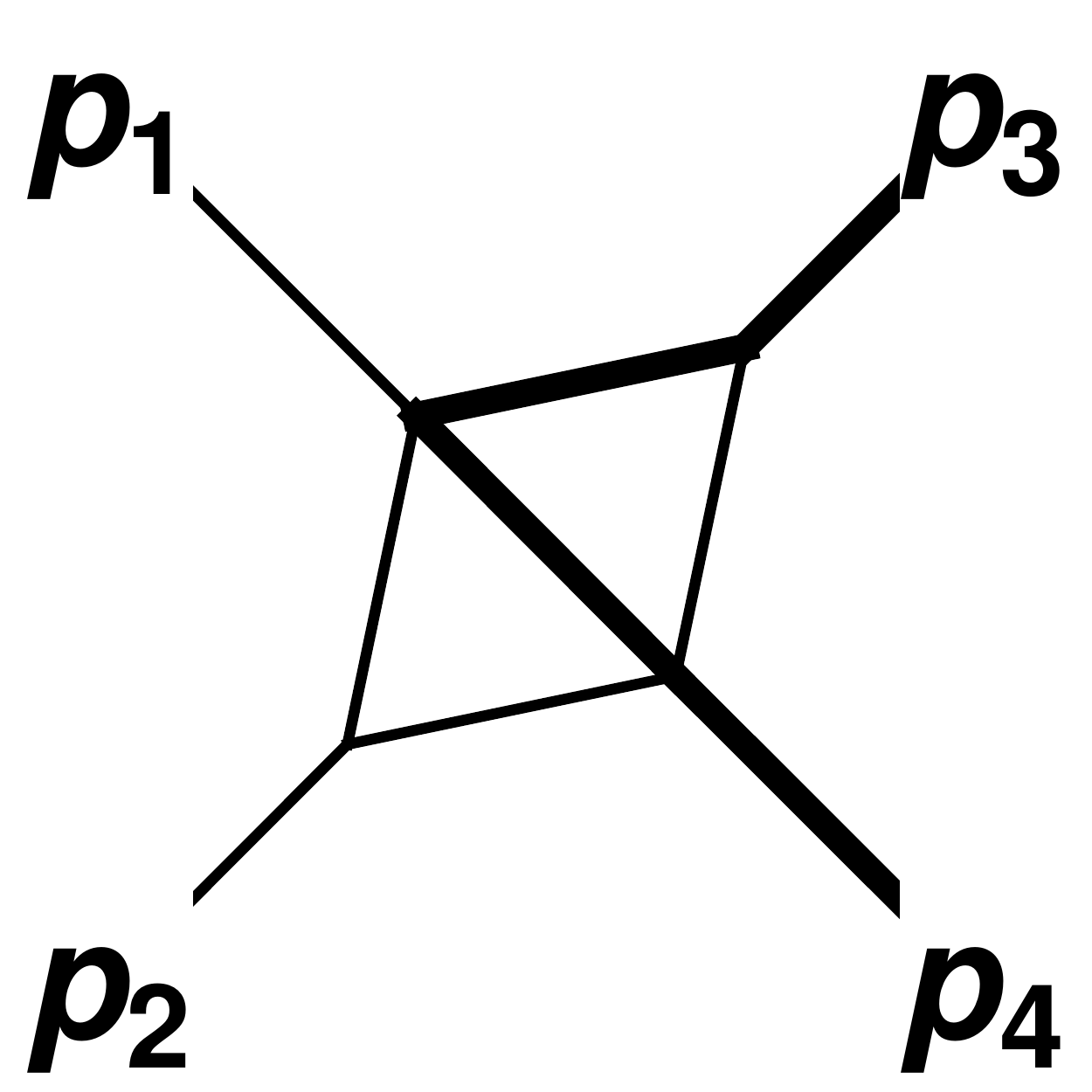}
  }
  \subfloat[$\mathcal{T}_{30}$]{%
    \includegraphics[width=0.15\textwidth]{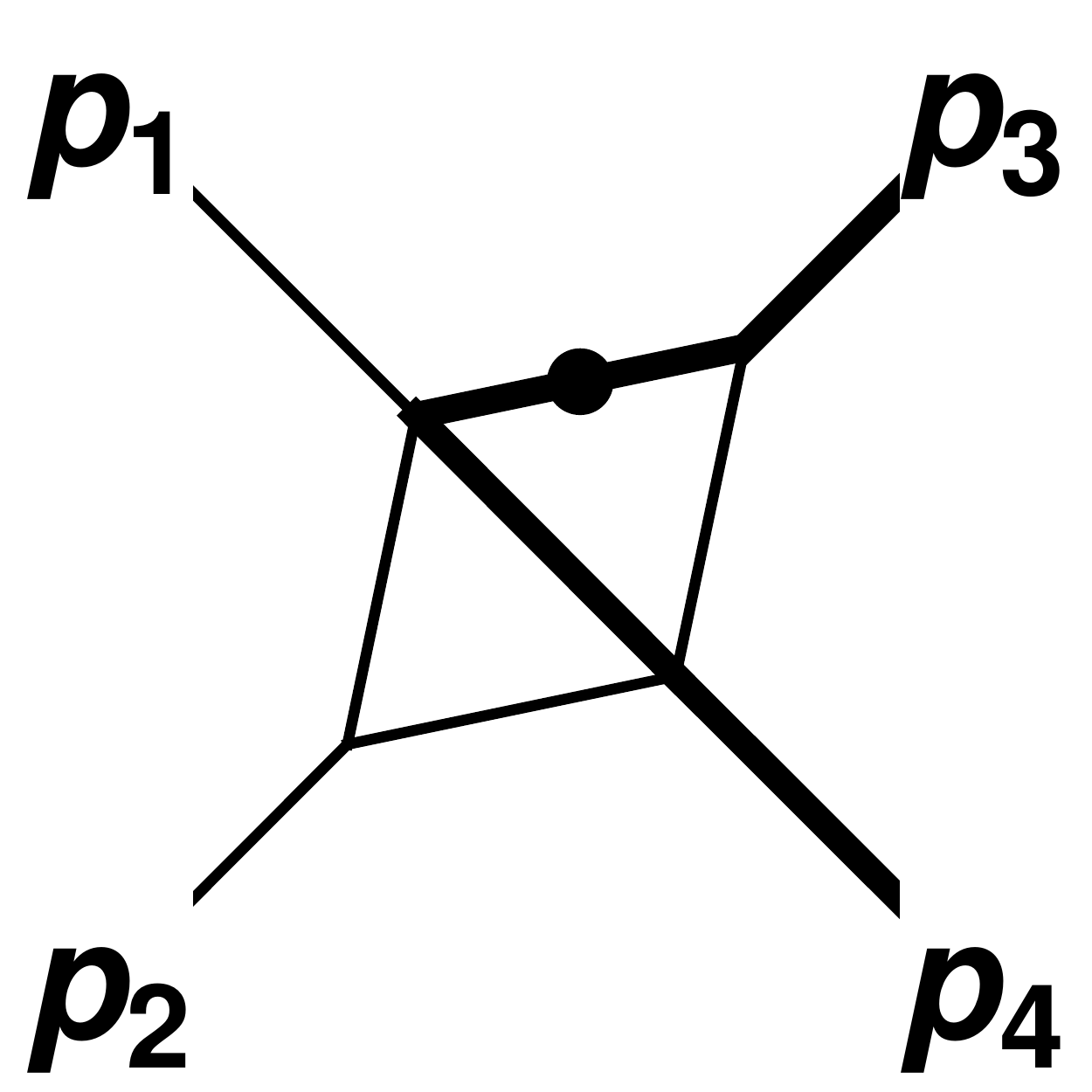}
  }
      \subfloat[$\mathcal{T}_{31}$]{%
    \includegraphics[width=0.15\textwidth]{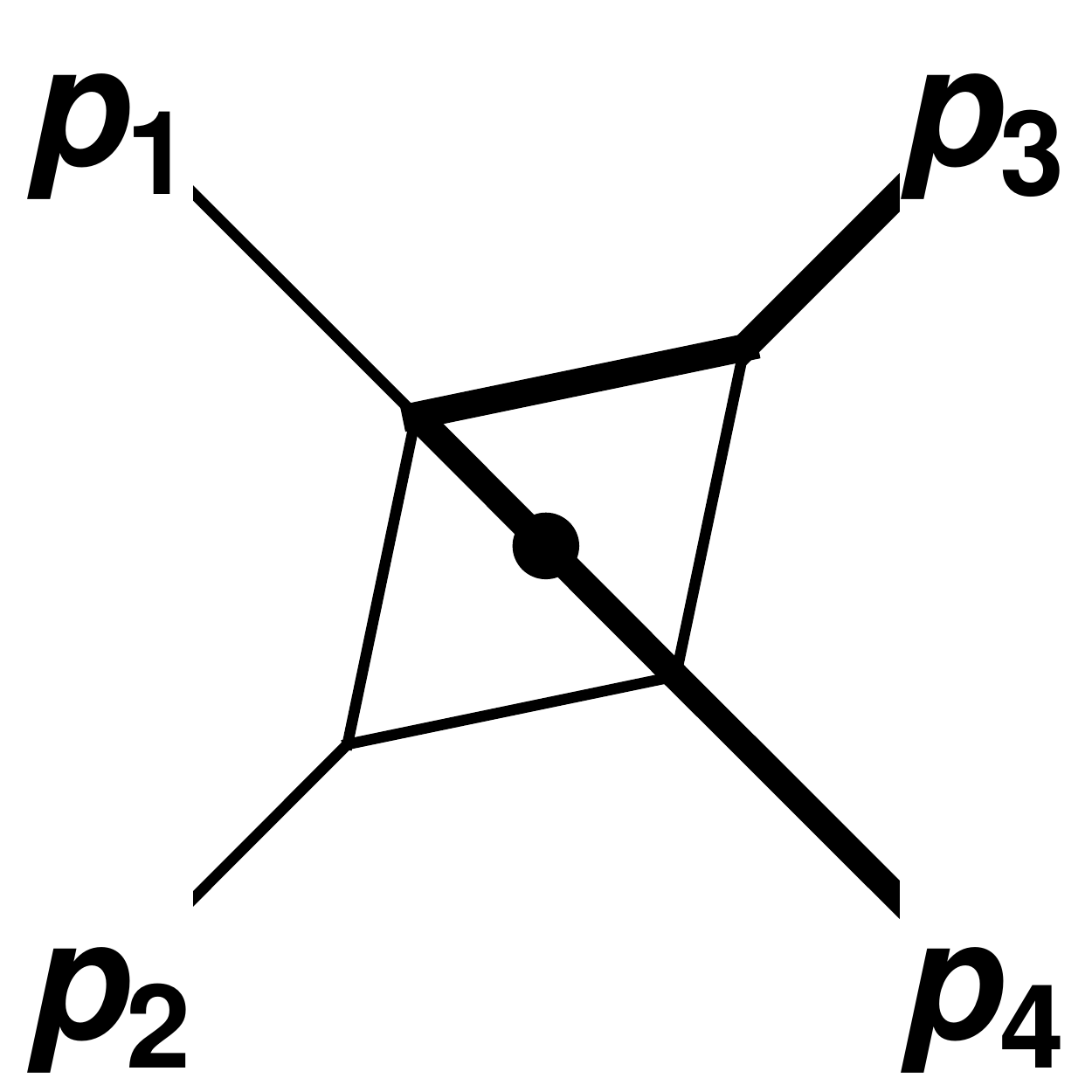}
  }\\
  \subfloat[$\mathcal{T}_{32}$]{%
    \includegraphics[width=0.15\textwidth]{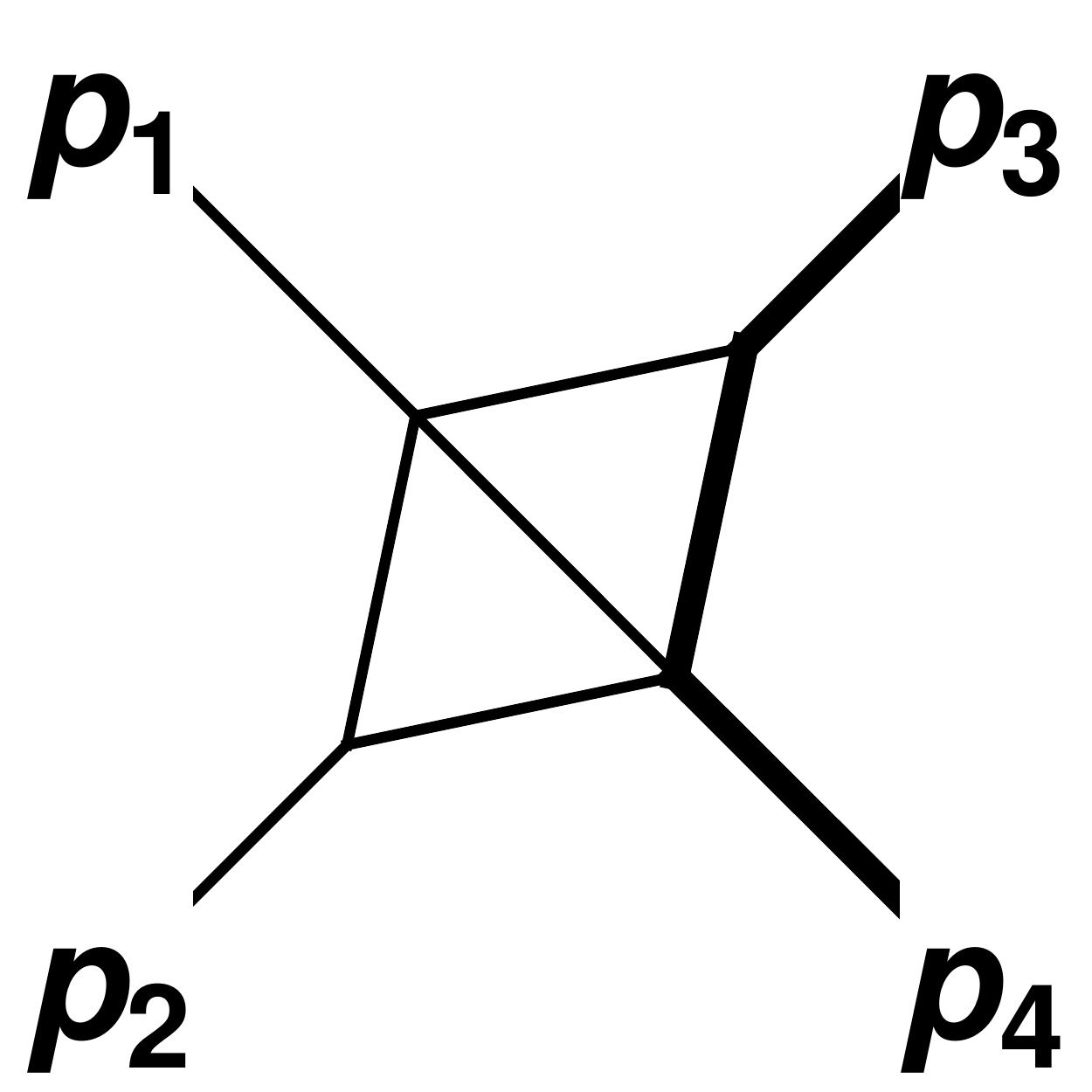}
  }
  \subfloat[$\mathcal{T}_{33}$]{%
    \includegraphics[width=0.15\textwidth]{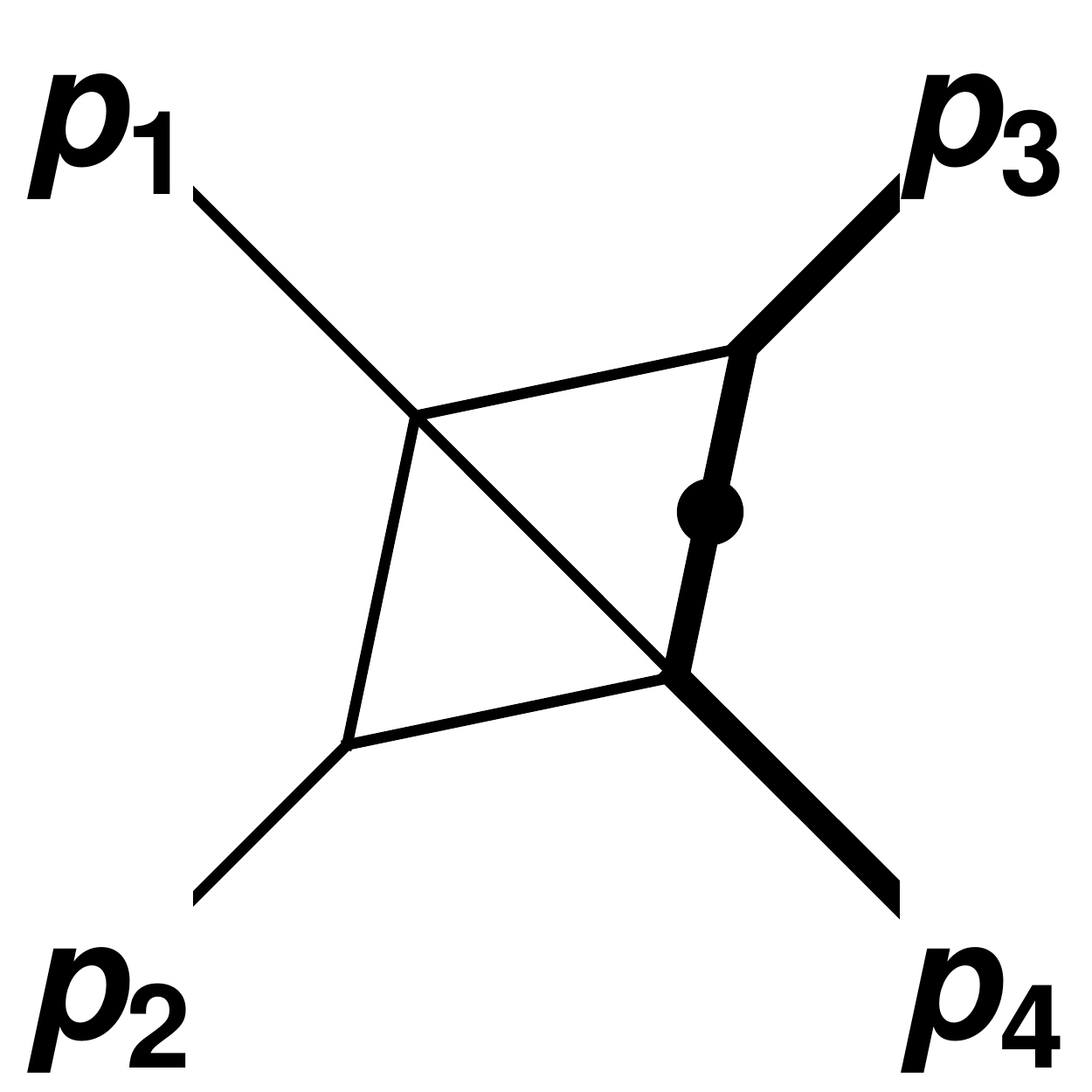}
  }
  \subfloat[$\mathcal{T}_{34}$]{%
    \includegraphics[width=0.15\textwidth]{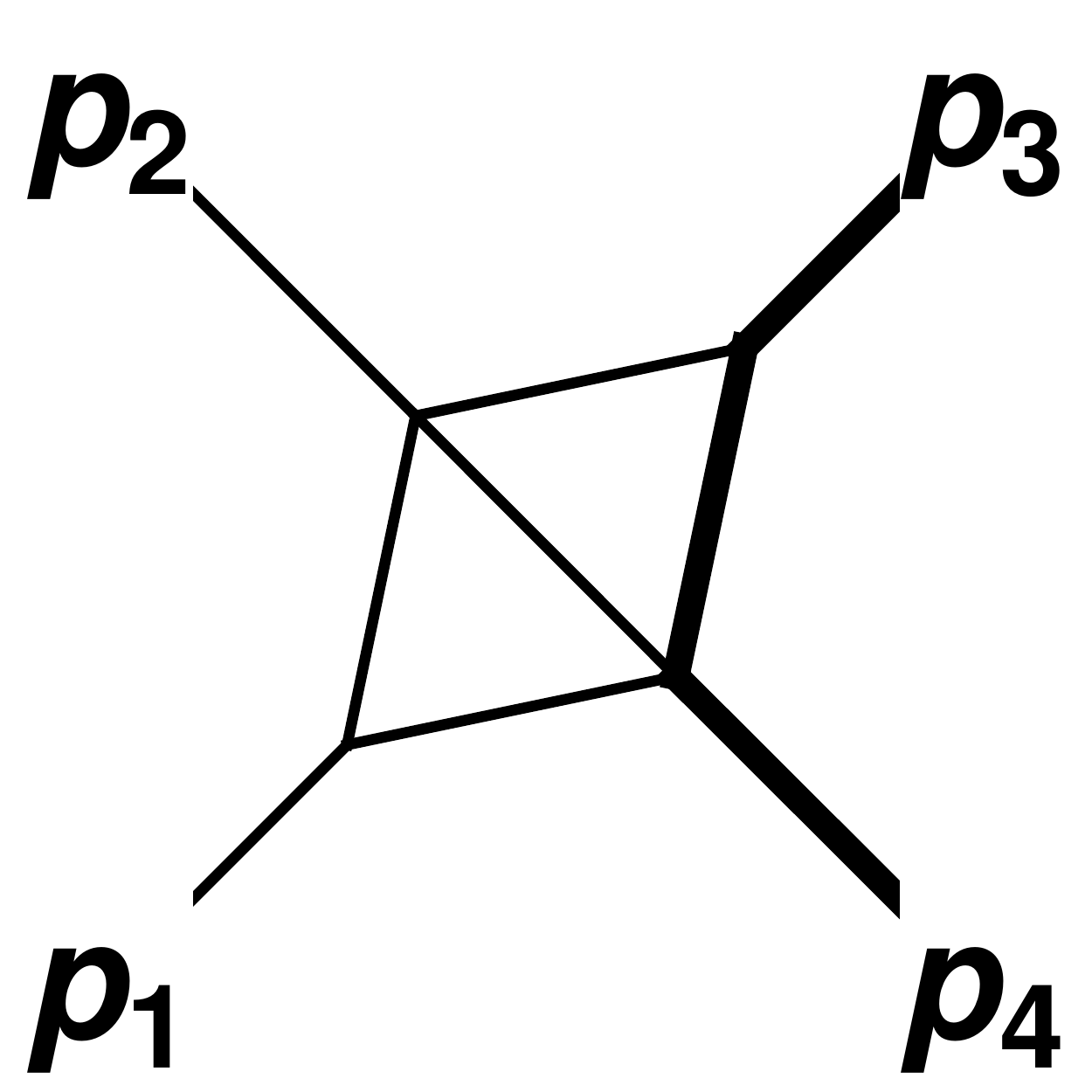}
  }
  \subfloat[$\mathcal{T}_{35}$]{%
    \includegraphics[width=0.15\textwidth]{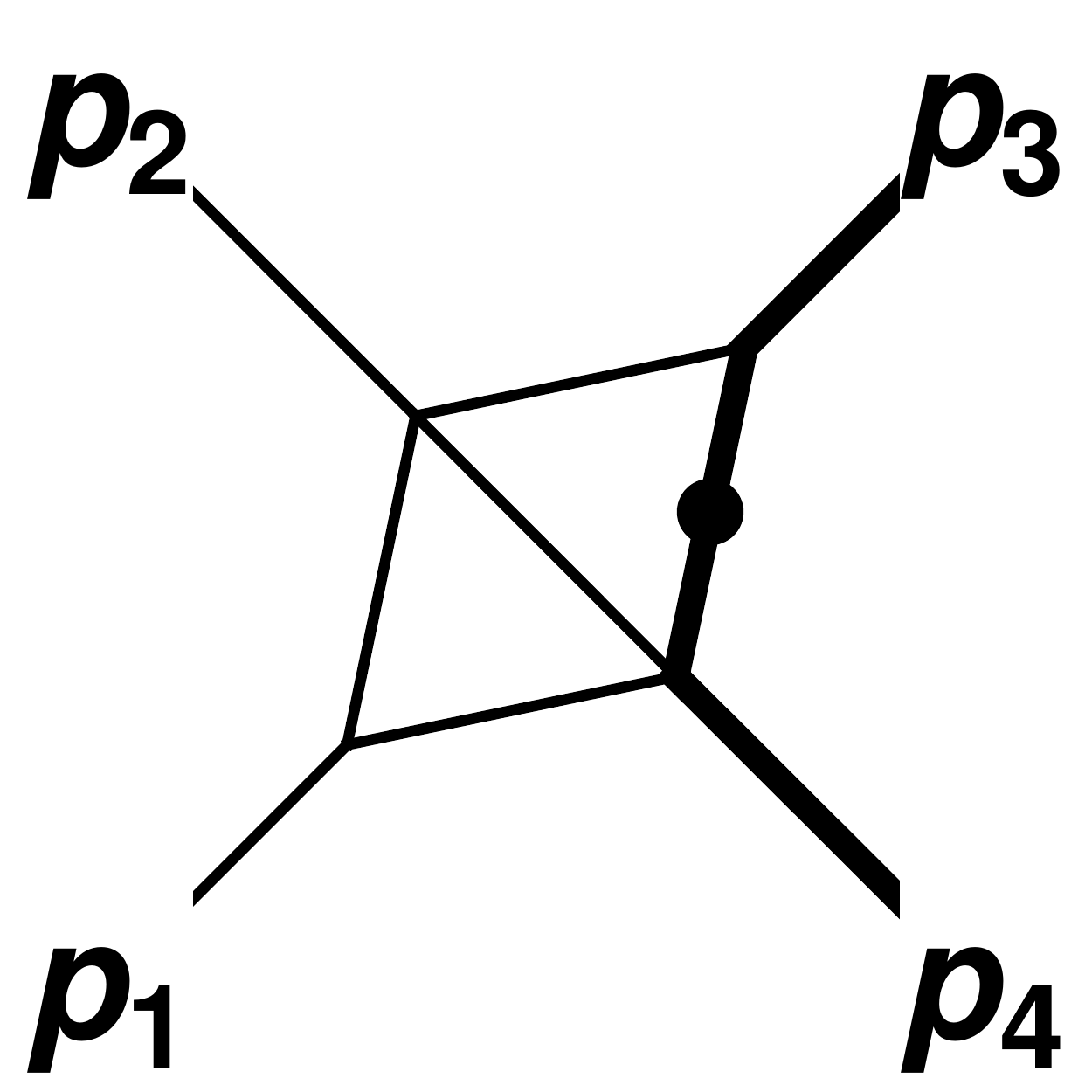}
  }
  \subfloat[$\mathcal{T}_{36}$]{%
    \includegraphics[width=0.15\textwidth]{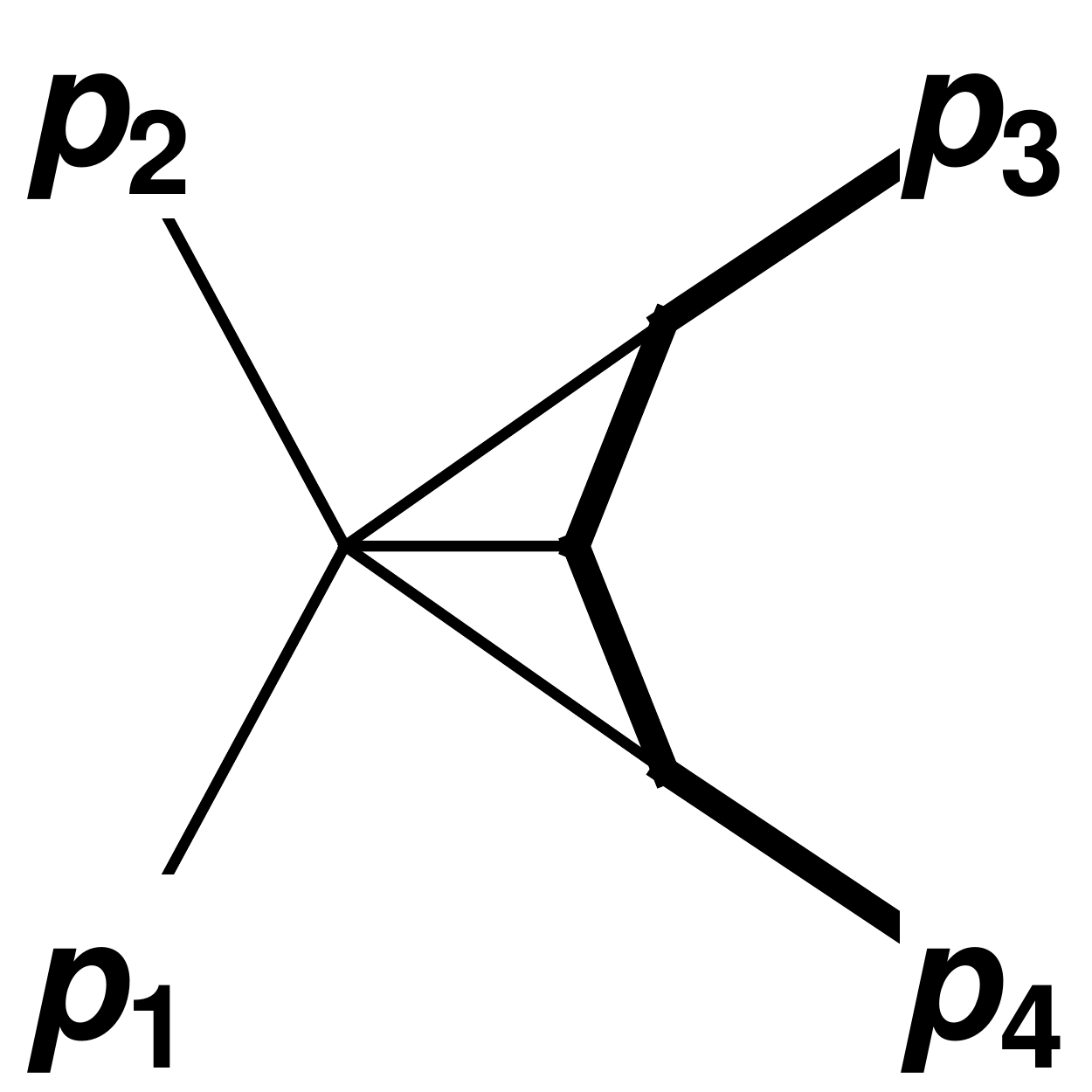}
  }
      \subfloat[$\mathcal{T}_{37}$]{%
    \includegraphics[width=0.15\textwidth]{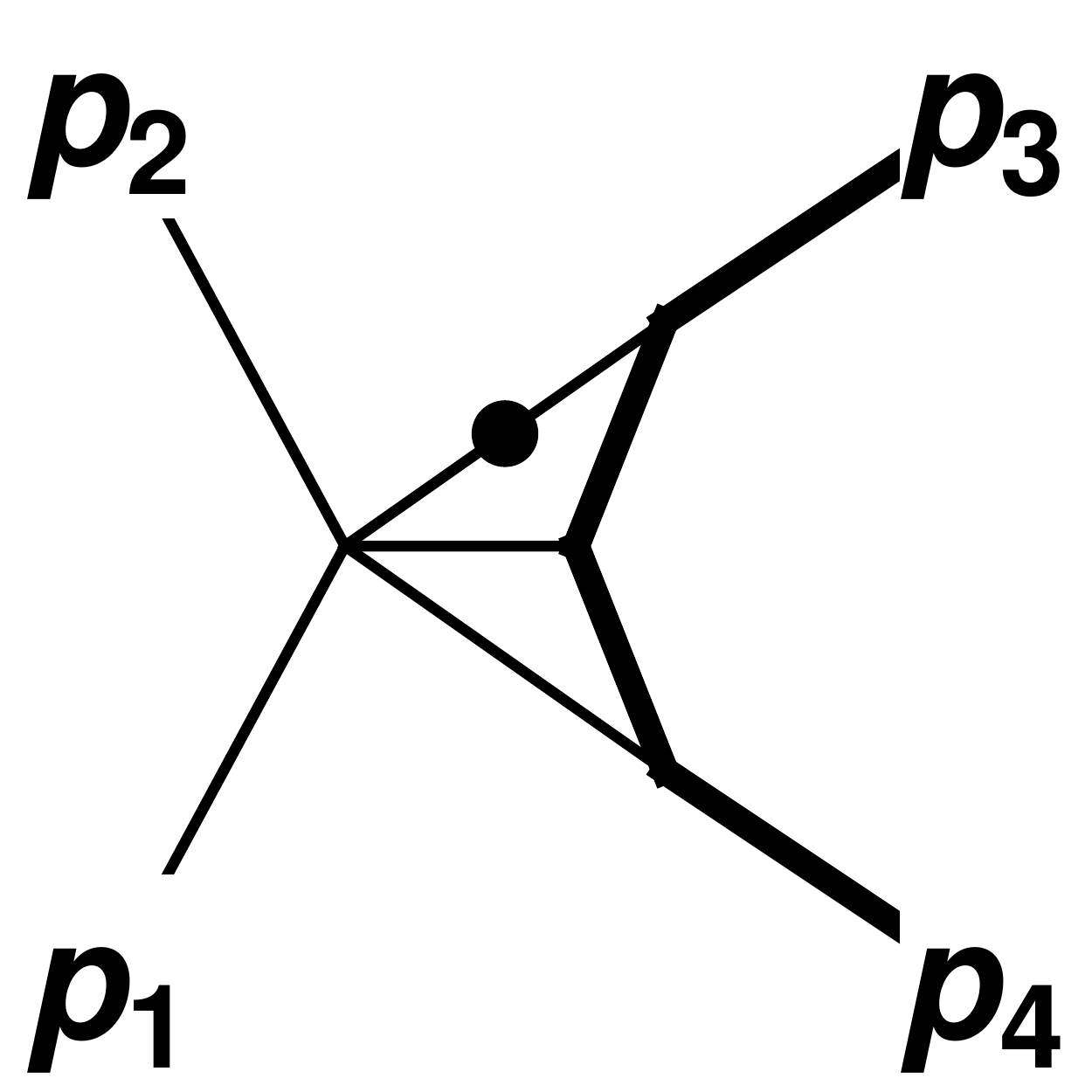}
  }\\
  \subfloat[$\mathcal{T}_{38}$]{%
    \includegraphics[width=0.15\textwidth]{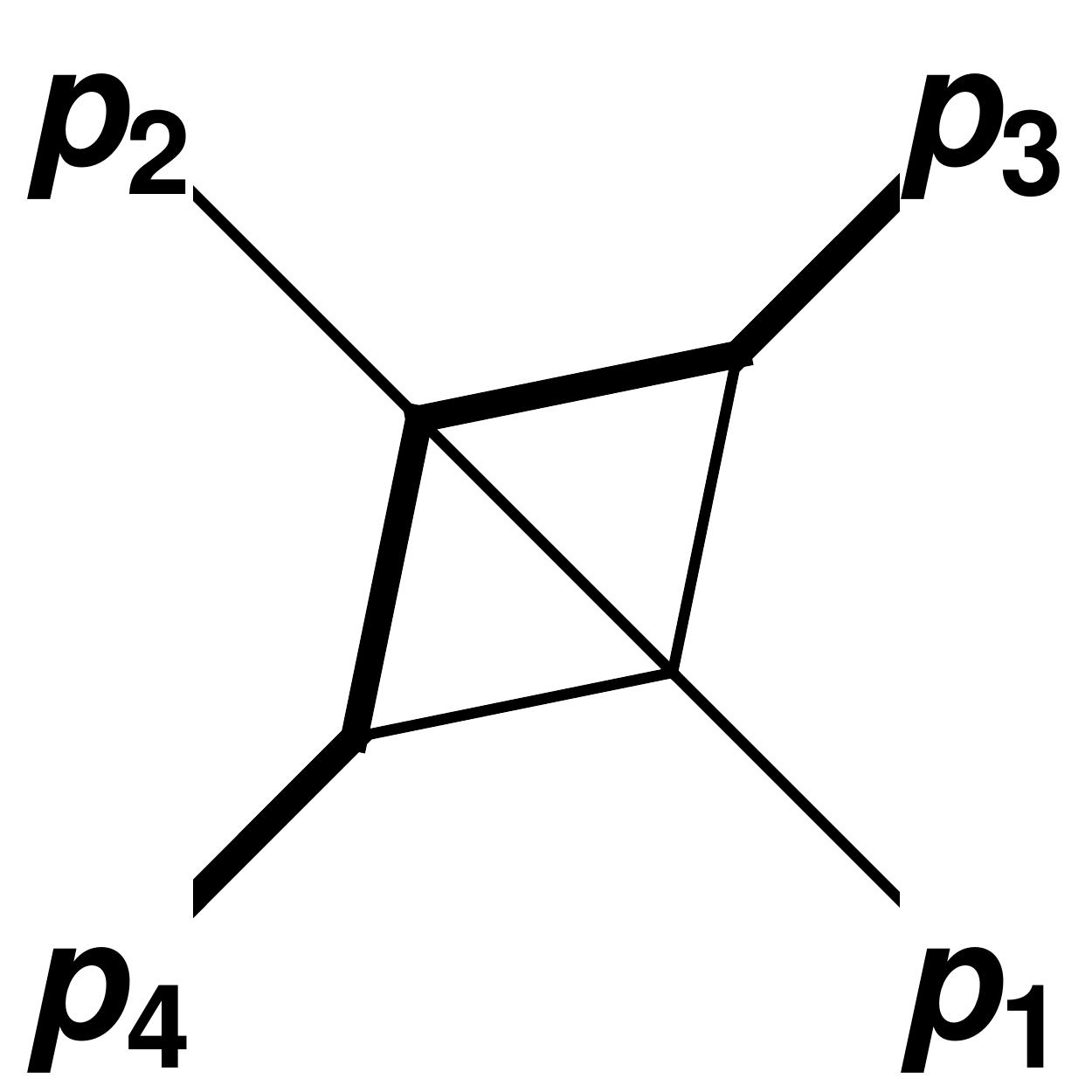}
  }
  \subfloat[$\mathcal{T}_{39}$]{%
    \includegraphics[width=0.15\textwidth]{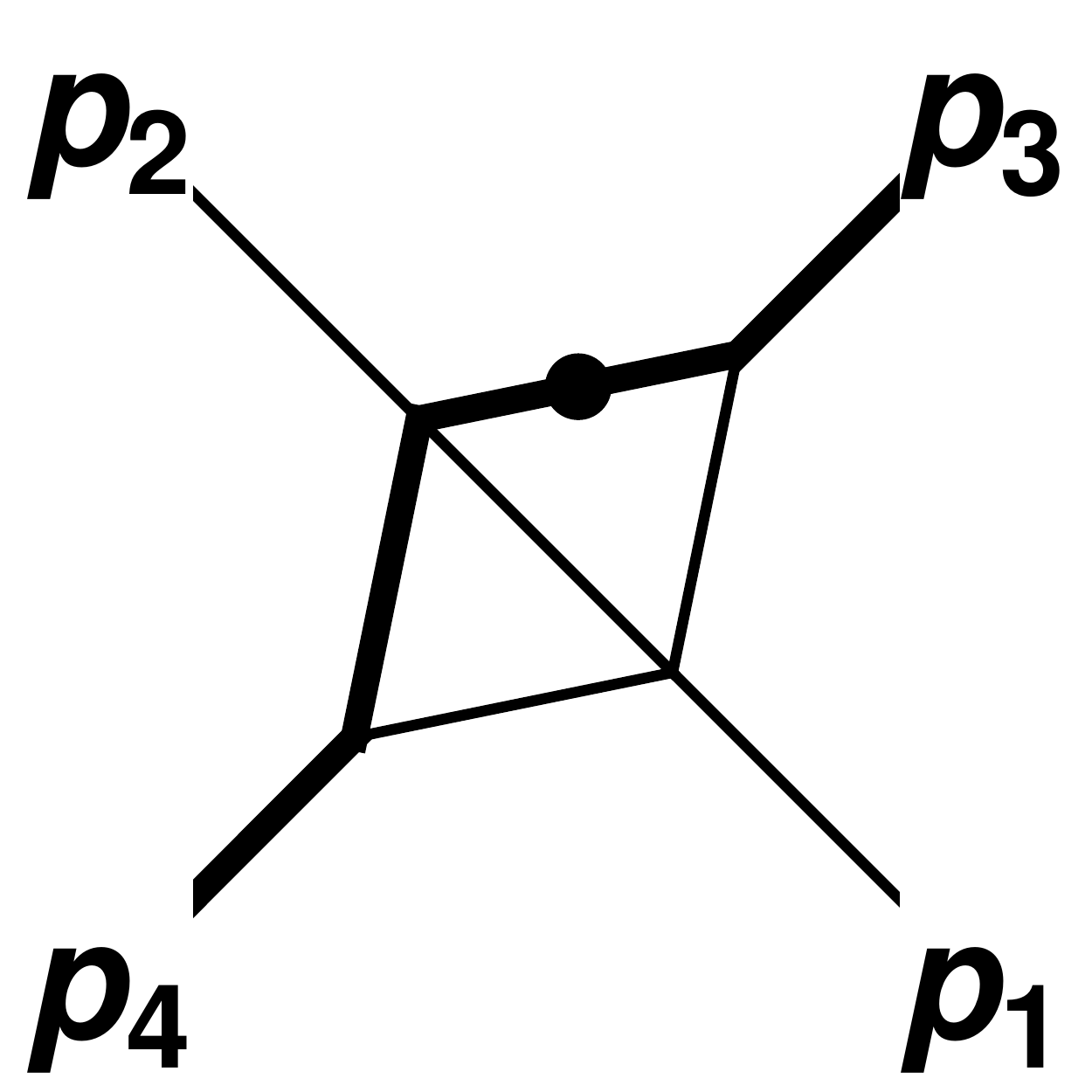}
  }
  \subfloat[$\mathcal{T}_{40}$]{%
    \includegraphics[width=0.15\textwidth]{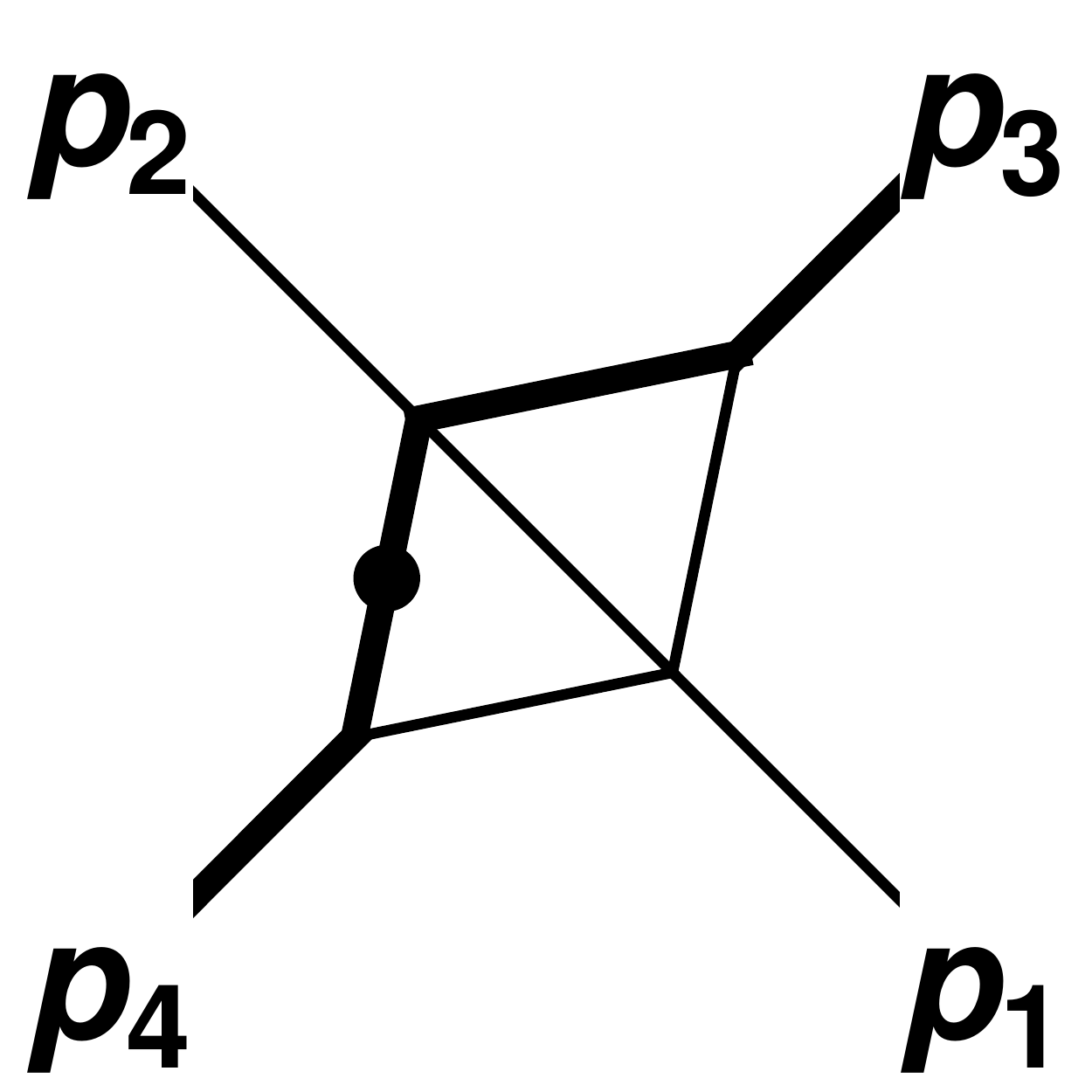}
  }
  \subfloat[$\mathcal{T}_{41}$]{%
    \includegraphics[width=0.15\textwidth]{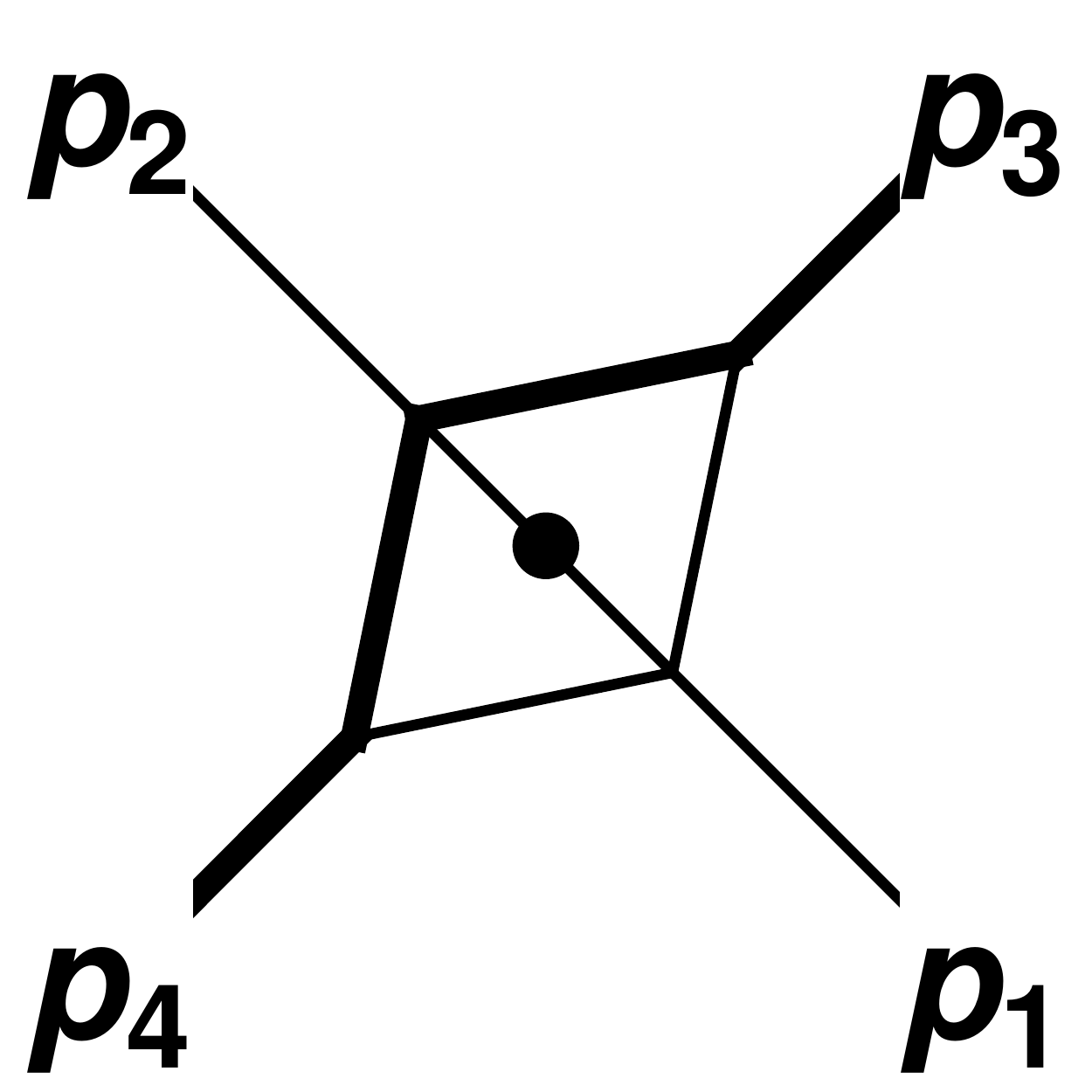}
  }
  \subfloat[$\mathcal{T}_{42}$]{%
    \includegraphics[width=0.15\textwidth]{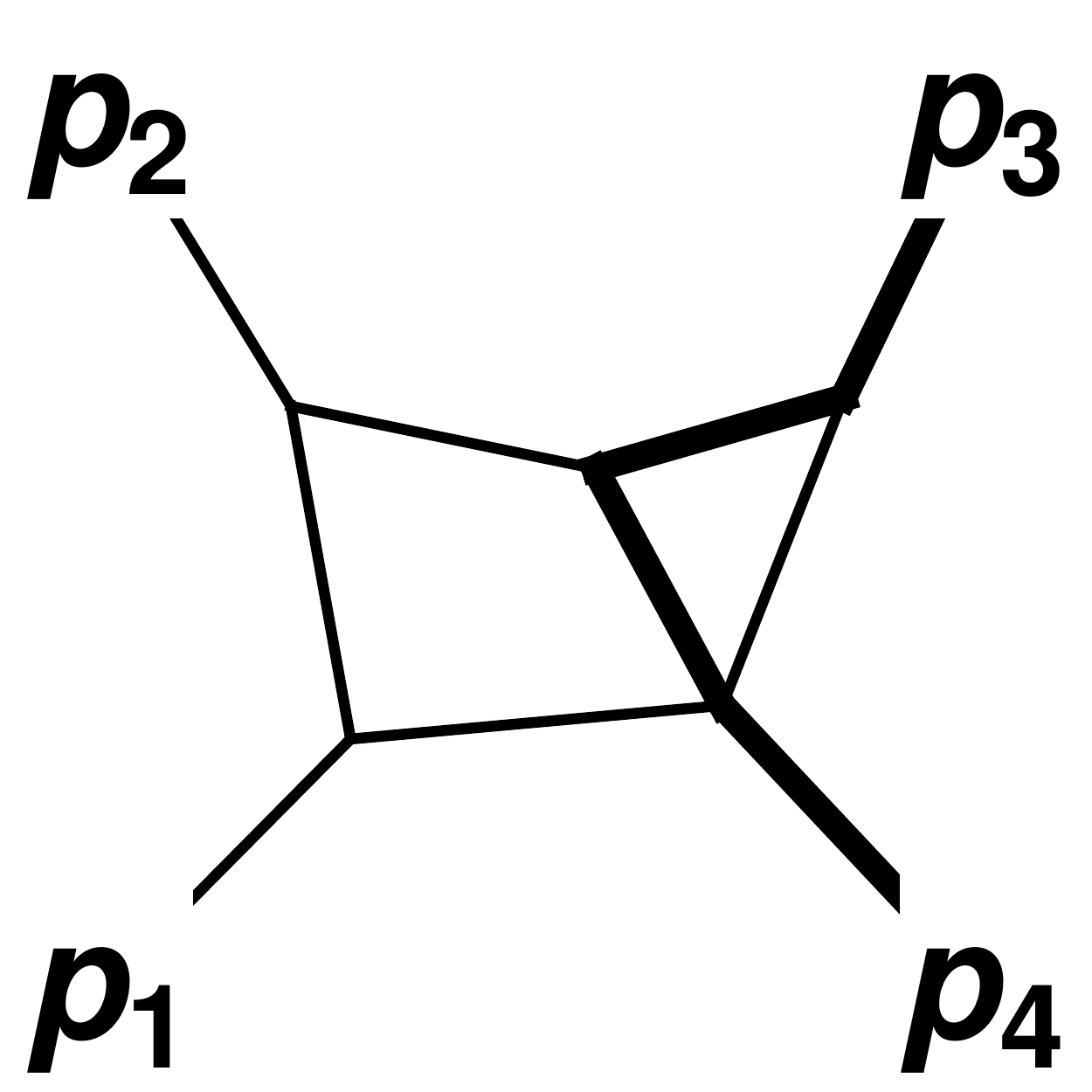}
  }
        \subfloat[$\mathcal{T}_{43}$]{%
    \includegraphics[width=0.15\textwidth]{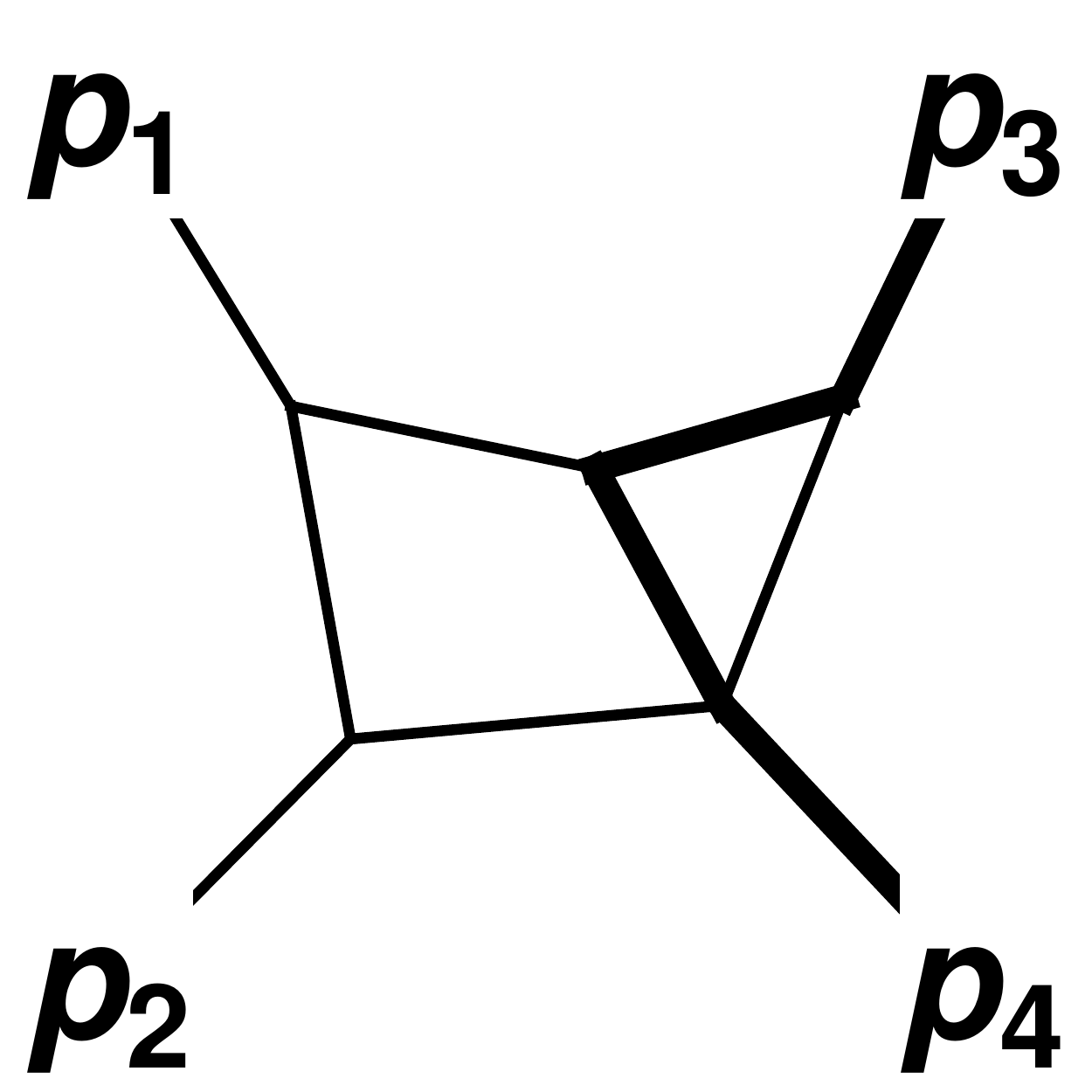}
  }\\
  \subfloat[$\mathcal{T}_{44}$]{%
    \includegraphics[width=0.15\textwidth]{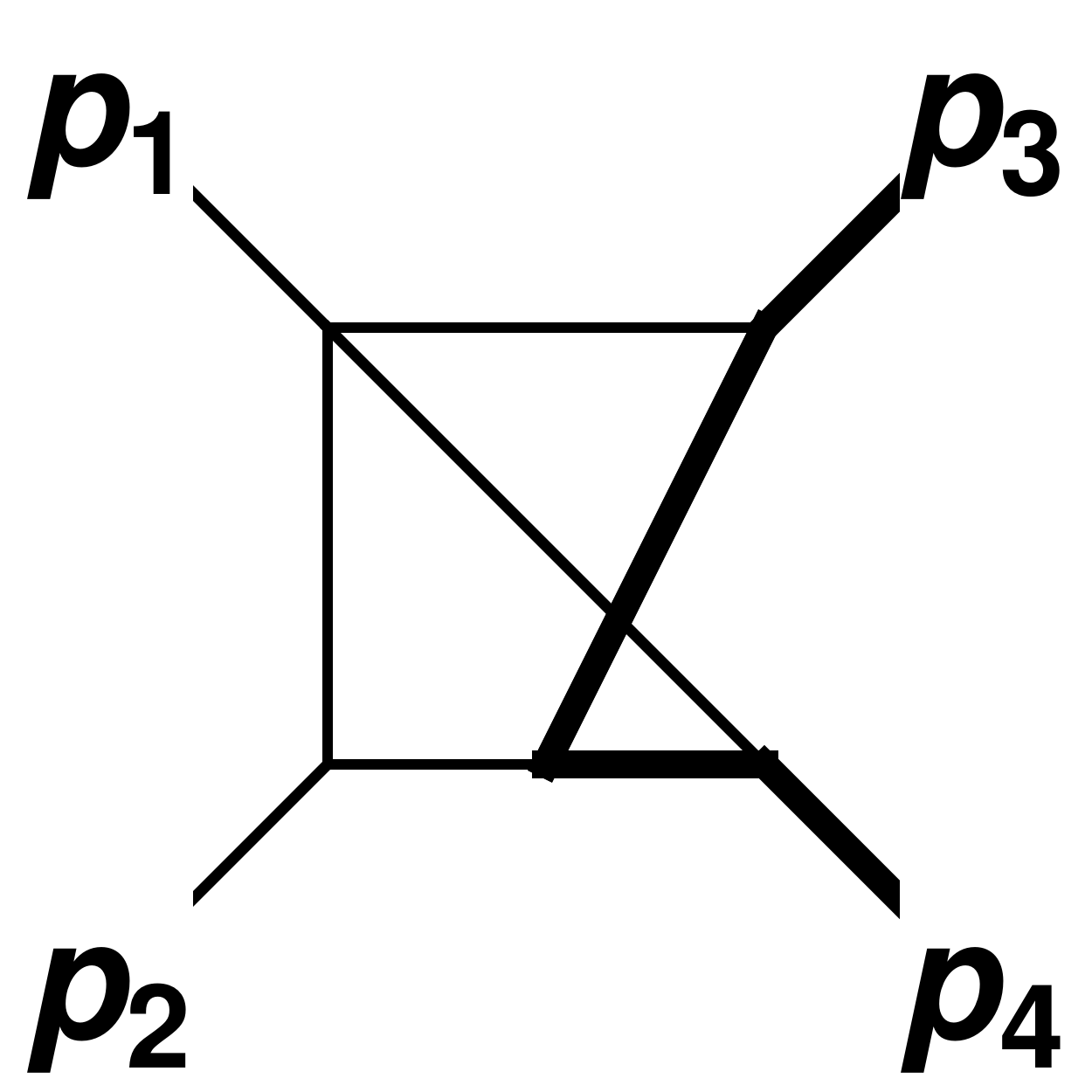}
  }
  \subfloat[$\mathcal{T}_{45}$]{%
    \includegraphics[width=0.15\textwidth]{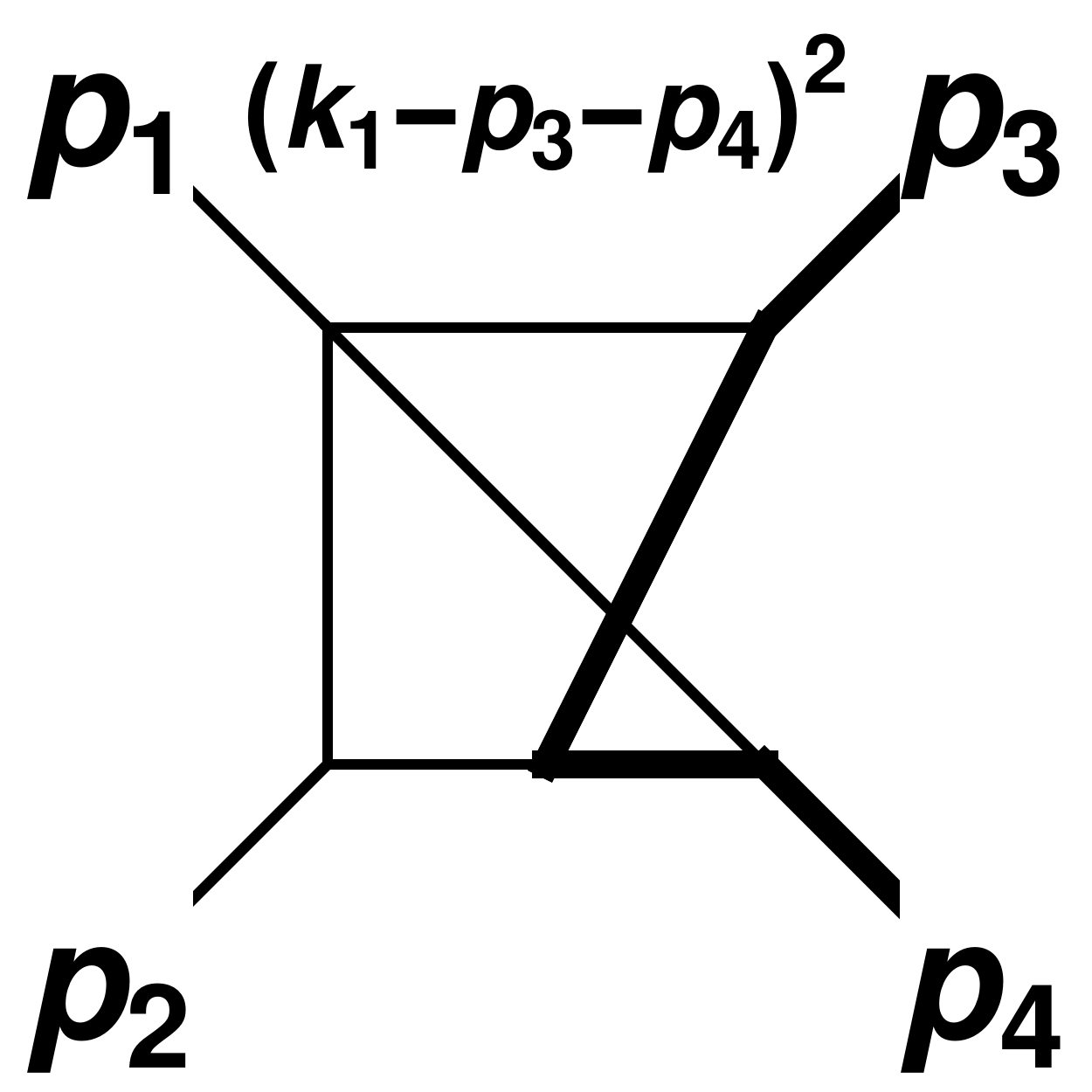}
  }
  \subfloat[$\mathcal{T}_{46}$]{%
    \includegraphics[width=0.15\textwidth]{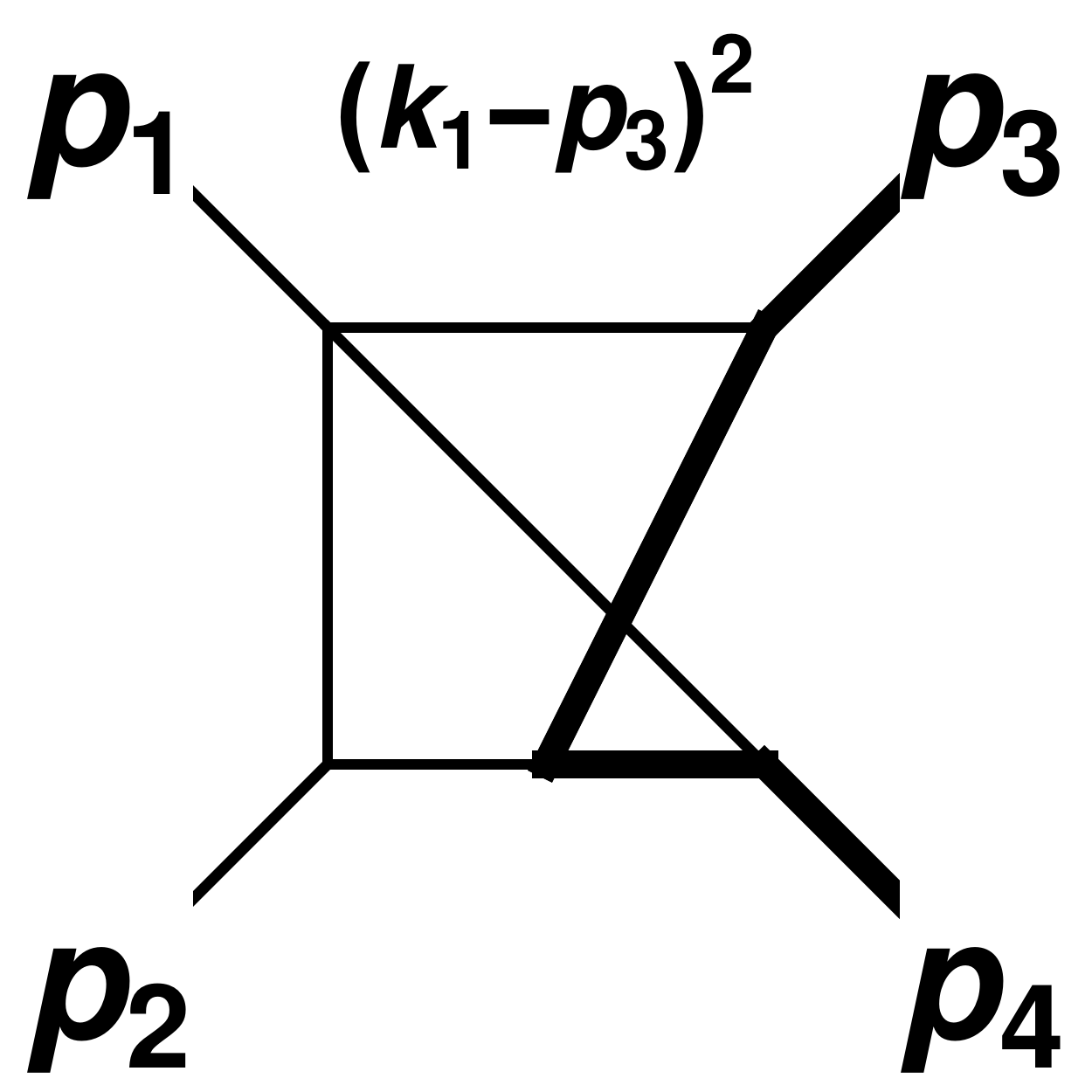}
  }
  \subfloat[$\mathcal{T}_{47}$]{%
    \includegraphics[width=0.15\textwidth]{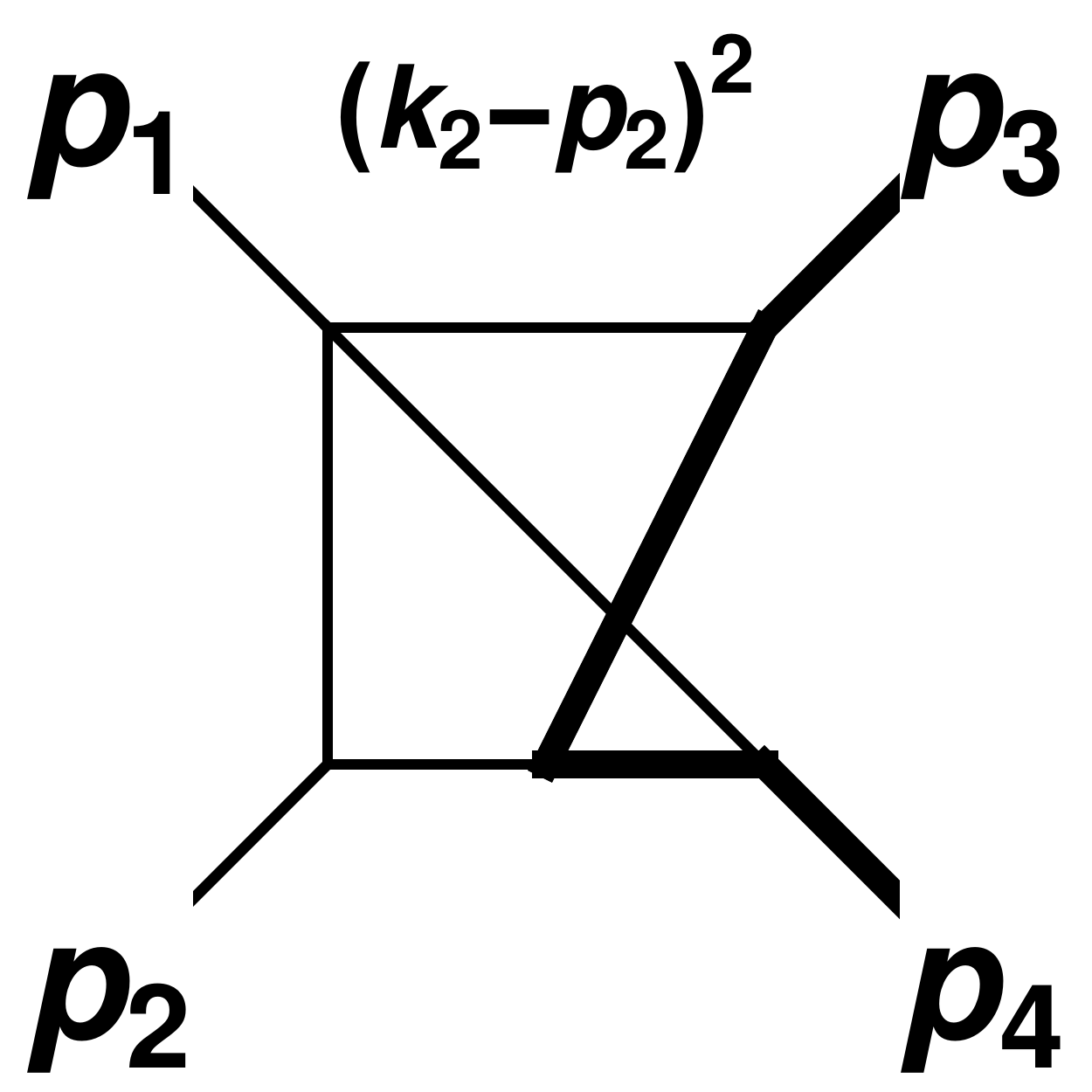}
  }
  \subfloat[$\mathcal{T}_{48}$]{%
    \includegraphics[width=0.15\textwidth]{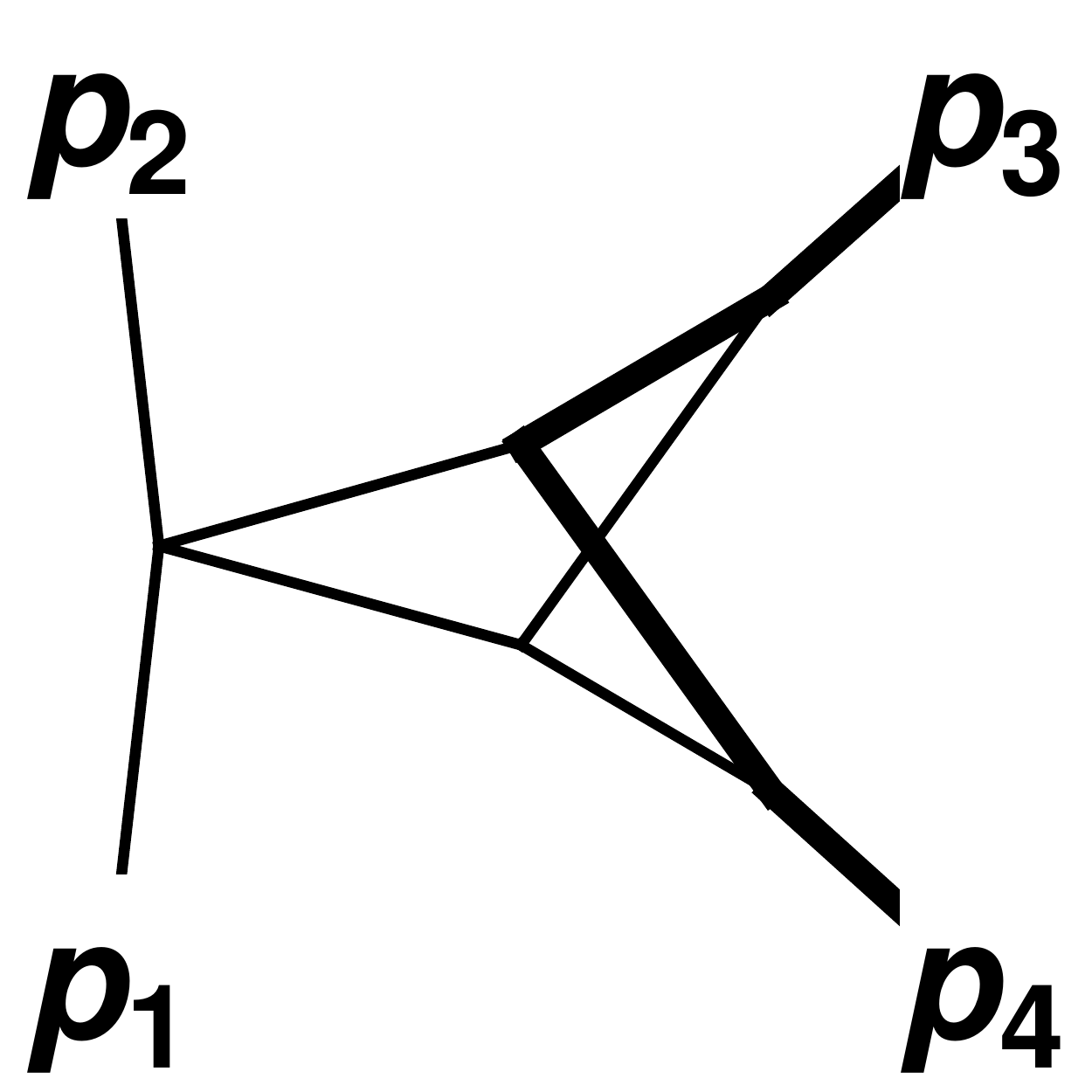}
  }\\
  \subfloat[$\mathcal{T}_{49}$]{%
    \includegraphics[width=0.15\textwidth]{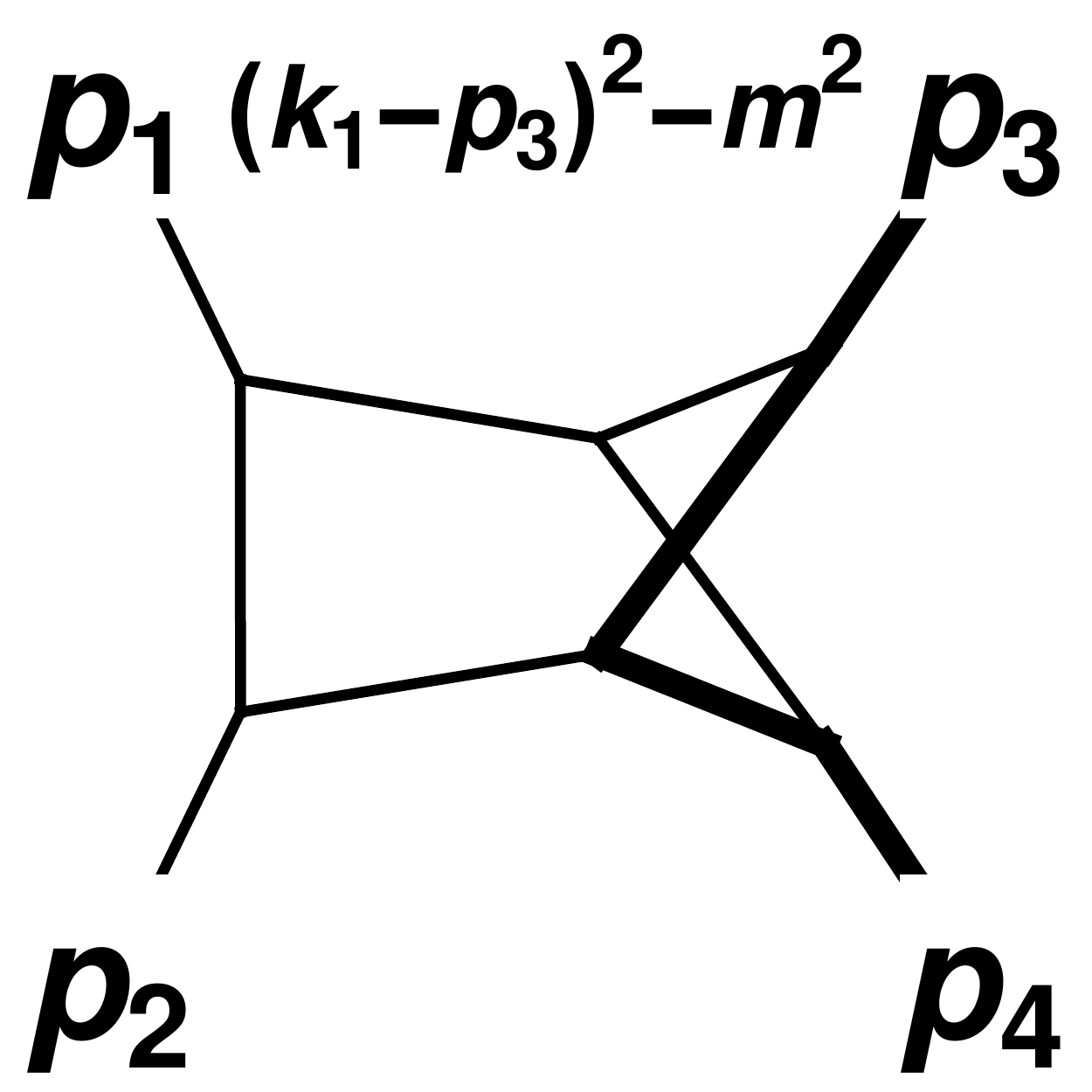}
  }
  \subfloat[$\mathcal{T}_{50}$]{%
    \includegraphics[width=0.15\textwidth]{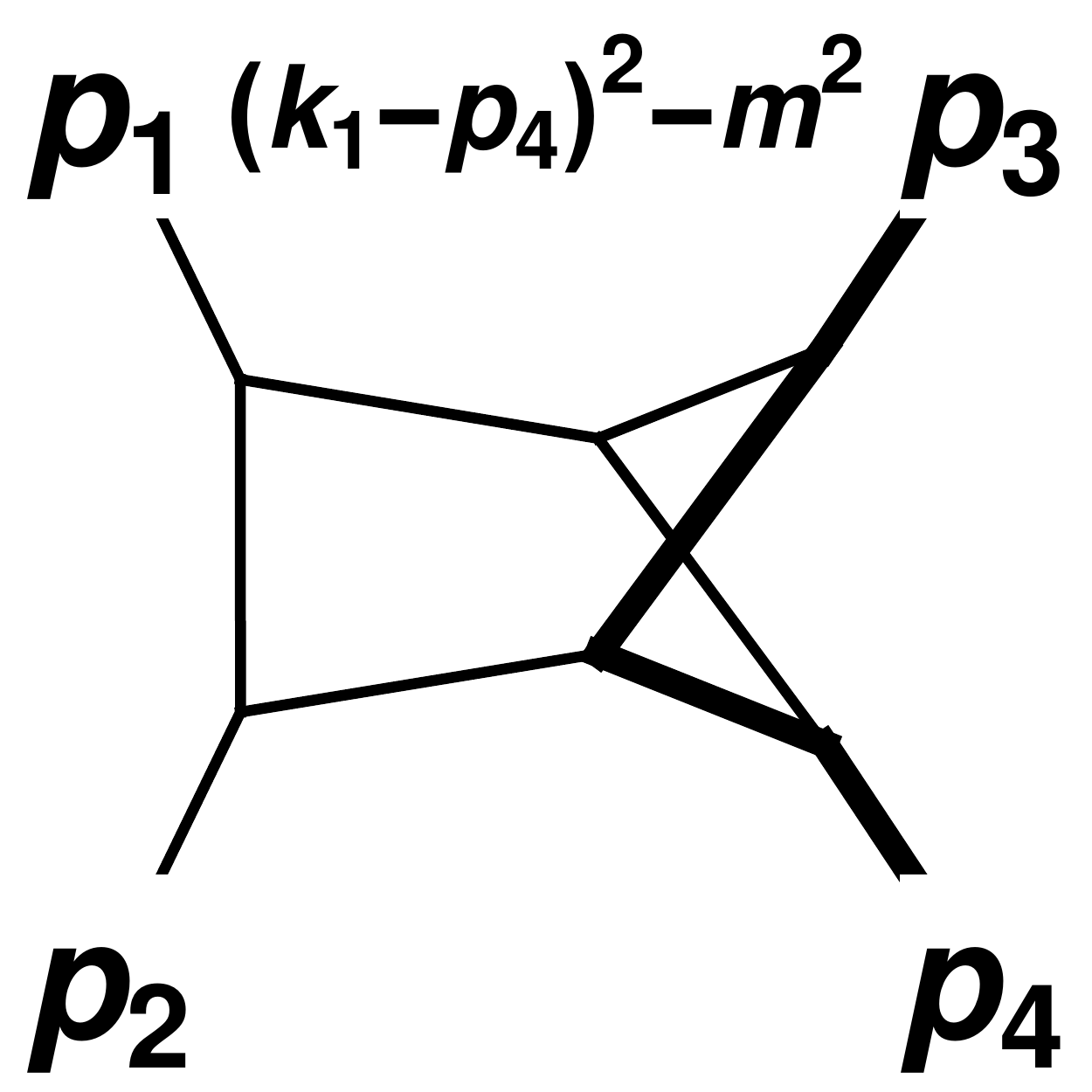}
  }
    \subfloat[$\mathcal{T}_{51}$]{%
    \includegraphics[width=0.15\textwidth]{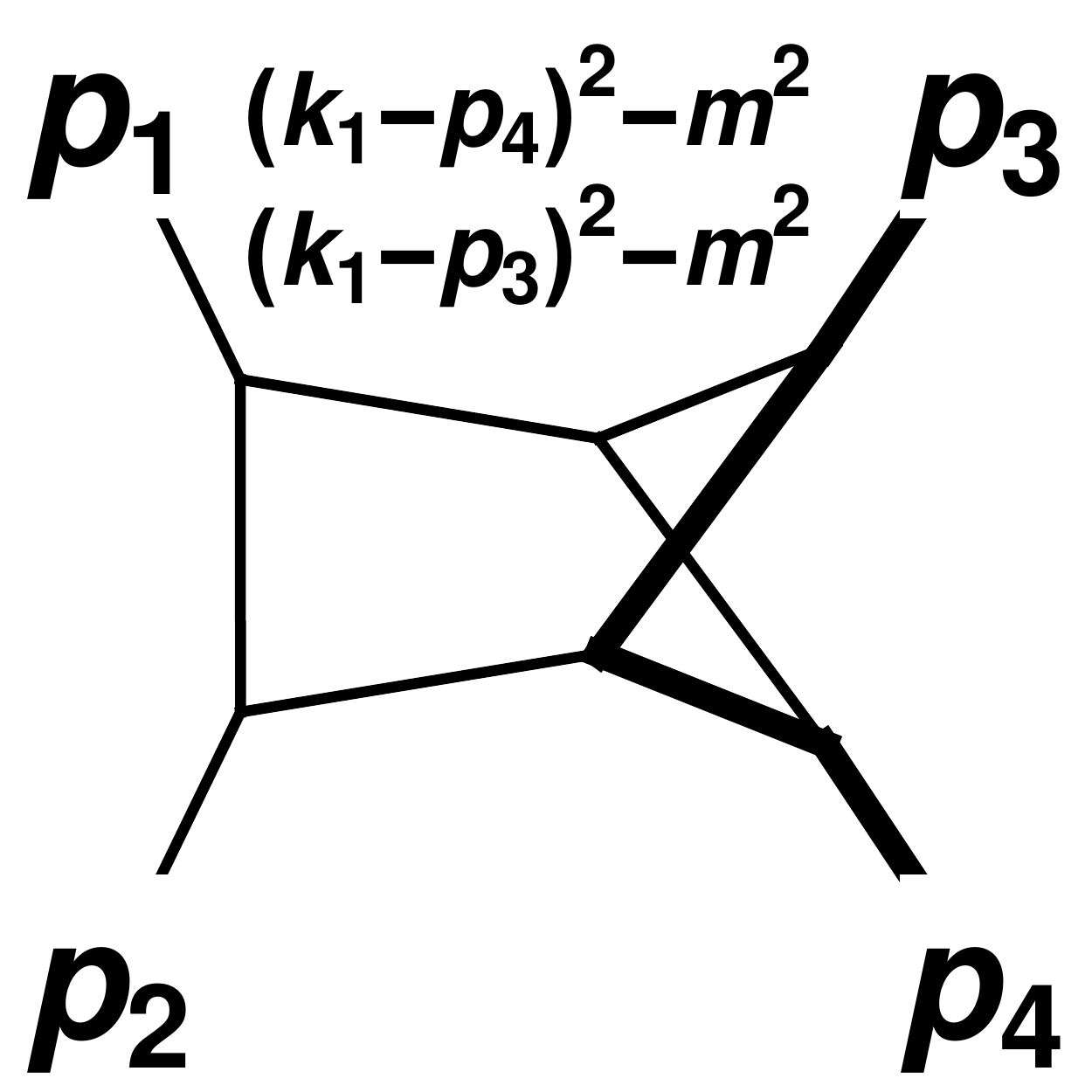}
  }
  \subfloat[$\mathcal{T}_{52}$]{%
    \includegraphics[width=0.15\textwidth]{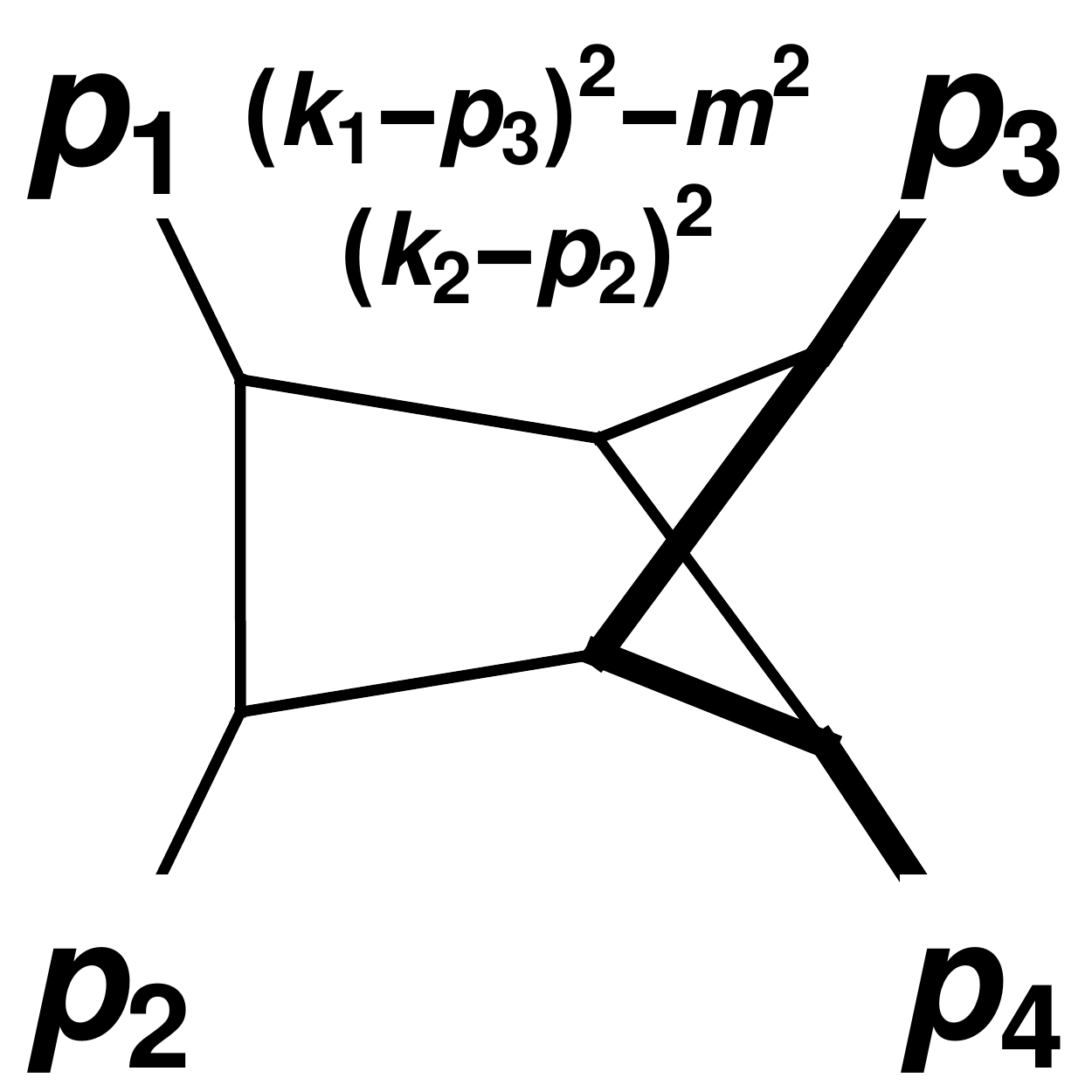}
  }
 \caption{
The remaining 27 MIs $\mathcal{T}_{26,\ldots,52}$ for the two-loop non-planar topology $N_2$ of figure~\ref{fig:npgraphs}. The conventions are the same as in figure~\ref{fig:MIsT6a}.}
 \label{fig:MIsT6b}
\end{figure}

%%%%%%%%%%%%%%%%%%
%%% Local Variables:
%%% TeX-master: "../main"
%%% End:

\subsection{Differential equations for master integrals}
\label{sec:subdeq}
We determine a canonical basis of MIs in $d=4-2\eps$ through the Magnus exponential algorithm described in~\cite{Argeri:2014qva,DiVita:2014pza}. As a starting point, we identify a set of 52 independent integrals $\FF_i$, whose DEQs depend linearly on
the dimensional regularization parameter $\eps$,
\begin{align*}
\FF_{1}&=\eps^2 \, \top{1}\,,  &
\FF_{2}&=\eps^2 \, \top{2}\,,  &
\FF_{3}&=\eps^2 \, \top{3}\,,  \\
\FF_{4}&=\eps^2 \, \top{4}\,,  &
\FF_{5}&=\eps^2 \, \top{5}\,,  &
\FF_{6}&=\eps^2 \, \top{6}\,,  \\
\FF_{7}&=\eps^2 \, \top{7}\,,  &
\FF_{8}&=\eps^2 \, \top{8}\,,  &
\FF_{9}&=\eps^2 \, \top{9}\,,  \\
\FF_{10}&=\eps^2 \, \top{10}\,,  &
\FF_{11}&=\eps^3 \, \top{11}\,,  &
\FF_{12}&=\eps^2 \, \top{12}\,,  \\
\FF_{13}&=\eps^3 \, \top{13}\,,  &
\FF_{14}&=\eps^2 \, \top{14}\,,  &
\FF_{15}&=\eps^3 \, \top{15}\,,  \\
\FF_{16}&=\eps^2 \, \top{16}\,,  &
\FF_{17}&=\eps^2 \, \top{17}\,,  &
\FF_{18}&=\eps^3\, \top{18}\,,  \\
\FF_{19}&=\eps^3\, \top{19}\,,  &
\FF_{20}&=\eps^4\, \top{20}\,,  &
\FF_{21}&=\eps^2 \, \top{21}\,, \\
\FF_{22}&=\eps^3 \, \top{22}\,, &
\FF_{23}&=\eps^2 \, \top{23}\,, &
\FF_{24}&=\eps^3 \, \top{24}\,, \\
\FF_{25}&=\eps^2 \, \top{25}\,, &
\FF_{26}&=\eps^4 \, \top{26}\,, &
\FF_{27}&=\eps^3 \, \top{27}\,, \\
\FF_{28}&=\eps^3 \, \top{28}\,, &
\FF_{29}&=\eps^4 \, \top{29}\,, &
\FF_{30}&=\eps^3 \, \top{30}\,, \\
\FF_{31}&=\eps^3 \, \top{31}\,, &
\FF_{32}&=\eps^4 \, \top{32}\,, &
\FF_{33}&=\eps^3 \, \top{33}\,, \\
\FF_{34}&=\eps^4 \, \top{34}\,, &
\FF_{35}&=\eps^3 \, \top{35}\,, &
\FF_{36}&=\eps^4 \, \top{36}\,, \\
\FF_{37}&=(1+2\eps)\eps^2 \, \top{37}\,, &
\FF_{38}&=\eps^4 \, \top{38}\,, &
\FF_{39}&=\eps^3 \, \top{39}\,, \\
\FF_{40}&=\eps^3 \, \top{40}\,, &
\FF_{41}&=\eps^3 \, \top{41}\,, &
\FF_{42}&=\eps^4 \, \top{42}\,, \\
\FF_{43}&=\eps^4\, \top{43}\,, &
\FF_{44}&=\eps^4 \, \top{44}\,, &
\FF_{45}&=\eps^4 \, \top{45}\,, \\
\FF_{46}&=\eps^4\, \top{46}\,, &
\FF_{47}&=\eps^4 \, \top{47}\,, &
\FF_{48}&=\eps^4 \, \top{48}\,, \\
\FF_{49}&=\eps^4\, \top{49}\,, &
\FF_{50}&=\eps^4 \, \top{50}\,, &
\FF_{51}&=\eps^4 \, \top{51}\,, \\
\FF_{52}&=\eps^4\, \top{52}\,, & &
\stepcounter{equation}\tag{\theequation}
\label{def:LBasisT6}
\end{align*}
where the $\mathcal{T}_i$ are the integrals graphically represented in figures~\ref{fig:MIsT6a} and~\ref{fig:MIsT6b}.

In particular, the numerators of integrals $\FF_{49,...,52}$, are found by following the ideas in~\cite{Henn:2014qga}, \ie by determining a set of canonical integrals through the inspection of their four-dimensional maximal-cuts.
To this aim, we first localize the integral over $k_2$, which corresponds to the non-planar part of the diagram specified by $\Den_{2,3,6,7}$.
By enforcing the additional constraints $k_1^2$=0 and $s=2k_1\cdot(p_3+p_4)$ (which originate from the cut of the propagators depending on $k_1$), we obtain
\begin{align}
\int d^4k_1\frac{N(k_1)}{D_1D_4D_5}\int d^4k_2\delta_2\delta_3\delta_6\delta_7
=\int d^4k_1\frac{N(k_1)}{D_1D_4D_5}\frac{1}{((k_1-p_3)^2-m^2)((k_1-p_4)^2-m^2)}\,,
\label{eq:boxcut}
\end{align}
where we have denoted $\delta_i=\delta(D_i)$ and assumed the integral to contain some arbitrary numerator depending on $k_1$.
From eq.~\eqref{eq:boxcut}, we observe that the maximal-cut of the one-loop subdiagram exposes two hidden propagators, $\Den_{10}=(k_1-p_3)^2-m^2$ and  $\Den_{11}=(k_1-p_4)^2-m^2$.
The latter, together with the residual uncut propagators $\Den_{1,4,5}$, form a one-loop pentagon integral, known to obey non-canonical DEQs around four dimensions. 
Therefore, we choose the numerator factor $N(k_1)$ such that it cancels one or both the hidden propagators, resulting in either a box or triangle integral, which both satisfy canonical DEQ. 
In this way, we determine three out of the four numerators corresponding to the integrals  $\FF_{49,...,51}$, as they are displayed in figure~\ref{fig:MIsT6b}.
The last numerator $\FF_{52}$ is defined to contain, besides the factor $D_{10}$, also the auxiliary denominator $D_9$, which, depending on $k_2$, ensures the linear independence from the other three basis integral of the sector, without spoiling the properties of the DEQs. 

Once a basis of MIs with $\eps$-linear DEQs has been determined, the integrals of eq.~\eqref{def:LBasisT6} can be rotated into a canonical basis of MIs $\GG_i$ by means of the Magnus exponential matrix. Through this procedure, we find that the integrals
\begin{align*}
% \begin{alignedat}{2}
  \GG_{1}&= \FF_1\,, &
  \GG_{2}&= -s \,  \FF_2\,, \nn
  \GG_{3}&= m^2\, \FF_3\,, &
  \GG_{4}&= t \FF_4 \,, \nn
  \GG_{5}&=-2 m^2 \, \FF_4 - (m^2-t) \, \FF_5\,, &
  \GG_{6}&= - u\,  \FF_6\,,  \nn
  \GG_{7}&= 2 m^2 \, \FF_6 - (u-m^2) \, \FF_7\,, &
  \GG_{8}&= -s\,\FF_{8}\,, \nn
  \GG_{9}&= -\frac{s}{2} \, \FF_{8} + \lambda_s \left( \frac{1}{2} \FF_8+ \FF_9 \right) \,, \quad \quad \quad   &
  \GG_{10}&= -s \, \FF_{10}\,, \nn 
  \GG_{11}&= -(m^2-t) \, \FF_{11}\,, &
  \GG_{12}&= - m^2 \left(m^2-t \right)\, \FF_{12}\,, \nn 
  \GG_{13}&= -(u-m^2)\, \FF_{13}\,,  &
  \GG_{14}&= -m^2 \left(u-m^2\right)\, \FF_{14}\,, \nn
  \GG_{15}&=\lambda_s \, \FF_{15}\,,  &
  \GG_{16}&= m^2 \lambda_s \,\FF_{16}\,, \nn 
  %  \end{alignedat}\\
  \GG_{17}&= \left(s- \lambda_s\right) \left(\frac{3}{2} \FF_{15}+ m^2 \, \FF_{16} \right) - m^2 \, s \, \FF_{17}  \,, & 
  %  \begin{alignedat}{2}
  \GG_{18}&= \lambda_s \FF_{18}\,, \nn 
  \GG_{19}&= \lambda_s \, \FF_{19}\,, &
  \GG_{20}&=\lambda_s \, \FF_{20}\,, \nn
  % \end{alignedat}\\
 \GG_{21}&= \rlap{$\displaystyle \left(s- \lambda_s \right)  \left( \FF_{15}+2\,m^2 \,\FF_{16}+ \FF_{18} -2 \, \FF_{20} \right)-m^2 \, s \, \FF_{21}\,$} && \nn
  %\begin{alignedat}{2}
  \GG_{22}&=  -s\,t\, \FF_{22}\,, &
  \GG_{23}&= m^2 \, s \left( \FF_{22} + (m^2-t) \,  \FF_{23} \right) \,, \nn
  \GG_{24}&= u \,s\, \FF_{24}\,,   &
  \GG_{25}&= -m^2 \, s \left( \FF_{24} -(u-m^2) \FF_{25} \right) \,, \nn
  \GG_{26}&= -(u-m^2)\, \FF_{26}\,,  &
  \GG_{27}&= -(m^2-t)\left( m^2 \FF_{27}+ \frac{2m^2+\lambda_s -s}{2}  \FF_{28} \right) \,, \nn
  \GG_{28}&=  -(m^2-t) \, \lambda_s\, \FF_{28} \,,&
  \GG_{29}&= - (m^2-t) \, \FF_{29} \,, \qquad\nn
  \GG_{30}&= -(u-m^2) \left(m^2 \FF_{30}+\frac{2m^2 + \lambda_s -s}{2} \FF_{31} \right) \,, &
  \GG_{31}&=- \lambda_s \, (u-m^2) \FF_{31} \,, \nn
    \GG_{32}&= -(m^2-t)\,\FF_{32}\,,  &
  \GG_{33}&= -m^2 \, s \, \FF_{33} \,,\nn
  \GG_{34}&= -(u-m^2)\,\FF_{34}\,,  &
  \GG_{35}&= -m^2 \, s \, \FF_{35} \,,\nn
  \GG_{36}&= \lambda_s \,\FF_{36}\,,  &
  \GG_{37}&= (\lambda_s-s)\left( \frac{1}{2} \FF_{19} -2 \FF_{36} \right)  -m^2 \, s \, \FF_{37} \,,\nn
  \GG_{38}&= \lambda_s\,\FF_{38}\,,  &
  \GG_{39}&= -m^2 \, (m^2-t) \, \FF_{39} \,,\nn
  \GG_{40}&= -m^2 (u-m^2)\,\FF_{40}\,,  &
  \GG_{41}&= ( m^2-t) \, (u-m^2) \, \FF_{41} \,,\nn
  \GG_{42}&= (m^2-t)\, s \, \FF_{42}\,,  &
  \GG_{43}&= (u-m^2) \,s \, \FF_{43} \,,\nn
  \GG_{44}&= -\sqrt{m^2(m^2-t)(u-m^2)(-s)} \,\FF_{44}\,,  & & \nn
  \GG_{45}&= \rlap{$\displaystyle - \lambda_s \FF_{38}- (m^2-t) \left((m^2-s) \FF_{44}+ \FF_{45}-\FF_{46}  \right) \,,$} && \nn
  \GG_{46}&= (u-m^2) \left( \FF_{38} +m^2 \FF_{44} - \FF_{46} \right)\,,  &
  \GG_{47}&= -(u-m^2) \FF_{34} - \lambda_s \left(\FF_{34}+ \FF_{44}- \FF_{47} \right) \,,\nn
  \GG_{48}&= -s \, \lambda_s \,  \FF_{48}\,,  &
  \GG_{49}&= (u-m^2)\, s  \, \FF_{49} \,,\nn
  \GG_{50}&= (m^2-t)\, s \, \FF_{50}\,,  &
  \GG_{51}&= -\lambda_s \, \FF_{38} + s (\FF_{38}-\FF_{51}) \,,\nn
%\end{alignedat}
\GG_{52}&= \rlap{$\displaystyle \sum_{i=2}^8c_{i,52} \FF_{i}+c_{10,52} \FF_{10}+c_{11,52} \FF_{11}+c_{13,52} \FF_{13} +c_{15,52} \FF_{15}+c_{17,52} \FF_{17} $} &&  \nn
&+ \rlap{$\displaystyle c_{18,52} \FF_{18}+\sum_{i=20}^{35}c_{i,52} \FF_{i}+\sum_{i=38}^{41}c_{i,52} \FF_{i}+\sum_{i=44}^{46}c_{i,52} \FF_{i}+\sum_{i=49}^{52}c_{i,52} \FF_{i} $} &&
\label{def:CanonicalBasisT6}
\stepcounter{equation}\tag{\theequation}
\end{align*}
obey canonical DEQs in both kinematic variables $w$ and $z$. In eq.~\eqref{def:CanonicalBasisT6} we have used the compact notation $\lambda_s=\sqrt{-s} \sqrt{4m^2-s}$. The analytic expression of the coefficients $c_{i,52}$, that are here omitted for brevity, can be found in appendix~\ref{sec:gMIcoeff}, as well as in the ancillary files of the \texttt{arXiv} version of the manuscript, which contain eq.~\eqref{def:LBasisT6} and eq.~\eqref{def:CanonicalBasisT6} in electronic format.

By organizing the 52 MIs into a vector $\GGvec(\eps,w,z)$ and by combining the DEQs in $w$ and $z$ obeyed by the latter into a single total differential, we can write the canonical DEQs compactly as
\begin{equation}
d \GGvec = \eps \dA \GGvec \, ,
\label{eq:canonicalDEQ}
\end{equation}
where $\dA$ is the 1-form
\begin{equation}
\dA = \sum_{i=1}^{12} \MM_i  \dlog(\eta_i) \, ,
\label{eq:dA}
\end{equation}
with $\MM_i$ being constant matrices (that are provided as ancillary files in the \texttt{arXiv} submission of this paper). The  {\it alphabet} of the DEQs, \ie the set of arguments $\eta_i$ of the $\dlog$-form in eq.~\eqref{eq:dA}, can be taken to be formed by the same 12 \emph{letters} that appear in the calculation of the MIs for the QED-like topology $A_1$ presented in ref.~\cite{DiVita:2018nnh}:
 \begin{align}
 \begin{alignedat}{2}
\eta_1 & =w\,,&\quad
\eta_2 & =1+w\,, \nn 
\eta_3 & =1-w\,, &\quad
\eta_4 & =z\,, \nn
\eta_5 & =1+z\,,&\quad
\eta_6 & =1-z\,,\nn
\eta_7 & =w+z  \,,&\quad
\eta_8 & =z-w  \,,\nn
\eta_9 & =z^2-w \,, &\quad
\eta_{10} & = 1-w+w^2-z^2\,,\nn
\eta_{11} & =1-3w+w^2+z^2 \,, &\quad
\eta_{12} & =z^2 - w^2 - w z^2 + w^2 \,z^2 \,.
\end{alignedat} \stepcounter{equation}\tag{\theequation}
\label{alphabet}
 \end{align}
For the MIs of the topology  $N_2$ under consideration, the matrix $\MM_{11}$ vanishes identically. Nevertheless, we adopted the above notation for consistency with ref.~\cite{DiVita:2018nnh}.\\
The analytic expression of the MIs is first derived in the region of the $wz$-plane where the whole alphabet is real and positive, 
\begin{equation}
  \label{eq:positivityx}
  0<w<1\land\sqrt{w}<z<\sqrt{1-w+w^2}\,,
 \end{equation} 
which corresponds to the unphysical kinematic region 
\begin{equation}
  t<0\,\land\,s<0\,.
\end{equation}
The analytic continuation to the full kinematic plane is discussed in section~\ref{sec:Analytic}.

By definition, the integrals given in eq.~\eqref{def:CanonicalBasisT6} are finite in the $\eps\to 0$ limit. Therefore, the vector $\GGvec(\eps,w,z)$ can be expanded as a Taylor series around $\epsilon=0$ as
\begin{align}
  \GGvec(\eps,w,z) =  \sum_{n=0}^{\infty}\GGvec^{(n)}(w,z)\eps^n\,.
  \label{eq:TayExp}
\end{align}
The $n$-th order coefficient of eq.~\eqref{eq:TayExp} is given by
\begin{align}
\GGvec^{(n)}(w,z) = \sum_{i=0}^n \Delta^{(n-i)} (w,z; w_0,z_0)  \GGvec^{(i)}(w_0,z_0),
\end{align}
where the $ \GGvec^{(i)}(w_0,z_0)$ are constant vectors and $\Delta^{(k)}$ is the operator
\begin{align}
\Delta^{(k)} (w,z; w_0,z_0) =\int_\gamma \dAk{k},\qquad \Delta^{(0)} (w,z; w_0,z_0) = 1\,,
\label{eq:delta}
\end{align}
that represents $k$ path-ordered iterated integrations of $\dA$ along a piecewise-smooth path $\gamma$ in the $wz$-plane.
As the roots of the rational alphabet given in eq.~\eqref{alphabet} are purely algebraic, the iterated integrals of eq.~\eqref{eq:delta} can be directly mapped into combinations of GPLs,
\begin{align}
  G({\vec a}_{n} ; x) &\equiv G(a_1, {\vec a}_{n-1} ; x) \equiv
  \int_0^x dt \frac{1}{t-a_1}
  G({\vec a}_{n-1};t) , \\
  G(\vec{0}_n;x)& \equiv \frac{1}{n!}\text{log}^n(x) \,.
\end{align}
The length $n$ of the vector  ${\vec a}_n$ is commonly referred to as the transcendental {\it weight} of $G({\vec a}_{n} ; x)$ and it amounts to the number of repeated integrations defining the GPL.
We found it convenient to determine the solution of the DEQs in terms of GPLs, up to $\mathcal O(\eps^5)$ terms, by integrating first in $w$ and then in $z$. Consequently, the GPLs that appear in the analytic expression of the MIs fall into two classes: GPLs with argument $w$ and weights drawn from the set
\begin{align}
  \left\{
  0\,,\;
  \pm 1\,,\;
  \pm z\,,\;
  z^2\,,\;
  \frac{z \left(z\pm\sqrt{4-3 z^2}\right)}{2 \left(z^2-1\right)}\,,\;
  \frac{1}{2} \left(1\pm\sqrt{4 z^2-3}\right)
  \right\}\,,
\end{align}
and GPLs with argument $z$ and with weights drawn from the set $  \{  0\,,\; \pm 1 \}$.
Due to the positivity of the alphabet, the combinations of GPLs that appear in the solution $\GGvec(\eps,w,z)$ are real-valued in the region defined by eq.~\eqref{eq:positivityx}. It therefore follows that all the imaginary parts associated to the MIs whose physical thresholds are overstepped in this region are made explicit in the integration constants $\GGvec^{(i)}(w_0,z_0)$.
\subsection{Boundary conditions}
In order to completely specify the analytic expression of the MIs, a suitable set of boundary conditions must be imposed on the general solution of the system of DEQs.
Boundary conditions are determined by imposing the regularity of the MIs at the pseudo-thresholds of the DEQs, that entails, order by order in $\eps$ expansion, a linear relation between the MIs. Such regularity conditions are complemented by the knowledge of the analytic expression of a limited number of input integrals in special kinematic configurations. 
In the case under study, the boundary constants of all the MIs can be expressed as combinations of GPLs of argument $1$ and complex weights,
the latter arising from the specific kinematic limits imposed on the alphabet of eq.~(\ref{alphabet}). 
With the help of \texttt{GiNaC}, we were able to numerically verify that, at each order in $\eps$, such combinations of constant GPLs are proportional to uniform combinations of the transcendental constants $\pi$, $\zeta_k$ and $\log 2$. 

The boundary constants of each integral have been determined as follows:
\begin{itemize}
\item the integrals $\GG_{1,...,7,10,...,14,32,...,37}$ are either common to the two-loop topologies discussed in reference~\cite{Mastrolia:2017pfy,DiVita:2018nnh} (to which we refer the reader for the discussion of the boundary value fixing procedure) or related to them by simple kinematic crossing, \ie by some interchange of the Mandelstam invariants;
\item the boundary constants of $\GG_{8,9,18}$ have been fixed by demanding finiteness of the MIs in the limit $s\rightarrow 0$;
\item the boundary constants of $\GG_{15,16,17,20,21,38,47}$ have been obtained by demanding finiteness of the MIs in the limit $s\rightarrow 4m^2$. Additional constraints for the integrals $\GG_{15,16,17,20,21}$ have been obtained by requiring their corresponding boundary constants to be real-valued;
\item the boundary constants of $\GG_{22,...,25,39,...,44,46}$ have been fixed by imposing the finiteness of the MIs in the limit $t \rightarrow m^2 \frac{\sqrt{4m^2-s}-\sqrt{-s}}{\sqrt{4m^2-s}+\sqrt{-s}}$;
\item the boundary constants of $\GG_{26,...,31}$ have been determined by demanding  the finiteness of the MIs the limits $ s \rightarrow 0$ and $t \rightarrow m^2 \frac{\sqrt{4m^2-s}-\sqrt{-s}}{\sqrt{4m^2-s}+\sqrt{-s}}$;
\item the boundary constants of $\GG_{49,...,52}$ have been obtained by requiring the finiteness of the MIs in the limits $ s \rightarrow 4m^2$ and $t \rightarrow m^2 \frac{\sqrt{4m^2-s}-\sqrt{-s}}{\sqrt{4m^2-s}+\sqrt{-s}}$;
\item the boundary constant of $\GG_{45}$ has been fixed by demanding the finiteness of the integral the limit $t \rightarrow m^2 \frac{\sqrt{4m^2-s}-\sqrt{-s}}{\sqrt{4m^2-s}+\sqrt{-s}}$;
\item the boundary constant of $\GG_{48}$ has been fixed by taking the massless limit, as described in~\cite{Mastrolia:2017pfy}.
\end{itemize}
We provide the analytic expressions of the MIs in electronic form in the ancillary files attached to the \texttt{arXiv} submission of the paper.

%%%%%%%%%%%%%%%%%%
%%% Local Variables:
%%% TeX-master: "../main"
%%% End:
\section{Analytic continuation}
\label{sec:Analytic}

The results of section~\ref{sec:DEQ} have been obtained in the unphysical region $s,t<0$. Therefore, the analytic continuation of such expressions to the $t\bar{t}$ production kinematics must be performed. In terms of the Mandelstam invariants, this region is defined by
\begin{align}
  s\geq 4m^2 \land m^2 - \frac{s}{2} - \frac12 \sqrt{s-4m^2}\sqrt{s} \leq t \leq m^2 - \frac{s}{2} + \frac12 \sqrt{s-4m^2}\sqrt{s}\,, \label{eq:phys_kin}
\end{align}
where the boundaries of the allowed interval for $t$ are in one-to-one correspondence, in the center-of-mass frame, with the minimum and maximum scattering angles of the $t\bar{t}$ pair with respect to the beam line. For completeness, we also quote the physical region for elastic scattering, corresponding to the crossed $t$-channel process:
\begin{align}
    t\geq m^2 \,\land\, -t\,\left(1-\frac{m^2}{t}\right)^2 \leq s\leq 0 \,\land\, 2m^2-t\leq u \leq \frac{m^4}{t} \,. \label{eq:phys_kin_tchannel}
\end{align}

In the case of non-planar four-point integrals, the analytic continuation of the GPLs is quite non trivial. As originally noted in~\cite{Gehrmann:2000zt,Gehrmann:2002zr}, thresholds associated with \emph{all} the Mandelstam invariants appear simultaneously, and $s,t,u$ should be treated as independent variables when discussing the analytic continuation of the expressions for the MIs. On the other hand, the approach we follow enforces the constraint $s+t+u=2m^2$ from the outset, yielding a system of DEQs in two variables, \eg $s$ and $t$. One way out could be to enforce the Mandelstam constraint only at a later stage (see \eg~\cite{Tausk:1999vh}), at the price of considerably complicating the problem to be solved due to the presence of an extra scale.

In this paper we address the analytic continuation in a different way by exploiting the iterated-path-integral nature of our canonical MIs, together with the so-called \emph{first-entry condition}~\cite{Gaiotto:2011dt,Abreu:2015zaa}, in order to devise an effective prescription. Our approach allows to analytically continue the MIs everywhere in the kinematic plane, and in particular to evaluate our results in the $t\bar{t}$ production region. As a byproduct of our analysis, we also obtain the analytic continuation of the MIs for $\mu e$ scattering, presented in~\cite{DiVita:2018nnh}, both to the di-muon production region and to the elastic scattering region.

\begin{figure}[t]
  \centering
  \includegraphics[width=.7\textwidth]{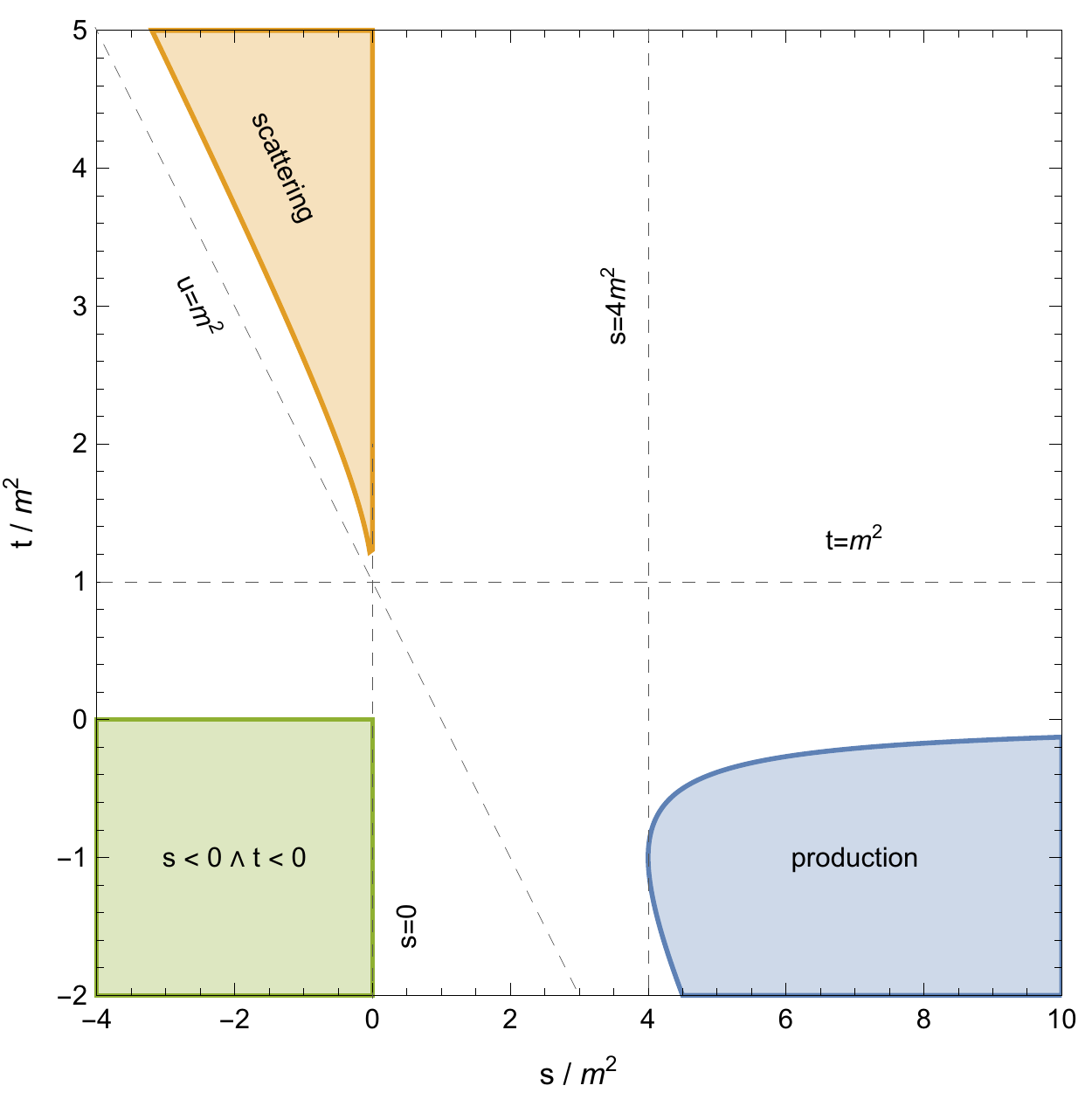}
  \caption{In the plot we show a representative portion of the kinematic phase space, parametrized in terms of $(s,t)$. The region in which we solved the system of differential equations, $s,t<0$, is marked in green. The physical region for the $s$-channel production process, given in eq.~\eqref{eq:phys_kin}, is highlighted in blue. In orange we also show the physical region for the $t$-channel process, given in eq.~\eqref{eq:phys_kin_tchannel}, which is relevant for $\mu e$ scattering. The dashed lines indicate the thresholds of the physical logarithms in eq.~\eqref{eq:gpl_theta1}-\eqref{eq:gpl_theta4}.}
  \label{fig:regions}
\end{figure}

We already observed in section \ref{sec:DEQ} that our canonical basis of MIs can be expressed, order by order in $\epsilon$, as a linear combination of GPLs and constants of uniform weight. From the first-entry condition it follows that only the innermost integration contributes to the discontinuities of the MIs.
Strong restrictions on the analytic structure of the MIs are imposed already at the level of the canonical DEQs (by the coefficient matrices in the $\dlog$ form), but knowledge of the boundary conditions is essential to fully pin it down. By inspection of our result (computed up to weight 4 in the region $s,t<0$), we find that the GPLs originating from the innermost integration are the following: $G_{0}(z),\, G_{0}(w),\, G_{1}(w),\,G_{z^2}(w)$.
This can be traced back to the fact that, of all the letters $\eta_i$ that appear in $\dA$ (see eq.~\eqref{alphabet}), only four contribute to the first integration, namely $\eta_{1,3,4,9}$.
Quite remarkably, one can build four combinations of the $\eta_{1,3,4,9}$ that correspond to simple functions of the Mandelstam invariants, whose logarithms exhibit the branch cuts expected from the normal thresholds of the four sunrise sub-topologies. If we define
\begingroup
\allowdisplaybreaks[0]
\begin{alignat}{2}
  \eta _1                              &\equiv \theta _1&&=\frac{\sqrt{4m^2-s}-\sqrt{-s}}{\sqrt{4m^2-s}+\sqrt{-s}}\,,  \\
  \eta _3^2 / \eta _9                  &\equiv\theta _2 &&=1-t/m^2\,, \\
  \eta _3^2 / \eta _1                  &\equiv\theta _3 &&=-s/m^2\,, \\
  (\eta _3 \eta _4)^2/(\eta _1 \eta _9)&\equiv\theta _4 &&=u/m^2-1\,,
\end{alignat}
\endgroup
where use has been made of the relation $s+t+u=2m^2$, then one can easily find the following GPL representations for the corresponding logarithmic differentials
\begin{align}
  \dlog \theta_1 ={}&\dG_0(w)\,, \label{eq:gpl_theta1}\\
  \dlog \theta_2     ={}&2 \dG_1(w)-2 \dG_0(z)-\dG_{z^2}(w)\,, \label{eq:gpl_theta2}\\
  \dlog \theta_3    ={}&2 \dG_1(w)-\dG_0(w)\,, \label{eq:gpl_theta3}\\
  \dlog \theta_4 ={}&2 \dG_1(w)-\dG_0(w)-\dG_{z^2}(w)\,. \label{eq:gpl_theta4}
\end{align}
We refer to $\log\theta_{1,2,3,4}$ as \emph{physical} logarithms. In figure~\ref{fig:regions} we show the physical regions for the $s$-channel production and the $t$-channel scattering processes, together with the region in which we solved the system of differential equations, and the thresholds of the physical logarithms.
For completeness we also give a more transparent rearrangement of the other letters:
\begin{alignat}{2}
\eta _{12}/(\eta _1 \eta _9)&\equiv \theta_5 &&=u/m^2\,, \label{eq:gpl_theta5}\\
\eta _{10}/\eta _9&\equiv \theta_6 &&=-t/m^2\,, \\
\eta _2^2/\eta _1&\equiv \theta_7 &&=4-s/m^2\,, \\
\eta _3^2 \eta _5 \eta _6 \eta _7 \eta _8/(\eta _1 \eta _9^2)&\equiv \theta_8 &&=1-t u/m^4\,, \\
\frac{\eta _6}{\eta _5}&\equiv \theta_{10} &&=\frac{1-\frac{\sqrt{u-m^2}}{\sqrt{m^2-t}}\sqrt{\frac{\sqrt{4m^2-s}-\sqrt{-s}}{\sqrt{4m^2-s}+\sqrt{-s}}}}{1+\frac{\sqrt{u-m^2}}{\sqrt{m^2-t}}\sqrt{\frac{\sqrt{4m^2-s}-\sqrt{-s}}{\sqrt{4m^2-s}+\sqrt{-s}}}}\,, \\
\frac{\eta _8}{\eta _7}&\equiv \theta_{11} &&=\frac{1-\frac{\sqrt{m^2-t}}{\sqrt{u-m^2}} \sqrt{\frac{\sqrt{4m^2-s}-\sqrt{-s}}{\sqrt{4m^2-s}+\sqrt{-s}}}}{1+\frac{\sqrt{m^2-t}}{\sqrt{u-m^2}} \sqrt{\frac{\sqrt{4m^2-s}-\sqrt{-s}}{\sqrt{4m^2-s}+\sqrt{-s}}}}\,, \\
\eta _{11}/\eta _9&\equiv \theta_{12} &&=2-t/m^2\,,\label{eq:gpl_theta12}
\end{alignat}

One can prove that, in the region $s,t<0$, eqs.~\eqref{eq:gpl_theta1}-\eqref{eq:gpl_theta4} also hold if the total differential operator is dropped, without adding any integration constants. By choosing a suitable analytic continuation prescription on the Mandelstam invariants one can then evaluate the integrated expressions in the full kinematic plane, in an unambiguous way. One can then check whether those expressions reproduce the imaginary parts of the corresponding physical logarithms.
The simple prescription we adopt is defined in two steps:
\begin{enumerate}
\item  As for the Mandelstam invariants, we express the real part of $u$ in terms of the real parts of $s$ and $t$, for which we use the standard Feynman prescription, but we give $u$ an independent prescription  (\ie \emph{before} using $u(s,t)=2m^2-s-t$ in the definition of $z$, eq.~\eqref{eq:varswz})
  \begin{align}
    s\to{}& s+\ieps\, \heaviside(s)\,, \label{eq:iepss}\\
    t\to{}& t+\ieps\, \heaviside(t-m^2),\,\label{eq:iepst}\\
    u(s,t)\to{}& 2m^2-s-t-\ieps\,, \label{eq:iepsu}
  \end{align}
  where $\ieps$ is an infinitesimal positive imaginary part, $\heaviside(x)$ is the Heaviside step function, and the presence of the constant $\ieps$ term in the last equation guarantees that the evaluation of the GPLs on spurious branch cuts (that are developed even for $s,t<0$) is always unambiguous.
  It can be shown, by repeated application of the identity
  \begingroup
    \allowdisplaybreaks[0]
  \begin{align}
    \log{a b} ={}& \log{a} + \log{b} + 2\pi i\left[\heaviside(-\Im a)\heaviside(-\Im b)\heaviside(\Im ab)\right.\nonumber\\
    {}& \qquad \qquad \qquad \qquad \left.- \heaviside(\Im a)\heaviside(\Im b)\heaviside(-\Im ab)\right]\,,
  \end{align}
  \endgroup
  that the above prescription is sufficient to reproduce, with the GPL representation in the variables $(w,z)$ of eqs~\eqref{eq:gpl_theta1}-\eqref{eq:gpl_theta4}, the physical logarithms $\log\theta_{1,2,3}$ (that, for instance, completely determine the analytic structure of the $s$-channel sunrise MIs and the $t$-channel sunrise MIs).
\item  It remains to be verified whether the above prescription allows to correctly reproduce also the imaginary part of $\log\theta_4$ (the one that, for instance, determines the discontinuity of the $u$-channel sunrise MIs across the one-particle branch cut). This is not guaranteed since, as stressed at the beginning of this section, we only have two independent variables at our disposal, while having to deal simultaneously with thresholds in all the three channels. The virtue of our prescription \eqref{eq:iepss}-\eqref{eq:iepsu} is that the representation of $\log\theta_4$ in terms of GPLs (corresponding to eq.~\eqref{eq:gpl_theta4})
  \begin{itemize}
  \item for $u>m^2$ is always on the physical Riemann sheet;
  \item for $u<m^2$ always ends up on the wrong side of the branch cut, \ie the imaginary part is always $-i\pi$ instead of $+i\pi$.
  \end{itemize}
  We can therefore apply, as a second step, a simple correction to the combination of GPLs corresponding to $\log\theta_{4}$ (and \emph{not} to the other three combinations) to bring our GPL representation for the MIs to the correct Riemann sheet. At weight one one can perform the replacement
  \begin{align}
    \label{eq:gpl_correction}
    2 G_{1}(w) - G_{0}(w) - G_{z^2}(w) \to 2 G_{1}(w) - G_{0}(w) - G_{z^2}(w) + 2\pi i \heaviside(m^2-u)\,,
  \end{align}
  which is then propagated iteratively to higher weights.
\end{enumerate}
All the explicit imaginary constants in our solution, as stated in section~\ref{sec:DEQ}, originate from the (iterated) integrations over $\dlog\theta_4$. Indeed we integrate our DEQs in the region where $s,t<0$, so that $u>2m^2$. The combinations of GPLs corresponding to such integrations (at any weight) are then always accompanied by a constant term, namely an additional $-i\pi$. Therefore, the net effect of the correction \eqref{eq:gpl_correction} is to flip the sign of the imaginary constants in our solution. In summary, our effective way of implementing the analytic continuation of the result for the full set of MIs is
\begin{enumerate}
\item to use the prescription on the Mandelstam invariants of equations~\eqref{eq:iepss}-\eqref{eq:iepsu},
\item[$2^{\,\prime}$.] to replace $i\pi \to -i\pi$ everywhere in the solution (amounting to the complex conjugation of all the integration constants), whenever the latter is evaluated for $u(s,t)<m^2$.
\end{enumerate}

Remarkably, once the analytic continuation of the four physical logarithms (\ie of the weight-1 contribution to the canonical MIs for the four sunrise topologies) is taken care of explicitly~\cite{Gehrmann:2002zr}, the first-entry condition guarantees that the analytic continuation of the \emph{full set of MIs at all weights} is also correctly obtained. In particular, it is \emph{not necessary} to make sure the GPL representation also reproduces the imaginary parts coming from the evaluation close to the branch cuts of the logarithms of the unphysical letters, eqs.~\eqref{eq:gpl_theta5}-\eqref{eq:gpl_theta12}. It is instead sufficient to always introduce an ``auxiliary'' $\ieps$ prescription in order to avoid the ambiguous evaluation \emph{on} such branch cuts (in our case this is inherited from \eqref{eq:iepss}-\eqref{eq:iepsu}). Our strategy for the analytic continuation has been validated by thorough numerical checks performed either with the help of \texttt{SecDec} or with the techniques outlined in section \ref{sec:Numerics}.

It is now clear that we can also obtain, by the same argument, the analytic continuation of the results for the MIs of the QED-like topology $A_1$ (see figure~\ref{fig:npgraphs}) presented in~\cite{DiVita:2018nnh}. The only difference with respect to the present case (besides a trivial relabeling of the Mandelstam invariants $s\leftrightarrow t$ to match the notations), is that the letter $\eta_{11}$ contributes to the $\dlog$ form with a non-zero coefficient matrix (but never appears in the first entry), and that $\eta_1=\theta_1$ is not a physical logarithm anymore (as expected due to the absence of a two-massive-particles normal threshold). Since the analytic continuations of $\log\theta_1$ and $\log\theta_3$ are not independent, in practice this difference does not change the situation, and the same procedure described above can then be used for the analytic continuation of the MIs from $s,t<0$ to the physical regions for the $\mu e\to \mu e$ and $e^+e^-\to \mu^+\mu^-$ processes, as confirmed also in this case by our numerical checks.

We stress that the method outlined above is fully general, as it only relies on the analyticity properties of the canonical MIs, and on their iterated-integral representation. It is in particular independent of the presence of massive propagators or massive external legs.

%%%%%%%%%%%%%%%%%%
%%% Local Variables:
%%% TeX-master: "../main"
%%% End:
\section{Numerical validation of the non-planar box integrals}
\label{sec:Numerics}
Using the analytic continuation as described in the previous section, the expressions for our MIs have been numerically evaluated in several points across the whole phase space, including the unphysical region 
$s,t < 0$ and the region relevant for $t\bar{t}$ production, eq.~\eqref{eq:phys_kin}.
In order to cross-check our analytic calculation, we numerically computed the MIs (or linear combinations of the latter) in some benchmark points with an alternative method, namely by integrating directly their Feynman-parameter representation.
In particular, the integrals $I_i$ with $i=1,\ldots,48$ were computed with the
package {\tt SecDec}.
For the most complex topologies, 
corresponding to the non-planar box integrals $I_i$ with
$i=49,\ldots,52$, we used \texttt{Reduze} to identify an alternative set of independent MIs that are {\it quasi finite}~\cite{vonManteuffel:2014qoa} in $d=6$.
On the one hand, the latter have been computed semi-numerically by
means of an in-house algorithm \cite{DiVita:2018nnh}. On the other hand, these integrals can be analytically related to our set of MIs by dimension-shifting identities~\cite{Tarasov:1996br,Lee:2009dh} and IBPs, implemented in \texttt{LiteRed}, therefore allowing for a numerical comparison.

The definition of the 4 non-planar 7-denominator MIs that are finite in $d=6$
dimensions, together with our results at the phase-space point
$s=-1/7$, $t=-1/3$, $m^2=1$, are collected in table~\ref{tab:numerics}.
In the next subsection, we use the first of those integrals as an example to describe our evaluation strategy. Throughout this section, we set $m^2=1$ and $u=2-s-t$.
\begin{table}
  \begin{center}
    \begin{tabular}{ |c | c | r |}
      \hline
      % {\bf graph} & $I^{[d]}-${\bf integral} & $ I^{[d=6-2\eps]}(s=-{1 \over 7}, t=-{1 \over 3}, m^2=1)$ \\
      {\bf graph} & $I^{[d]}-${\bf integral} & $ I^{[d=6-2\eps]}(s=-1/7,\, t=-1/3)$ \\
      \hline
      \hline
      \hline
      \raisebox{-19pt}{\includegraphics[scale=0.13]{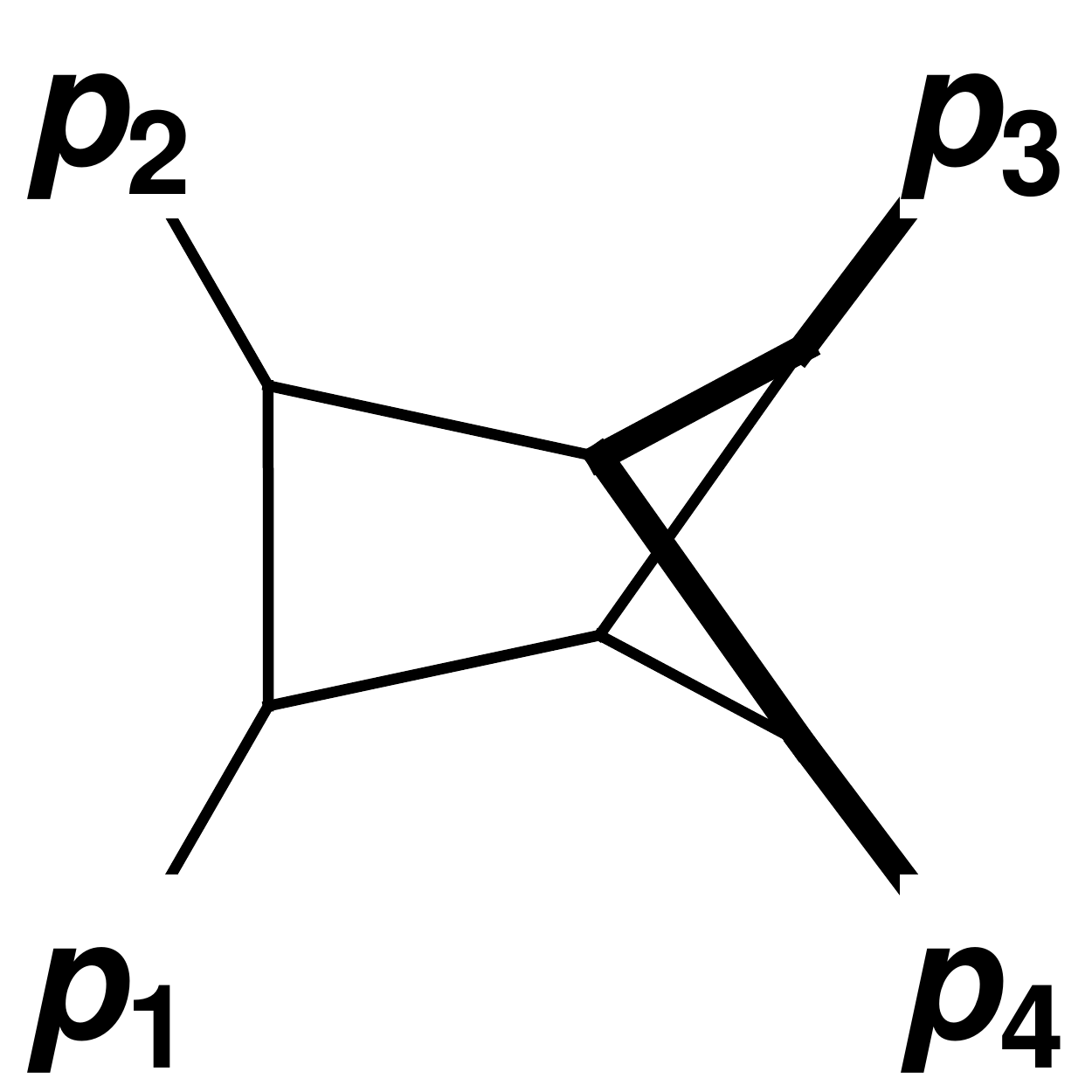}} 
                  & $I^{[d]}(1,1,1,1,1,1,1,0,0)$ & $-2.073498 - i\, 0.4872188$ \\
      \hline
      \raisebox{-19pt}{\includegraphics[scale=0.13]{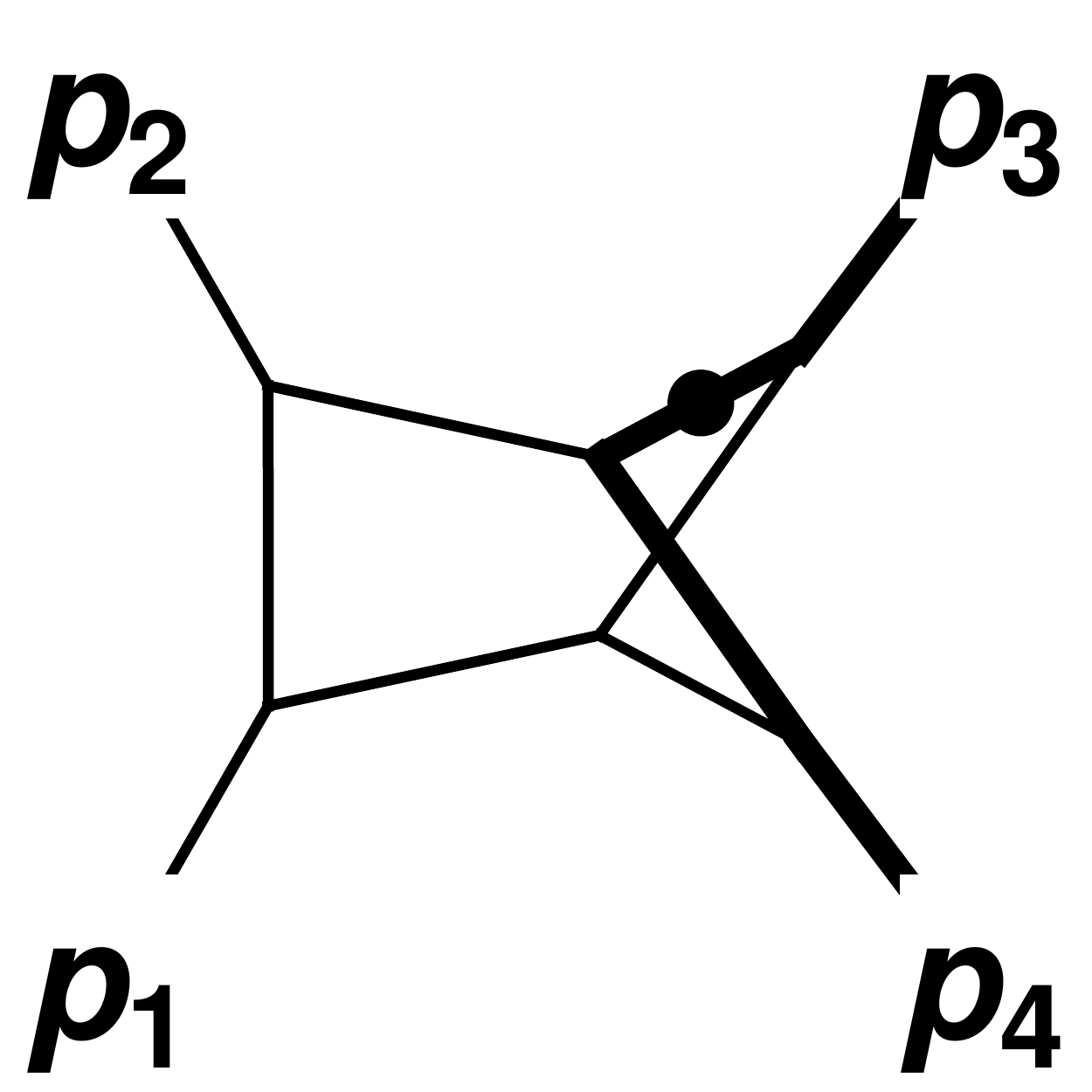}} 
                  & $I^{[d]}(1,2,1,1,1,1,1,0,0)$ & $1.63816 + i\, 1.72217$ \\
      \hline
      \raisebox{-19pt}{\includegraphics[scale=0.13]{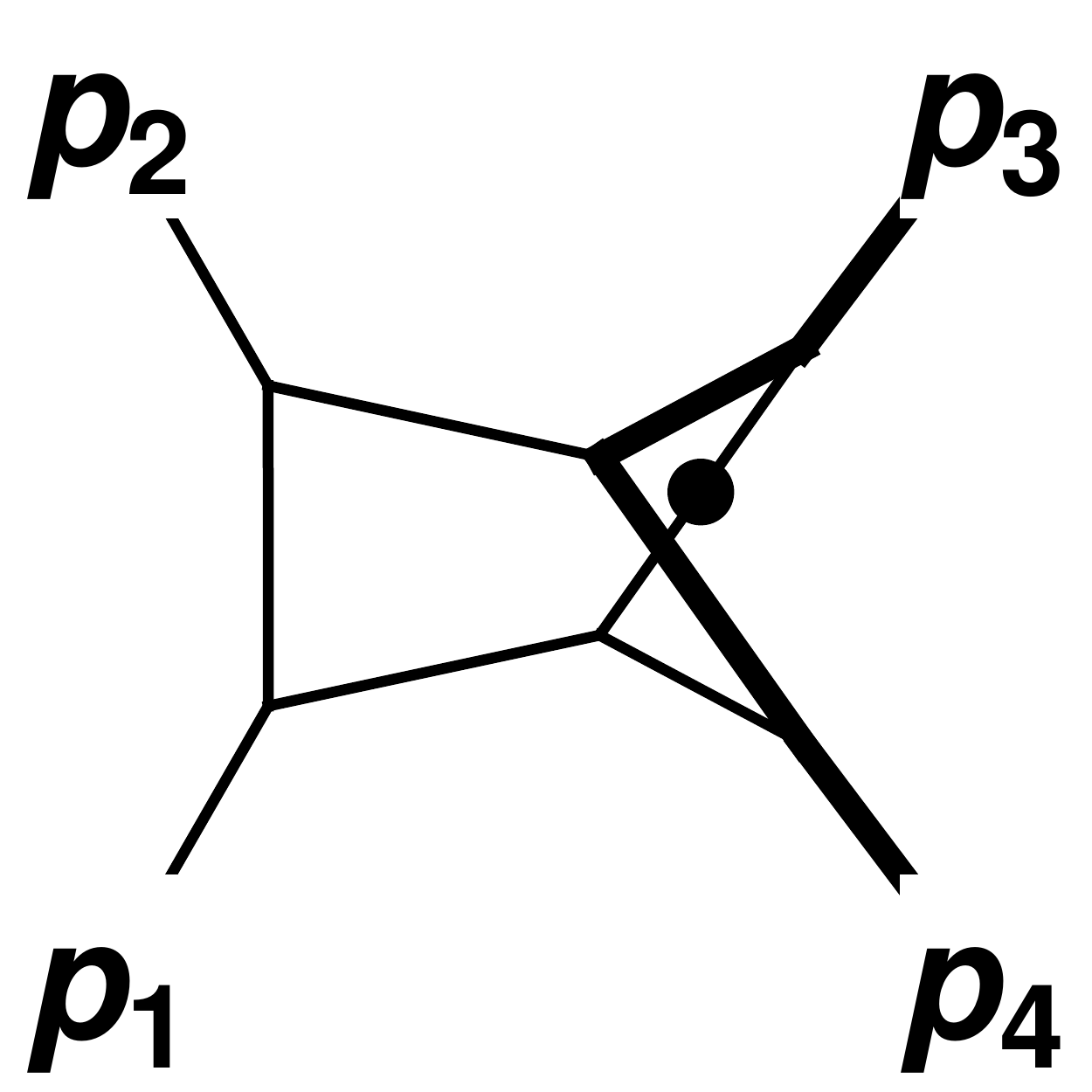}} 
                  & $I^{[d]}(1,1,2,1,1,1,1,0,0)$ & $18.8765 + i\, 12.4507$ \\
      \hline
      \raisebox{-19pt}{\includegraphics[scale=0.13]{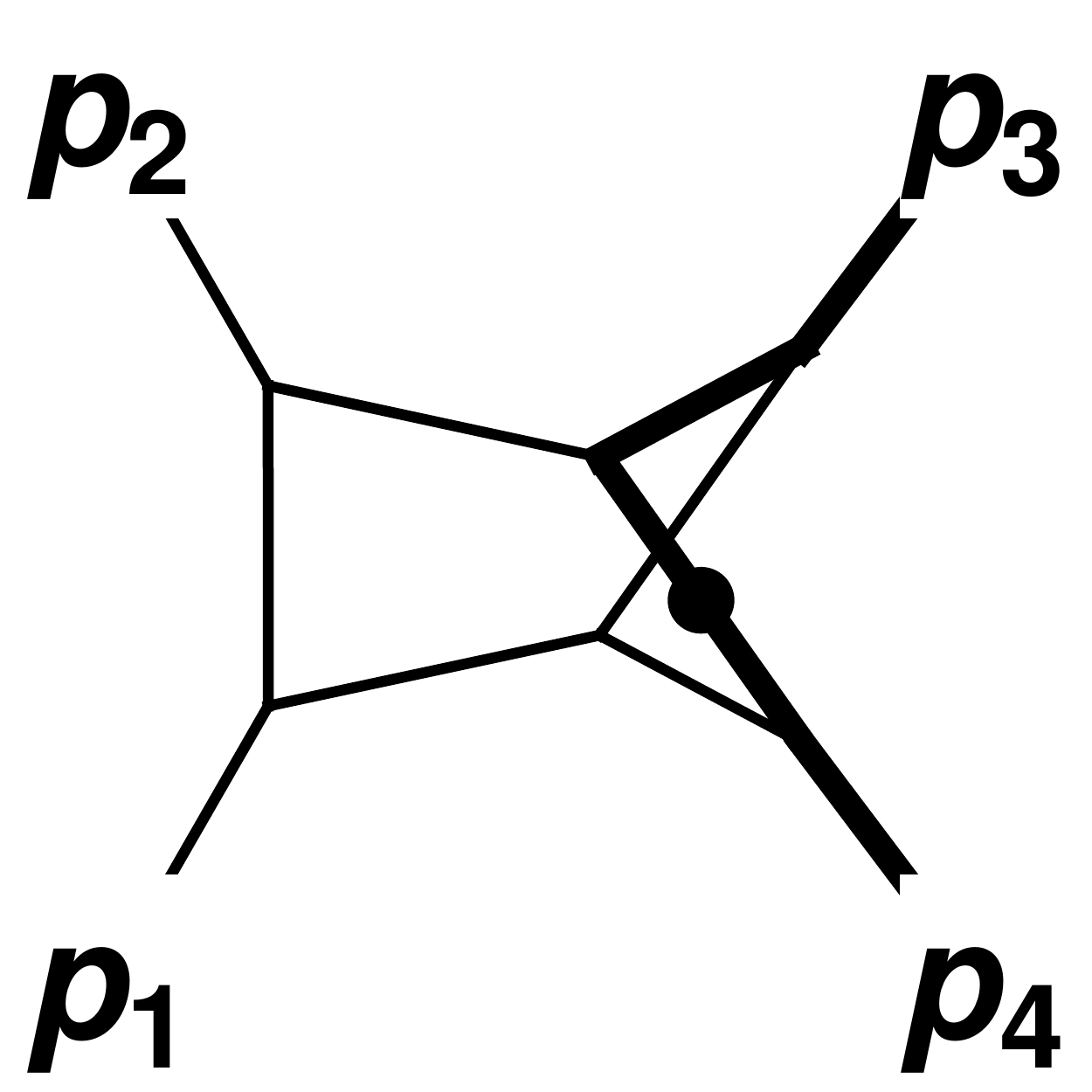}} 
                  & $I^{[d]}(1,1,1,1,1,2,1,0,0)$ & $1.827588 + i\, 1.121664$ \\
      \hline
    \end{tabular}
  \end{center}
  \caption{Numerical results for our set of quasi-finite non-planar MIs belonging to the 7-denominator topologies ($m^2=1$ and $u=2-s-t$).}
  \label{tab:numerics}
\end{table}

\subsection{The non-planar box in $d=6$ dimensions}

As an example, we consider the non-planar scalar integral
\begin{equation}
\raisebox{-19pt}{\includegraphics[scale=0.13]{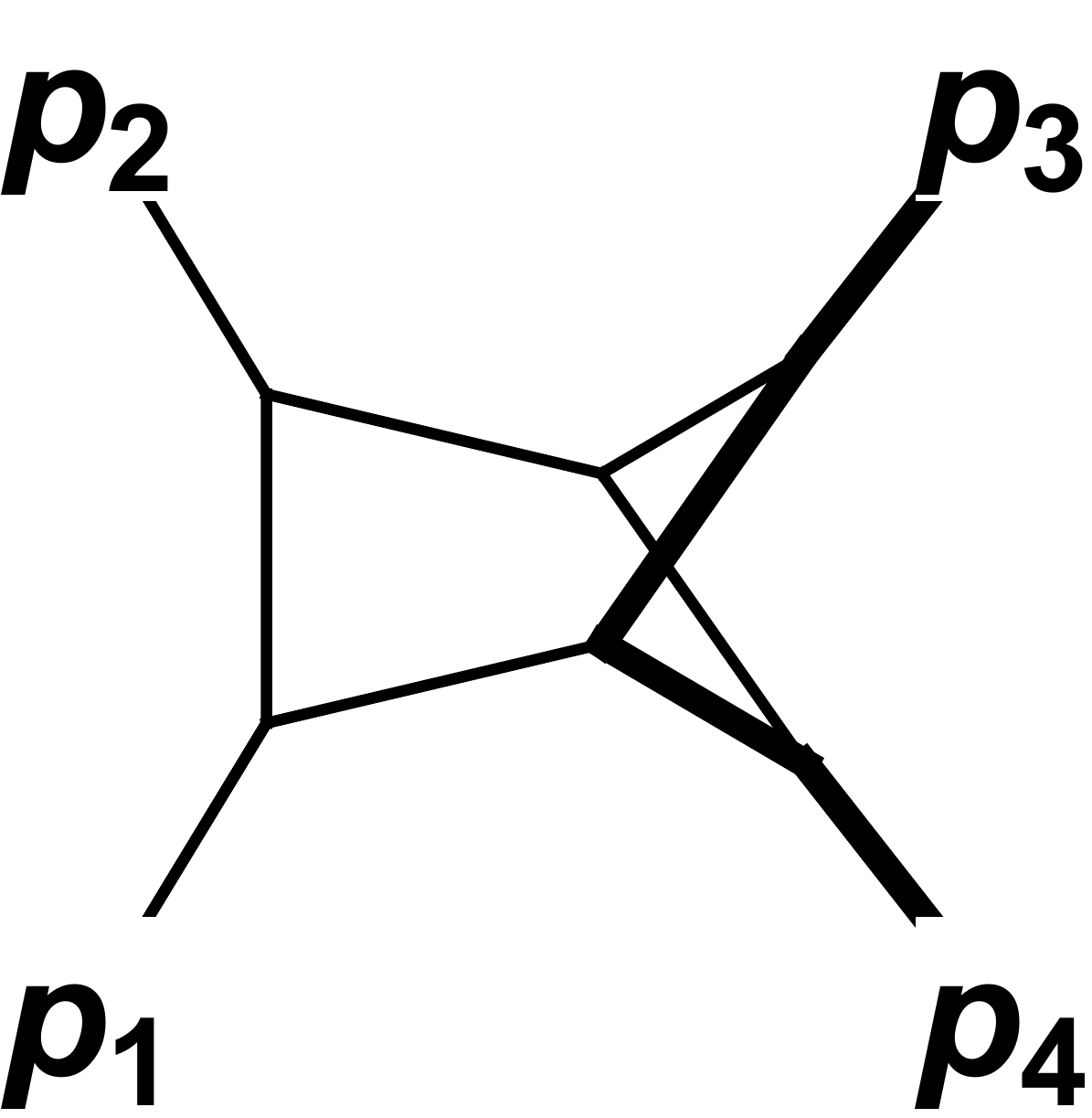}} 
=I^{[d]}(1,1,1,1,1,1,1,0,0)\ ,
\end{equation}
and we describe its numerical evaluation, which we carry out in two
steps.

\subsubsection{Analytic integrations}
As a first step, we introduce the Feynman parametrization.
By keeping track of the Feynman prescription $+i\feyneps$,
the integral can be written as 
\begin{equation}
\raisebox{-19pt}{\includegraphics[scale=0.13]{figures/Numerics/npGraph}}
= \Gamma(7) \int \widetilde{\dd^dk_1} \widetilde{\dd^dk_2} \int_0^1 dx_1 {\ldots} \int_0^1 dx_7 \,
\frac{\delta(1-x_{1234567})}{D_{tot}^7}\,,
\end{equation}
where
\begin{equation}
D_{tot}=x_1 D_1 + x_2 D_2 +x_3 D_4 +x_4 D_5 +x_5 D_6 +x_6 D_3 +x_7 D_7
+i\feyneps
\, .
\end{equation}
Next, we integrate over  $k_1$ and $k_2$, 
%%%%%%%%%%%%%%%%%%%%%%%%%%%%%%%%%%%%%%%%%%%%%%%%%%%%%%%%%%%
\begin{equation}
\label{inte71}
\Gamma_\eps^2
\raisebox{-19pt}{\includegraphics[scale=0.13]{figures/Numerics/npGraph}} 
=-\Gamma(7-d)\int_0^1 dx_1 {\ldots} \int_0^1 dx_7 
\frac{\delta(1-x_{1234567})}{A_0^{\frac{3d}{2}-7}\Delta^{7-d}} \,,
\end{equation}
%%%%%%%%%%%%%%%%%%%%%%%%%%%%%%%%%%%%%%%%%%%%%%%%%%%%%%%%%%%%%%%%%%%%%%%%%%%%%%
\begin{align}
A_0&= x_{34}x_{56} + x_5 x_6 + x_{346} x_7 +  x_2 x_{3457} + x_1
x_{2567} \,, \cr
\Delta&=
x_2^2 x_{3457} + x_1 x_{25}^2 + x_2 x_5 (2 x_{34} + x_5) + x_5^2 x_{346}  
+(t-1) (x_3 (-x_5 x_6 + x_2 x_7)
\cr
&
+s (x_2 (-x_4 x_5 + x_3 x_7) - x_1 (x_4 x_{256} + x_{46} x_7))
-i \feyneps A_0 \,,
\end{align}
where we used the notation 
$x_{i_1 i_2 {\ldots} i_n} =x_{i_1}+x_{i_2}+{\ldots}+x_{i_n}$ and the abbreviation $\Gamma_\eps \equiv \Gamma(1+\eps)$.
Now we perform as many analytic integrations as possible.
In particular, we eliminate the $\delta$-function by integrating over $x_3$, 
and we make the change of variables
$x_6 \to x_{26}-x_2$, $x_7 \to x_{57}-x_5$.
In this way, the polynomial $\Delta$ becomes linear in $x_4$,
and eq.~\eqref{inte71} is rewritten as
\begin{align}
\label{inte6}
\Gamma_\eps^2
\raisebox{-19pt}{\includegraphics[scale=0.13]{figures/Numerics/npGraph}} 
&=-\Gamma(7-d)
\int_0^1 dx_{26}
\int_0^{1-x_{26}} \frac{dx_{57}}{A^{\frac{3d}{2}-7}}
\int_0^{1-x_{2657}} \hspace{-1em} dx_1
\int_0^{x_{26}} \hspace{-0.5em} dx_2  
\int_0^{x_{57}} dx_5 
\times \cr &\phantom{=}
\int_0^{1-x_{12657}}  
\frac{dx_4}{(C_{41} x_4+C_{40})^{7-d}} \,,
\end{align}
where
\begin{align}
A&=x_{2657} (1 - x_{2657}) + x_{26} x_{57}\,, \cr
C_{41}&=
(t-1) (x_{26} x_5 - x_2 x_{57}) - s (x_2 x_{57} + x_1 x_{2657})
\,,
\cr
C_{40}&=
x_2^2 (1 - x_{26}) + x_5^2 (1 - x_{57}) + 2 x_2 x_5 (1- x_{2657})
+(t-1)(1-x_{12657}) (x_{26} x_5 - x_2 x_{57})
\cr
&\phantom{=}
+s (x_{57} - x_5)\left(x_2 (1-x_{2657})-x_1 x_{26}\right)
-i\feyneps A
\,.
\end{align}
The integral over $x_4$ in eq.~\eqref{inte6} is finite for $d\to 6$. In this limit, we get   
\begin{equation}
\label{inte5}
\Gamma_\eps^2
\raisebox{-19pt}{\includegraphics[scale=0.13]{figures/Numerics/npGraph}} 
\stackrel{d \to 6}{=}
-
\int_0^1 dx_{26}
\int_0^{1-x_{26}} \frac{dx_{57}}{A^{2}}
\int_0^{x_{57}} dx_5
\int_0^{x_{26}} \hspace{-0.5em} dx_2  
\int_0^{1-x_{2657}} \frac{dx_1}{f_2-g_2 x_1}
\ln \left(\frac{g_2 x_1^2+g_1 x_1 +g_0}{f_3 x_1 +P_6}\right) \,,
\end{equation}
with
\begin{align}
f_1&=(t-1) x_{26} x_5 - (s+t-1) x_2 x_{57} -  s x_{2657} (1-x_{2657})\,, \cr
f_2&=(t-1) x_{26} x_5 - (s+t-1) x_2 x_{57}\,, \cr
f_3&=(s+t-1) x_{26} x_5 + x_2 x_{57} - (t x_2 + s x_{26}) x_{57}\,, \cr
%%P_6&=C_{40}(x_1=0) \,, \cr
P_6&=x_2^2 (1 - x_{26}) + x_5^2 (1 - x_{57}) + (2-s) x_2 x_5 (1- x_{2657})
-f_2 (1-x_{2657}) -i\feyneps A \,,\cr
g_0&=P_6+f_2 (1-x_{2657})\,, \cr
g_1&=s (x_{26} x_5 + x_2 x_{57} - A) \,, \cr
g_2&=s x_{2657}\,, \cr
%%g_2&(-P_2^2)+g_1(-P_2)+g_0=g_2(-P_4^2)+g_1(-P_4)+g_0=0 \,, \cr   
P_2&=\frac{g_1+\sqrt{g_1^2-4g_0 g_2}}{2 g_2} \,, \cr
P_4&=\frac{g_1-\sqrt{g_1^2-4g_0 g_2}}{2 g_2} \,, \cr
P_1&=P_2+ 1-x_{2657}  \,, \cr
P_3&=P_4+ 1-x_{2657}  \,,  \cr
P_5&=P_6+f_3 (1-x_{2657})=s x_{2657} P_1 P_3  \,. \cr 
\end{align}
Finally, we integrate over $x_1$ and reduce eq.~\eqref{inte5} to 
\begin{align}
\label{inte4}
\Gamma_\eps^2
\raisebox{-19pt}{\includegraphics[scale=0.13]{figures/Numerics/npGraph}} 
\stackrel{d \to 6}{=}
&-
\int_0^1 \frac{dx_{26}}{s}
\int_0^{1-x_{26}} \frac{dx_{57}}{x_{2657} \; A^{2}}
\int_0^{x_{26}} dx_2  
\int_0^{x_{57}} dx_5
\ \times \cr
 \biggl(
&
 {\mathrm{Li}}_2\left(\frac{Q_1}{R_1}\right)
-{\mathrm{Li}}_2\left(\frac{Q_2}{R_1}\right)
+{\mathrm{Li}}_2\left(\frac{Q_3}{R_2}\right)
-{\mathrm{Li}}_2\left(\frac{Q_4}{R_2}\right)
-{\mathrm{Li}}_2\left(\frac{Q_5}{R_3}\right)
+{\mathrm{Li}}_2\left(\frac{Q_6}{R_3}\right)
\biggr) \,, \cr
\end{align}
where 
\begin{align}
Q_1=Q_3=&f_1 \,, \quad Q_2=Q_4=f_2 \,, \quad Q_5=f_1 f_3 \,, \quad
Q_6=f_2 f_3 \,, \cr
%R_{i}&= Q_{2i-j}+P_{2i-j} s x_{2657}, \quad \forall i,j,  \ i=1,{\ldots},3\ ,\;j=0,1
R_{1}&= Q_{1}+P_{1} s x_{2657}= Q_{2}+P_{2} s x_{2657}\;, \cr
R_{2}&= Q_{3}+P_{3} s x_{2657}= Q_{4}+P_{4} s x_{2657}\;, \cr
R_{3}&= Q_{5}+P_{5} s x_{2657}= Q_{6}+P_{6} s x_{2657}= R_1 R_2 \;. 
\end{align}
Differently from the case of the non-planar integral of the family $A_1$, which was evaluated with a similar strategy in~\cite{DiVita:2018nnh}, the integrand contains 6 distinct dilogarithms, 
whose arguments depend of the algebraic functions
$P_1$, $P_2$, $P_3$, $P_4$, $R_1$ and $R_2$ , which contain the square root of the same polynomial.
\subsubsection{Numerical integrations} 
We rescale the four remaining integration variables in
eq.~\eqref{inte4}
in order to map them  onto a four-dimensional hypercube of unit side,
\begin{equation}
x_{26}=t_1\;,\ 
x_{57}=(1-x_{26})t_2  \; ,\ 
x_{2}=x_{26}t_3  \; , \ 
x_{5}=x_{57}t_4  \; .
\end{equation}
In this way, the new variables $t_i$ have to be integrated over $[0,1]$.
The six dilogarithms that
appear in eq.~\eqref{inte4} 
have branch points, which correspond to the hypersurfaces defined by the equations
\begin{equation}
\label{disc1}
R_i(t_1,t_2,t_3,t_4)=0\,, \quad P_j(t_1,t_2,t_3,t_4)=0, \quad
i=1,{\ldots} 3 \,, \quad j=1,{\ldots} 6 \,.
\end{equation}
In order to obtain a fast convergence of the numerical integration,
the integrands must be sampled carefully 
in the neighborhood of these branch points.
We start from the integration over $t_4$.  We split the
integration interval 
at the $N_4(t_1,t_2,t_3)$ solutions $z_{4j}(t_1,t_2,t_3)$ of
eq.~\eqref{disc1}, which satisfy $0 \le z_{4j}\le 1$.
\begin{equation}
\int_0^1 dt_4 = \sum_{j=0}^{N_4-1}
\int_{z_{4j}(t_1,t_2,t_3)}^{z_{4,j+1}(t_1,t_2,t_3)} dt_4\,, 
z_{40}=0 \;, \ z_{4{N_4}}=1\,.
\end{equation}
Now we consider the integration over $t_3$. We split the integration
interval at the $N_3(t_1,t_2)$  zeros $z_{3j}(t_1,t_2)$
of the discriminants
(polynomials in $(t_1,t_2,t_3))$  that appear in the zeros $z_{4j}$ and
which satisfy $0 \le z_{3j} \le 1$. 
These are the points where the hypersurfaces of eq.~\eqref{disc1} are tangent to the
hyperplane $t_4=\text{constant}$,
\begin{equation}
\int_0^1 dt_3 = \sum_{j=0}^{{N_3}-1} \int_{z_{3j}(t_1,t_2)}^{z_{3,j+1}(t_1,t_2)} dt_3\,, 
\quad z_{30}=0 \;,\  z_{3{N_3}}=1\,.
\end{equation}
Next, we consider the integration over $t_2$. We split the integration
interval at the $N_2(t_1)$ zeros $z_{2j}(t_1)$ of the discriminants
(polynomials in $(t_1,t_2))$  that appear in the zeros $z_{3j}$
and which satisfy $0 \le z_{2j} \le 1$, 
\begin{equation}
\int_0^1 dt_2 = \sum_{j=0}^{{N_2}-1} \int_{z_{2j}(t_1)}^{z_{2,j+1}(t_1)} dt_2\,, 
\quad z_{20}=0 \;, \ z_{2{N_2}}=1\,.
\end{equation}
We proceed in a similar way for the last integration over $t_1$, using
\begin{equation}
\int_0^1 dt_1 = \sum_{j=0}^{{N_1}-1} \int_{z_{1j}}^{z_{1,j+1}} dt_1\,, 
\quad z_{10}=0 \;, \ z_{1{N_1}}=1\,.
\end{equation}
\\
In order to carry out the integration over a generic interval $[t_a,t_b]$, 
we perform the change
of variables $t_i \to u_i$, with
\begin{equation}
t_i=t_{ai}+\frac{e^{u_i^3}}{e^{u_i^3}+1} (t_{bi}-t_{ai}) \,,
\quad i=1,{\ldots},4 \,.
\end{equation}
This change of variable maps $t_a \to -\infty$ and $t_b \to \infty$,
so that possible singularities at the endpoints are easily managed.
The variable $u_i$ must be integrated in $(-\infty,\infty)$: in
practice we truncate the integration domain to $(-M,+M)$, with $M$ suitably large
according to the desired precision of calculation: for instance, for 16-digits
computation, the value $M=4$ is adequate. 
Finally, the integration interval in $u$ is subdivided in
$2^n$ subintervals, and Gauss-Legendre integration over 16
points in each subinterval is employed.
All the singularities in the integrands are integrable logarithmic
singularities. Therefore, we can safely set a very small
value of $\feyneps$, like $10^{-30}$,  adequate for 16-digit
calculations.  

By using $16$ subdivisions in each interval and in every variable, we
find that our integral, in the phase space point $s=-1/7$, $t=-1/3$, $m^2=1$,
has the value

\begin{equation}
\raisebox{-19pt}{\includegraphics[scale=0.13]{figures/Numerics/npGraph}} 
\stackrel{d \to 6}{=}
-2.073498-i\,0.487218 \,. \\ 
\end{equation}

We adopt a similar procedure for the other integrals shown in
table~\ref{tab:numerics}.  
In all cases, after the analytic
integrations the integrands, in the $d \to 6$ limit,
are found to contain combinations of logarithms of ratios of the
polynomials $P_i$. Henceforth, the decomposition
of the integration domain, as well as the numerical integration are
carried out along the same lines described above 
as for the scalar box
integral.

%%%%%%%%%%%%%%%%%%
%%% Local Variables:
%%% TeX-master: "../main"
%%% End:

\section{Conclusions}

In this paper, we presented the analytic expressions of the master integrals for 
a set of non-planar two-loop Feynman graphs, with two
quark currents exchanging massless gauge bosons. 
Our results are the last missing ingredients required
for the analytic evaluation of the double-virtual contribution to the
scattering amplitude for the process $q {\bar q} \to t {\bar t}$ at
NNLO in QCD, which was so far known only numerically. The present computation completes the calculation of all the required master integrals, hence proving that the analytic evaluation of such amplitude is
indeed feasible.

The two-loop integrals were evaluated by means of the differential
equations method, which, combined with the ideas of the Magnus exponential matrix
and of the canonical basis, yielded a representation of the master integrals in terms of generalized polylogarithms.
The canonical systems of differential equations was conveniently solved in a non-physical
region. Subsequently, we studied,
in the presence of massive internal lines and of a non-trivial structure of branch cuts, the
analytic continuation of the two-loop functions
to the physical region relevant for the process under consideration.

The results of this paper represent an important milestone of the research
program dedicated to the evaluation of integrals originating from the
planar and non-planar two-loop four-point diagrams that contribute to 
both QED and QCD processes, which include, beside the top-pair production at
hadron colliders, also di-muon production at lepton
colliders, as well as muon-electron elastic scattering, which is the
investigation target of the novel experiment {\sc MUonE}, recently proposed at CERN.

%%%%%%%%%%%%%%%%%%
%%% Local Variables:
%%% TeX-master: "../main"
%%% End:

\section*{Acknowledgments}
We wish to thank Roberto Bonciani, Matteo Becchetti, Valerio Casconi,
Andrea Ferroglia, Simone Lavacca and Andreas von Manteuffel, for mutual
comparisons on the master integrals. We acknowledge Massimo Passera for
stimulating discussions. 
T.G., P.M., and U.S. wish to thank the organizers of {\it Amplitudes 2018},
and the hospitality of SLAC National Lab, where this work was
initiated. We wish to thank the organizers of the workshop {\it Theory
  for muon-electron scattering @ 10ppm} in Zurich, where part of this
work was carried out.

The work of S.L. and P.M. has been supported by the Supporting TAlent in
ReSearch at Padova University 
(UniPD STARS Grant 2017 ``Diagrammalgebra'').

This work of T.G. and A.P. has been supported by the Swiss National
Science Foundation under grant number 200020-175595.

The work of U.S. is supported in part by the U.S. National Science
Foundation under Grant No. NSF-PHY-1719690 and NSF-PHY-1652066.
\appendix
\section{The canonical master integral $\GG_{52}$}
\label{sec:gMIcoeff}
In this appendix, we provide the expression of the kinematic coefficients that appear in the definition of canonical basis defined by eq.~\eqref{def:CanonicalBasisT6}.
The coefficients of the integral $\GG_{52}$ are given by
\begin{align}
c_{2,52}&=\frac{\lambda_s-s}{2} \, , & 
c_{3,52}&= \frac{7m^2\, (m^2-t) \, \lambda_s } {4 (u-m^2) \, s} \, , \nn
c_{4,52}&= -\frac{(3m^2+4t)(\lambda_s-s)}{8 s}\, , &
c_{5,52}&= - \frac{3(m^2-t)(\lambda_s-s)}{16 s}\, ,  \nn
c_{6,52}& \rlap{ $ \displaystyle =- \frac{1}{24 (u-m^2) s} \left[ \left( \lambda_s-s \right) \left( 33m^4-45m^2t+12t^2+25m^2s-4st-16s^2 \right) \right. $ } \nn
& \quad \left. +14(m^2-t)s(m^2+2u)  \right]  \, , \nn
c_{7,52}&=\frac{(9m^2-9t+5s)\lambda_s+5s(u-m^2)}{48s}\, , &
c_{8,52}&= -\frac{\lambda_s-s}{4} \, ,\nn
c_{10,52}&= -\frac{(2m^2-2t+s)\lambda_s+s(u-m^2)}{4(u-m^2)}\, , &
c_{11,52}&= -\frac{(m^2-t)(\lambda_s-s)}{4s}\, , \nn
c_{13,52}&= \frac{(u-m^2)(\lambda_s-s)}{4s}\, , &
c_{15,52}&= \lambda_s-s \, ,\nn
c_{17,52}&=-m^2(\lambda_s-s)\, , &
c_{18,52}&= -\frac{\lambda_s-s}{2}\, , \nn
c_{20,52}&= \lambda_s-s\, , &
c_{21,52}&= m^2\frac{\lambda_s-s}{2}\, , \nn
c_{22,52}&= (m^2-2t)\frac{\lambda_s-s}{6}\, ,&
c_{23,52}&= m^2(m^2-t)\frac{\lambda_s-s}{6}\, ,\nn
c_{24,52}&= -\frac{(3m^2-4t+2s)\lambda_s-s(m^2-2u)}{6}\, ,&
c_{25,52}&= -\frac{(5m^2-5t+s)\lambda_s+s(u-m^2)}{6}\, ,\nn
c_{26,52}&= \frac{(m^2-2t+u)\lambda_s-s(u-m^2)}{4s}\, ,&
c_{27,52}&= m^2 \frac{(m^2-t)(\lambda_s-s)}{4s}\, ,\nn
c_{28,52}&= \frac{(m^2-t)(2m^2-s)(\lambda_s-s)}{8s}\, , &
c_{29,52}&= - \frac{3(m^2-t)(\lambda_s-s)}{4s}\, , \nn
c_{30,52}&=  -m^2 \frac{(m^2-t+s)\lambda_s+s(u-m^2)}{4s}\, , &
c_{31,52}&=  - (2m^2-s) \frac{(m^2-t+s)\lambda_s+s(u-m^2)}{8s}\, , \nn
c_{32,52}&=  (m^2-t) \frac{(2m^2-2t+s)\lambda_s+s(u-m^2)}{(u-m^2)s}\, ,&
c_{33,52}&= \frac{(3m^2-3t+s)\lambda_s+s(u-m^2)}{2(u-m^2)}\, ,\nn
c_{34,52}&= \frac{(t-u)\lambda_s+s(u-m^2)}{s}\, ,  &
c_{35,52}&= m^2 \frac{\lambda_s-s}{2}\, , \nn
c_{38,52}&= -(m^2-t)\frac{\lambda_s-s}{2s}\, , &
c_{39,52}&= m^2 (m^2-t) \frac{\lambda_s-s}{4s}\, , \nn
c_{40,52}&= m^2 \frac{(m^2-u+2s)\lambda_s+3s(u-m^2)}{4s}\, ,&
c_{41,52}&= (m^2-t) \frac{(m^2-u+s)\lambda_s+2s(u-m^2)}{4s}\, , \nn
c_{44,52}&= -\frac{(\lambda_s-s)(2m^2u+st)}{2s}\, , &
c_{45,52}&= - (m^2-t)\frac{\lambda_s-s}{2s}\, ,\nn
c_{46,52}&= (u-t) \frac{\lambda_s-s}{2s}\, , &
c_{49,52}&= -\frac{(3m^2-u)\lambda_s+u-m^2}{2}\, , \nn
c_{50,52}&= (m^2-t) \frac{\lambda_s-s}{2}\, , &
c_{51,52}&= \frac{\lambda_s-s}{2}\, , \nn
c_{52,52}&= \lambda_s\, . & & 
\end{align}

%%% Local Variables:
%%% TeX-master: "main"
%%% End:

\bibliographystyle{JHEP}
\bibliography{references}

\providecommand{\href}[2]{#2}\begingroup\raggedright\begin{thebibliography}{10}

\bibitem{Aad:2015mbv}
{\scshape ATLAS} collaboration, \emph{{Measurements of top-quark pair
  differential cross-sections in the lepton+jets channel in $pp$ collisions at
  $\sqrt{s}=8$ TeV using the ATLAS detector}},
  \href{https://doi.org/10.1140/epjc/s10052-016-4366-4}{\emph{Eur. Phys. J.}
  {\bfseries C76} (2016) 538}
  [\href{https://arxiv.org/abs/1511.04716}{{\ttfamily 1511.04716}}].

\bibitem{Aaboud:2017fha}
{\scshape ATLAS} collaboration, \emph{{Measurements of top-quark pair
  differential cross-sections in the lepton+jets channel in $pp$ collisions at
  $\sqrt{s}=13$ TeV using the ATLAS detector}},
  \href{https://doi.org/10.1007/JHEP11(2017)191}{\emph{JHEP} {\bfseries 11}
  (2017) 191} [\href{https://arxiv.org/abs/1708.00727}{{\ttfamily
  1708.00727}}].

\bibitem{Khachatryan:2015oqa}
{\scshape CMS} collaboration, \emph{{Measurement of the differential cross
  section for top quark pair production in pp collisions at $\sqrt{s} =
  8\,\text {TeV} $}},
  \href{https://doi.org/10.1140/epjc/s10052-015-3709-x}{\emph{Eur. Phys. J.}
  {\bfseries C75} (2015) 542}
  [\href{https://arxiv.org/abs/1505.04480}{{\ttfamily 1505.04480}}].

\bibitem{Sirunyan:2018wem}
{\scshape CMS} collaboration, \emph{{Measurement of differential cross sections
  for the production of top quark pairs and of additional jets in lepton+jets
  events from pp collisions at $\sqrt{s} =$ 13 TeV}},
  \href{https://doi.org/10.1103/PhysRevD.97.112003}{\emph{Phys. Rev.}
  {\bfseries D97} (2018) 112003}
  [\href{https://arxiv.org/abs/1803.08856}{{\ttfamily 1803.08856}}].

\bibitem{Czakon:2013goa}
M.~Czakon, P.~Fiedler and A.~Mitov, \emph{{Total Top-Quark Pair-Production
  Cross Section at Hadron Colliders Through $O(\alpha_S^4)$}},
  \href{https://doi.org/10.1103/PhysRevLett.110.252004}{\emph{Phys. Rev. Lett.}
  {\bfseries 110} (2013) 252004}
  [\href{https://arxiv.org/abs/1303.6254}{{\ttfamily 1303.6254}}].

\bibitem{Czakon:2014xsa}
M.~Czakon, P.~Fiedler and A.~Mitov, \emph{{Resolving the Tevatron Top Quark
  Forward-Backward Asymmetry Puzzle: Fully Differential
  Next-to-Next-to-Leading-Order Calculation}},
  \href{https://doi.org/10.1103/PhysRevLett.115.052001}{\emph{Phys. Rev. Lett.}
  {\bfseries 115} (2015) 052001}
  [\href{https://arxiv.org/abs/1411.3007}{{\ttfamily 1411.3007}}].

\bibitem{Czakon:2015owf}
M.~Czakon, D.~Heymes and A.~Mitov, \emph{{High-precision differential
  predictions for top-quark pairs at the LHC}},
  \href{https://doi.org/10.1103/PhysRevLett.116.082003}{\emph{Phys. Rev. Lett.}
  {\bfseries 116} (2016) 082003}
  [\href{https://arxiv.org/abs/1511.00549}{{\ttfamily 1511.00549}}].

\bibitem{Czakon:2016dgf}
M.~Czakon, D.~Heymes and A.~Mitov, \emph{{Dynamical scales for multi-TeV
  top-pair production at the LHC}},
  \href{https://doi.org/10.1007/JHEP04(2017)071}{\emph{JHEP} {\bfseries 04}
  (2017) 071} [\href{https://arxiv.org/abs/1606.03350}{{\ttfamily
  1606.03350}}].

\bibitem{Catani:2019iny}
S.~Catani, S.~Devoto, M.~Grazzini, S.~Kallweit, J.~Mazzitelli and H.~Sargsyan,
  \emph{{Top-quark pair hadroproduction at next-to-next-to-leading order in
  QCD}}, \href{https://doi.org/10.1103/PhysRevD.99.051501}{\emph{Phys. Rev.}
  {\bfseries D99} (2019) 051501}
  [\href{https://arxiv.org/abs/1901.04005}{{\ttfamily 1901.04005}}].

\bibitem{Baernreuther:2013caa}
P.~Barnreuther, M.~Czakon and P.~Fiedler, \emph{{Virtual amplitudes and
  threshold behaviour of hadronic top-quark pair-production cross sections}},
  \href{https://doi.org/10.1007/JHEP02(2014)078}{\emph{JHEP} {\bfseries 02}
  (2014) 078} [\href{https://arxiv.org/abs/1312.6279}{{\ttfamily 1312.6279}}].

\bibitem{Chen:2017jvi}
L.~Chen, M.~Czakon and R.~Poncelet, \emph{{Polarized double-virtual amplitudes
  for heavy-quark pair production}},
  \href{https://doi.org/10.1007/JHEP03(2018)085}{\emph{JHEP} {\bfseries 03}
  (2018) 085} [\href{https://arxiv.org/abs/1712.08075}{{\ttfamily
  1712.08075}}].

\bibitem{Bonciani:2008az}
R.~Bonciani, A.~Ferroglia, T.~Gehrmann, D.~Maitre and C.~Studerus,
  \emph{{Two-Loop Fermionic Corrections to Heavy-Quark Pair Production: The
  Quark-Antiquark Channel}},
  \href{https://doi.org/10.1088/1126-6708/2008/07/129}{\emph{JHEP} {\bfseries
  07} (2008) 129} [\href{https://arxiv.org/abs/0806.2301}{{\ttfamily
  0806.2301}}].

\bibitem{Bonciani:2009nb}
R.~Bonciani, A.~Ferroglia, T.~Gehrmann and C.~Studerus, \emph{{Two-Loop Planar
  Corrections to Heavy-Quark Pair Production in the Quark-Antiquark Channel}},
  \href{https://doi.org/10.1088/1126-6708/2009/08/067}{\emph{JHEP} {\bfseries
  08} (2009) 067} [\href{https://arxiv.org/abs/0906.3671}{{\ttfamily
  0906.3671}}].

\bibitem{Bonciani:2010mn}
R.~Bonciani, A.~Ferroglia, T.~Gehrmann, A.~von Manteuffel and C.~Studerus,
  \emph{{Two-Loop Leading Color Corrections to Heavy-Quark Pair Production in
  the Gluon Fusion Channel}},
  \href{https://doi.org/10.1007/JHEP01(2011)102}{\emph{JHEP} {\bfseries 01}
  (2011) 102} [\href{https://arxiv.org/abs/1011.6661}{{\ttfamily 1011.6661}}].

\bibitem{Bonciani:2013ywa}
R.~Bonciani, A.~Ferroglia, T.~Gehrmann, A.~von Manteuffel and C.~Studerus,
  \emph{{Light-quark two-loop corrections to heavy-quark pair production in the
  gluon fusion channel}},
  \href{https://doi.org/10.1007/JHEP12(2013)038}{\emph{JHEP} {\bfseries 12}
  (2013) 038} [\href{https://arxiv.org/abs/1309.4450}{{\ttfamily 1309.4450}}].

\bibitem{Abelof:2014fza}
G.~Abelof, A.~Gehrmann-De~Ridder, P.~Maierhofer and S.~Pozzorini, \emph{{NNLO
  QCD subtraction for top-antitop production in the $ q\overline{q} $
  channel}}, \href{https://doi.org/10.1007/JHEP08(2014)035}{\emph{JHEP}
  {\bfseries 08} (2014) 035} [\href{https://arxiv.org/abs/1404.6493}{{\ttfamily
  1404.6493}}].

\bibitem{Abelof:2015lna}
G.~Abelof, A.~Gehrmann-De~Ridder and I.~Majer, \emph{{Top quark pair production
  at NNLO in the quark-antiquark channel}},
  \href{https://doi.org/10.1007/JHEP12(2015)074}{\emph{JHEP} {\bfseries 12}
  (2015) 074} [\href{https://arxiv.org/abs/1506.04037}{{\ttfamily
  1506.04037}}].

\bibitem{vonManteuffel:2017hms}
A.~von Manteuffel and L.~Tancredi, \emph{{A non-planar two-loop three-point
  function beyond multiple polylogarithms}},
  \href{https://doi.org/10.1007/JHEP06(2017)127}{\emph{JHEP} {\bfseries 06}
  (2017) 127} [\href{https://arxiv.org/abs/1701.05905}{{\ttfamily
  1701.05905}}].

\bibitem{Adams:2018bsn}
L.~Adams, E.~Chaubey and S.~Weinzierl, \emph{{Planar Double Box Integral for
  Top Pair Production with a Closed Top Loop to all orders in the Dimensional
  Regularization Parameter}},
  \href{https://doi.org/10.1103/PhysRevLett.121.142001}{\emph{Phys. Rev. Lett.}
  {\bfseries 121} (2018) 142001}
  [\href{https://arxiv.org/abs/1804.11144}{{\ttfamily 1804.11144}}].

\bibitem{Adams:2018kez}
L.~Adams, E.~Chaubey and S.~Weinzierl, \emph{{Analytic results for the planar
  double box integral relevant to top-pair production with a closed top loop}},
  \href{https://doi.org/10.1007/JHEP10(2018)206}{\emph{JHEP} {\bfseries 10}
  (2018) 206} [\href{https://arxiv.org/abs/1806.04981}{{\ttfamily
  1806.04981}}].

\bibitem{Chen:2019zoy}
L.-B. Chen and J.~Wang, \emph{{Master integrals of a planar double-box family
  for top-quark pair production}},
  \href{https://doi.org/10.1016/j.physletb.2019.03.030}{\emph{Phys. Lett.}
  {\bfseries B792} (2019) 50}
  [\href{https://arxiv.org/abs/1903.04320}{{\ttfamily 1903.04320}}].

\bibitem{Calame:2015fva}
C.~M. Carloni~Calame, M.~Passera, L.~Trentadue and G.~Venanzoni, \emph{{A new
  approach to evaluate the leading hadronic corrections to the muon $g$-2}},
  \href{https://doi.org/10.1016/j.physletb.2015.05.020}{\emph{Phys. Lett.}
  {\bfseries B746} (2015) 325}
  [\href{https://arxiv.org/abs/1504.02228}{{\ttfamily 1504.02228}}].

\bibitem{Abbiendi:2016xup}
G.~Abbiendi et~al., \emph{{Measuring the leading hadronic contribution to the
  muon $g$-2 via $\mu e$ scattering}},
  \href{https://doi.org/10.1140/epjc/s10052-017-4633-z}{\emph{Eur. Phys. J.}
  {\bfseries C77} (2017) 139}
  [\href{https://arxiv.org/abs/1609.08987}{{\ttfamily 1609.08987}}].

\bibitem{Mastrolia:2017pfy}
P.~Mastrolia, M.~Passera, A.~Primo and U.~Schubert, \emph{{Master integrals for
  the NNLO virtual corrections to $\mu e$ scattering in QED: the planar
  graphs}}, \href{https://doi.org/10.1007/JHEP11(2017)198}{\emph{JHEP}
  {\bfseries 11} (2017) 198}
  [\href{https://arxiv.org/abs/1709.07435}{{\ttfamily 1709.07435}}].

\bibitem{DiVita:2018nnh}
S.~Di~Vita, S.~Laporta, P.~Mastrolia, A.~Primo and U.~Schubert, \emph{{Master
  integrals for the NNLO virtual corrections to $μe$ scattering in QED: the
  non-planar graphs}},
  \href{https://doi.org/10.1007/JHEP09(2018)016}{\emph{JHEP} {\bfseries 09}
  (2018) 016} [\href{https://arxiv.org/abs/1806.08241}{{\ttfamily
  1806.08241}}].

\bibitem{Henn:2013pwa}
J.~M. Henn, \emph{{Multiloop integrals in dimensional regularization made
  simple}},
  \href{https://doi.org/10.1103/PhysRevLett.110.251601}{\emph{Phys.Rev.Lett.}
  {\bfseries 110} (2013) 251601}
  [\href{https://arxiv.org/abs/1304.1806}{{\ttfamily 1304.1806}}].

\bibitem{Argeri:2014qva}
M.~Argeri, S.~Di~Vita, P.~Mastrolia, E.~Mirabella, J.~Schlenk et~al.,
  \emph{{Magnus and Dyson Series for Master Integrals}},
  \href{https://doi.org/10.1007/JHEP03(2014)082}{\emph{JHEP} {\bfseries 1403}
  (2014) 082} [\href{https://arxiv.org/abs/1401.2979}{{\ttfamily 1401.2979}}].

\bibitem{DiVita:2014pza}
S.~Di~Vita, P.~Mastrolia, U.~Schubert and V.~Yundin, \emph{{Three-loop master
  integrals for ladder-box diagrams with one massive leg}},
  \href{https://doi.org/10.1007/JHEP09(2014)148}{\emph{JHEP} {\bfseries 09}
  (2014) 148} [\href{https://arxiv.org/abs/1408.3107}{{\ttfamily 1408.3107}}].

\bibitem{Bonciani:2016ypc}
R.~Bonciani, S.~Di~Vita, P.~Mastrolia and U.~Schubert, \emph{{Two-Loop Master
  Integrals for the mixed EW-QCD virtual corrections to Drell-Yan scattering}},
  \href{https://doi.org/10.1007/JHEP09(2016)091}{\emph{JHEP} {\bfseries 09}
  (2016) 091} [\href{https://arxiv.org/abs/1604.08581}{{\ttfamily
  1604.08581}}].

\bibitem{DiVita:2017xlr}
S.~Di~Vita, P.~Mastrolia, A.~Primo and U.~Schubert, \emph{{Two-loop master
  integrals for the leading QCD corrections to the Higgs coupling to a $W$ pair
  and to the triple gauge couplings $ZWW$ and $\gamma^*WW$}},
  \href{https://doi.org/10.1007/JHEP04(2017)008}{\emph{JHEP} {\bfseries 04}
  (2017) 008} [\href{https://arxiv.org/abs/1702.07331}{{\ttfamily
  1702.07331}}].

\bibitem{Primo:2018zby}
A.~Primo, G.~Sasso, G.~Somogyi and F.~Tramontano, \emph{{Exact Top Yukawa
  corrections to Higgs boson decay into bottom quarks}},
  \href{https://doi.org/10.1103/PhysRevD.99.054013}{\emph{Phys. Rev.}
  {\bfseries D99} (2019) 054013}
  [\href{https://arxiv.org/abs/1812.07811}{{\ttfamily 1812.07811}}].

\bibitem{Tkachov:1981wb}
F.~V. Tkachov, \emph{{A Theorem on Analytical Calculability of Four Loop
  Renormalization Group Functions}},
  \href{https://doi.org/10.1016/0370-2693(81)90288-4}{\emph{Phys. Lett.}
  {\bfseries 100B} (1981) 65}.

\bibitem{Chetyrkin:1981qh}
K.~Chetyrkin and F.~Tkachov, \emph{{Integration by Parts: The Algorithm to
  Calculate beta Functions in 4 Loops}},
  \href{https://doi.org/10.1016/0550-3213(81)90199-1}{\emph{Nucl.Phys.}
  {\bfseries B192} (1981) 159}.

\bibitem{Laporta:2001dd}
S.~Laporta, \emph{{High precision calculation of multiloop Feynman integrals by
  difference equations}},
  \href{https://doi.org/10.1016/S0217-751X(00)00215-7}{\emph{Int.J.Mod.Phys.}
  {\bfseries A15} (2000) 5087}
  [\href{https://arxiv.org/abs/hep-ph/0102033}{{\ttfamily hep-ph/0102033}}].

\bibitem{Barucchi:1973zm}
G.~Barucchi and G.~Ponzano, \emph{{Differential equations for one-loop
  generalized feynman integrals}},
  \href{https://doi.org/10.1063/1.1666327}{\emph{J. Math. Phys.} {\bfseries 14}
  (1973) 396}.

\bibitem{Kotikov:1990kg}
A.~Kotikov, \emph{{Differential equations method: New technique for massive
  Feynman diagrams calculation}},
  \href{https://doi.org/10.1016/0370-2693(91)90413-K}{\emph{Phys.Lett.}
  {\bfseries B254} (1991) 158}.

\bibitem{Remiddi:1997ny}
E.~Remiddi, \emph{{Differential equations for Feynman graph amplitudes}},
  {\emph{Nuovo Cim.} {\bfseries A110} (1997) 1435}
  [\href{https://arxiv.org/abs/hep-th/9711188}{{\ttfamily hep-th/9711188}}].

\bibitem{Gehrmann:1999as}
T.~Gehrmann and E.~Remiddi, \emph{{Differential equations for two loop four
  point functions}},
  \href{https://doi.org/10.1016/S0550-3213(00)00223-6}{\emph{Nucl. Phys.}
  {\bfseries B580} (2000) 485}
  [\href{https://arxiv.org/abs/hep-ph/9912329}{{\ttfamily hep-ph/9912329}}].

\bibitem{Goncharov:polylog}
A.~Goncharov, \emph{{Polylogarithms in arithmetic and geometry}},
  {\emph{Proceedings of the International Congree of Mathematicians} {\bfseries
  1,2} (1995) 374}.

\bibitem{Remiddi:1999ew}
E.~Remiddi and J.~Vermaseren, \emph{{Harmonic polylogarithms}},
  \href{https://doi.org/10.1142/S0217751X00000367}{\emph{Int.J.Mod.Phys.}
  {\bfseries A15} (2000) 725}
  [\href{https://arxiv.org/abs/hep-ph/9905237}{{\ttfamily hep-ph/9905237}}].

\bibitem{Gehrmann:2001pz}
T.~Gehrmann and E.~Remiddi, \emph{{Numerical evaluation of harmonic
  polylogarithms}},
  \href{https://doi.org/10.1016/S0010-4655(01)00411-8}{\emph{Comput.Phys.Commun.}
  {\bfseries 141} (2001) 296}
  [\href{https://arxiv.org/abs/hep-ph/0107173}{{\ttfamily hep-ph/0107173}}].

\bibitem{Vollinga:2004sn}
J.~Vollinga and S.~Weinzierl, \emph{{Numerical evaluation of multiple
  polylogarithms}},
  \href{https://doi.org/10.1016/j.cpc.2004.12.009}{\emph{Comput.Phys.Commun.}
  {\bfseries 167} (2005) 177}
  [\href{https://arxiv.org/abs/hep-ph/0410259}{{\ttfamily hep-ph/0410259}}].

\bibitem{Lee:2019lno}
R.~N. Lee and K.~T. Mingulov, \emph{{Master integrals for two-loop $C$-odd
  contribution to $e^+e^-\to \ell^+\ell^-$ process}},
  \href{https://arxiv.org/abs/1901.04441}{{\ttfamily 1901.04441}}.

\bibitem{Maierhoefer2018a}
P.~Maierhofer, J.~Usovitsch and P.~Uwer, \emph{Kira—a feynman integral
  reduction program},
  \href{https://doi.org/10.1016/j.cpc.2018.04.012}{\emph{Comput. Phys. Commun.}
  {\bfseries 230} (2018) 99}
  [\href{https://arxiv.org/abs/1705.05610}{{\ttfamily 1705.05610}}].

\bibitem{Lee:2012cn}
R.~N. Lee, \emph{{Presenting LiteRed: a tool for the Loop InTEgrals
  REDuction}},  \href{https://arxiv.org/abs/1212.2685}{{\ttfamily 1212.2685}}.

\bibitem{Lee:2013mka}
R.~N. Lee, \emph{{LiteRed 1.4: a powerful tool for reduction of multiloop
  integrals}}, \href{https://doi.org/10.1088/1742-6596/523/1/012059}{\emph{J.
  Phys. Conf. Ser.} {\bfseries 523} (2014) 012059}
  [\href{https://arxiv.org/abs/1310.1145}{{\ttfamily 1310.1145}}].

\bibitem{vonManteuffel:2012np}
A.~von Manteuffel and C.~Studerus, \emph{{Reduze 2 - Distributed Feynman
  Integral Reduction}},  \href{https://arxiv.org/abs/1201.4330}{{\ttfamily
  1201.4330}}.

\bibitem{vonManteuffel:2014qoa}
A.~von Manteuffel, E.~Panzer and R.~M. Schabinger, \emph{{A quasi-finite basis
  for multi-loop Feynman integrals}},
  \href{https://doi.org/10.1007/JHEP02(2015)120}{\emph{JHEP} {\bfseries 02}
  (2015) 120} [\href{https://arxiv.org/abs/1411.7392}{{\ttfamily 1411.7392}}].

\bibitem{Bauer:2000cp}
C.~W. Bauer, A.~Frink and R.~Kreckel, \emph{{Introduction to the GiNaC
  framework for symbolic computation within the C++ programming language}},
  \href{https://arxiv.org/abs/cs/0004015}{{\ttfamily cs/0004015}}.

\bibitem{Borowka:2015mxa}
S.~Borowka, G.~Heinrich, S.~P. Jones, M.~Kerner, J.~Schlenk and T.~Zirke,
  \emph{{SecDec-3.0: numerical evaluation of multi-scale integrals beyond one
  loop}}, \href{https://doi.org/10.1016/j.cpc.2015.05.022}{\emph{Comput. Phys.
  Commun.} {\bfseries 196} (2015) 470}
  [\href{https://arxiv.org/abs/1502.06595}{{\ttfamily 1502.06595}}].

\bibitem{Becchetti:2019tjy}
M.~Becchetti, R.~Bonciani, V.~Casconi, A.~Ferroglia, S.~Lavacca and A.~von
  Manteuffel, \emph{{Master Integrals for the two-loop, non-planar QCD
  corrections to top-quark pair production in the quark-annihilation channel}},
   \href{https://arxiv.org/abs/1904.10834}{{\ttfamily 1904.10834}}.

\bibitem{Collins2016}
J.~C. Collins and J.~A.~M. Vermaseren, \emph{Axodraw version 2},
  \href{https://arxiv.org/abs/1606.01177}{{\ttfamily 1606.01177}}.

\bibitem{Gehrmann:2001ck}
T.~Gehrmann and E.~Remiddi, \emph{{Two loop master integrals for $\gamma^{*}
  \to$ 3 jets: The Nonplanar topologies}},
  \href{https://doi.org/10.1016/S0550-3213(01)00074-8}{\emph{Nucl. Phys.}
  {\bfseries B601} (2001) 287}
  [\href{https://arxiv.org/abs/hep-ph/0101124}{{\ttfamily hep-ph/0101124}}].

\bibitem{Bonciani:2003te}
R.~Bonciani, P.~Mastrolia and E.~Remiddi, \emph{{Vertex diagrams for the QED
  form-factors at the two loop level}},
  \href{https://doi.org/10.1016/j.nuclphysb.2004.08.009}{\emph{Nucl.Phys.}
  {\bfseries B661} (2003) 289}
  [\href{https://arxiv.org/abs/hep-ph/0301170}{{\ttfamily hep-ph/0301170}}].

\bibitem{Bonciani:2003hc}
R.~Bonciani, P.~Mastrolia and E.~Remiddi, \emph{{Master integrals for the two
  loop QCD virtual corrections to the forward backward asymmetry}},
  \href{https://doi.org/10.1016/j.nuclphysb.2004.04.011}{\emph{Nucl. Phys.}
  {\bfseries B690} (2004) 138}
  [\href{https://arxiv.org/abs/hep-ph/0311145}{{\ttfamily hep-ph/0311145}}].

\bibitem{Bonciani:2008wf}
R.~Bonciani and A.~Ferroglia, \emph{{Two-Loop QCD Corrections to the
  Heavy-to-Light Quark Decay}},
  \href{https://doi.org/10.1088/1126-6708/2008/11/065}{\emph{JHEP} {\bfseries
  11} (2008) 065} [\href{https://arxiv.org/abs/0809.4687}{{\ttfamily
  0809.4687}}].

\bibitem{Asatrian:2008uk}
H.~M. Asatrian, C.~Greub and B.~D. Pecjak, \emph{{NNLO corrections to $ {\bar
  B} \to X_u \ell {\bar \nu}$ in the shape-function region}},
  \href{https://doi.org/10.1103/PhysRevD.78.114028}{\emph{Phys. Rev.}
  {\bfseries D78} (2008) 114028}
  [\href{https://arxiv.org/abs/0810.0987}{{\ttfamily 0810.0987}}].

\bibitem{Beneke:2008ei}
M.~Beneke, T.~Huber and X.~Q. Li, \emph{{Two-loop QCD correction to
  differential semi-leptonic b $\to$ u decays in the shape-function region}},
  \href{https://doi.org/10.1016/j.nuclphysb.2008.11.019}{\emph{Nucl. Phys.}
  {\bfseries B811} (2009) 77}
  [\href{https://arxiv.org/abs/0810.1230}{{\ttfamily 0810.1230}}].

\bibitem{Bell:2008ws}
G.~Bell, \emph{{NNLO corrections to inclusive semileptonic B decays in the
  shape-function region}},
  \href{https://doi.org/10.1016/j.nuclphysb.2008.12.018}{\emph{Nucl. Phys.}
  {\bfseries B812} (2009) 264}
  [\href{https://arxiv.org/abs/0810.5695}{{\ttfamily 0810.5695}}].

\bibitem{Huber:2009se}
T.~Huber, \emph{{On a two-loop crossed six-line master integral with two
  massive lines}},
  \href{https://doi.org/10.1088/1126-6708/2009/03/024}{\emph{JHEP} {\bfseries
  03} (2009) 024} [\href{https://arxiv.org/abs/0901.2133}{{\ttfamily
  0901.2133}}].

\bibitem{Manteuffel2013}
A.~von Manteuffel and C.~Studerus, \emph{Massive planar and non-planar double
  box integrals for light-nf contributions to gg $\to $tt},
  \href{https://doi.org/10.1007/JHEP10(2013)037}{\emph{JHEP} {\bfseries 10}
  (2013) 037} [\href{https://arxiv.org/abs/1306.3504}{{\ttfamily 1306.3504}}].

\bibitem{Henn:2014qga}
J.~M. Henn, \emph{{Lectures on differential equations for Feynman integrals}},
  \href{https://doi.org/10.1088/1751-8113/48/15/153001}{\emph{J. Phys.}
  {\bfseries A48} (2015) 153001}
  [\href{https://arxiv.org/abs/1412.2296}{{\ttfamily 1412.2296}}].

\bibitem{Gehrmann:2000zt}
T.~Gehrmann and E.~Remiddi, \emph{{Two loop master integrals for $\gamma^*\to$
  3 jets: The Planar topologies}},
  \href{https://doi.org/10.1016/S0550-3213(01)00057-8}{\emph{Nucl.Phys.}
  {\bfseries B601} (2001) 248}
  [\href{https://arxiv.org/abs/hep-ph/0008287}{{\ttfamily hep-ph/0008287}}].

\bibitem{Gehrmann:2002zr}
T.~Gehrmann and E.~Remiddi, \emph{{Analytic continuation of massless two loop
  four point functions}},
  \href{https://doi.org/10.1016/S0550-3213(02)00569-2}{\emph{Nucl.Phys.}
  {\bfseries B640} (2002) 379}
  [\href{https://arxiv.org/abs/hep-ph/0207020}{{\ttfamily hep-ph/0207020}}].

\bibitem{Tausk:1999vh}
J.~Tausk, \emph{{Nonplanar massless two loop Feynman diagrams with four
  on-shell legs}},
  \href{https://doi.org/10.1016/S0370-2693(99)01277-0}{\emph{Phys.Lett.}
  {\bfseries B469} (1999) 225}
  [\href{https://arxiv.org/abs/hep-ph/9909506}{{\ttfamily hep-ph/9909506}}].

\bibitem{Gaiotto:2011dt}
D.~Gaiotto, J.~Maldacena, A.~Sever and P.~Vieira, \emph{{Pulling the straps of
  polygons}}, \href{https://doi.org/10.1007/JHEP12(2011)011}{\emph{JHEP}
  {\bfseries 12} (2011) 011} [\href{https://arxiv.org/abs/1102.0062}{{\ttfamily
  1102.0062}}].

\bibitem{Abreu:2015zaa}
S.~Abreu, R.~Britto and H.~Gronqvist, \emph{{Cuts and coproducts of massive
  triangle diagrams}},
  \href{https://doi.org/10.1007/JHEP07(2015)111}{\emph{JHEP} {\bfseries 07}
  (2015) 111} [\href{https://arxiv.org/abs/1504.00206}{{\ttfamily
  1504.00206}}].

\bibitem{Tarasov:1996br}
O.~V. Tarasov, \emph{{Connection between Feynman integrals having different
  values of the space-time dimension}},
  \href{https://doi.org/10.1103/PhysRevD.54.6479}{\emph{Phys. Rev.} {\bfseries
  D54} (1996) 6479} [\href{https://arxiv.org/abs/hep-th/9606018}{{\ttfamily
  hep-th/9606018}}].

\bibitem{Lee:2009dh}
R.~N. Lee, \emph{{Space-time dimensionality D as complex variable: Calculating
  loop integrals using dimensional recurrence relation and analytical
  properties with respect to D}},
  \href{https://doi.org/10.1016/j.nuclphysb.2009.12.025}{\emph{Nucl. Phys.}
  {\bfseries B830} (2010) 474}
  [\href{https://arxiv.org/abs/0911.0252}{{\ttfamily 0911.0252}}].

\end{thebibliography}\endgroup

\end{document}